\documentclass[%
 reprint,
 groupedaddress,
 amsmath,amssymb,
 aps,
 pra
]{revtex4-1}

\usepackage[pdftex]{graphicx, color}
\usepackage[pdftex]{hyperref}
\usepackage{bm}
\usepackage{braket}
\let\Re\relax
\let\Im\relax
\DeclareMathOperator{\Re}{Re}
\DeclareMathOperator{\Im}{Im}

\begin{document}

\title{Unconventional superconducting gap structure protected by space group symmetry}

\author{Shuntaro Sumita}
\email[]{s.sumita@scphys.kyoto-u.ac.jp}
\affiliation{%
 Department of Physics, Kyoto University, Kyoto 606-8502, Japan
}%

\author{Youichi Yanase}
\affiliation{%
 Department of Physics, Kyoto University, Kyoto 606-8502, Japan
}%


\date{\today}

\begin{abstract}
Recent superconducting gap classifications based on space group symmetry have revealed nontrivial gap structures that were not shown by point group symmetry.
First, we review a comprehensive classification of symmetry-protected line nodes within the range of centrosymmetric space groups.
Next, we show an additional constraint; line nodes peculiar to nonsymmorphic systems appear only for primitive or orthorhombic base-centered Bravais lattice.
Then, we list useful classification tables of 59 primitive or orthorhombic base-centered space groups for the superconducting gap structures.
Furthermore, our gap classification reveals the \textit{$j_z$-dependent point nodes (gap opening)} appearing on a 3- or 6-fold axis, which means that the presence (absence) of point nodes depends on the Bloch-state angular momentum $j_z$.
We suggest that this unusual gap structure is realized in a heavy-fermion superconductor UPt$_3$, using a group-theoretical analysis and a numerical calculation.
The calculation demonstrates that a Bloch phase contributes to $j_z$ as \textit{effective orbital angular momentum} by site permutation.
We also discuss superconducting gap structures in MoS$_2$, SrPtAs, UBe$_{13}$, and PrOs$_4$Sb$_{12}$.
\end{abstract}


\maketitle


\section{Introduction}
Classification of a superconducting gap is one of the central subjects in the research field of unconventional superconductivity.
The momentum dependence of the superconducting gap is closely related to the symmetry of superconductivity and the pairing mechanism.
Since the superconducting gap structure can be identified by various experiments~\cite{Matsuda2006, Sakakibara2016}, combined studies of the superconducting gap by theory and experiment may clarify the characteristics of superconductivity.
Most of the previous studies have been based on the classification of an order parameter by the crystal point group~\cite{Volovik1984, Volovik1985, Anderson1984}, which was summarized by Sigrist and Ueda (called the Sigrist-Ueda method in this paper)~\cite{Sigrist-Ueda}.
However, their classification may not provide a precise result of the superconducting gap.

The results of the Sigrist-Ueda method may not be precise because of the following two reasons.
First, the order parameter obtained by the method does not appropriately indicate the superconducting gap structure.
Second, the space group symmetry is not taken into account in the method.
Since a space group is given by the combination of a point group and a translation group, it provides us with more information than the point group.
The difficulties of the Sigrist-Ueda method are resolved by directly classifying the superconducting gap on the basis of the space group symmetry.
Indeed, several studies have shown that the space group symmetry ensures the unconventional gap structures beyond the results of the Sigrist-Ueda method~\cite{Yarzhemsky1992, Norman1995, Yarzhemsky1998, Yarzhemsky2000, Yarzhemsky2003, Yarzhemsky2008, Micklitz2009, Kobayashi2016, Yanase2016, Nomoto2016_PRL, Micklitz2017_PRB, Nomoto2017, Micklitz2017_PRL, Sumita2017}.

In 1985, Blount showed that no symmetry-protected line node exists in odd-parity superconductors~\cite{Blount1985}.
The Sigrist-Ueda method is consistent with Blount's theorem.
After that, however, some studies showed a counter-example, namely, a line node in nonsymmorphic systems, and indeed suggested a line node protected by a nonsymmorphic space group symmetry in UPt$_3$~\cite{Norman1995, Micklitz2009, Kobayashi2016, Yanase2016, Nomoto2016_PRL, Micklitz2017_PRB}.
At present, it is known that Blount's theorem holds only in symmorphic crystals.
The essence is that the nonsymmorphic symmetry causes the difference in the group-theoretical representation of gap functions between the basal planes (BPs) and the zone faces (ZFs) in the Brillouin zone (BZ).
Although the Sigrist-Ueda method appropriately implies the gap functions on the BPs, it may fail to show those on the ZFs.
Indeed, such unconventional gap structures on the ZFs have recently been revealed in various superconductors, not only UPt$_3$~\cite{Norman1995, Micklitz2009, Kobayashi2016, Yanase2016, Nomoto2016_PRL, Micklitz2017_PRB}, but also UCoGe~\cite{Nomoto2017}, UPd$_2$Al$_3$~\cite{Nomoto2017, Micklitz2017_PRL}, and Sr$_2$IrO$_4$~\cite{Sumita2017}.

Regarding point nodes, on the other hand, Weyl nodes in superconductors, namely point nodes protected by a nontrivial topological number, have been intensively investigated~\cite{Daido2016, Yanase2016, Yanase2017, Kozii2016, Venderbos2016, Sau2012, Goswami2015, Fischer2014, Meng2012, Yang2014, Volovik2017}.
However, there are only a few and less known results about point nodes connected with crystal symmetry~\cite{Yarzhemsky1992, Yarzhemsky1998, Yarzhemsky2000, Yarzhemsky2003, Yarzhemsky2008}.

In this paper, we classify unconventional line nodes and point nodes beyond the results of the Sigrist-Ueda method using the group-theoretical analysis of the superconducting gap.
First, we review the results of symmetry-protected line nodes~\cite{Norman1995, Micklitz2009, Kobayashi2016, Yanase2016, Nomoto2016_PRL, Micklitz2017_PRB, Nomoto2017, Micklitz2017_PRL, Sumita2017}, clarifying the condition for the existence of line nodes protected by nonsymmorphic symmetry.
Next we show our original and useful results;
nonsymmorphic-symmetry-protected line nodes appear only on the ZF of a primitive or orthorhombic base-centered Bravais lattice.
We classify all space groups under the additional constraint.
Second, we consider the gap structures on high-symmetry $n$-fold ($n = 2$, $3$, $4$, and $6$) axes in the BZ, and we elucidate the symmetry-protected point nodes.
Surprisingly, the analysis shows the existence of point nodes depending on the Bloch-state angular momentum $j_z$, and we suggest such \textit{$j_z$-dependent point nodes} in UPt$_3$, MoS$_2$, SrPtAs, UBe$_{13}$, and PrOs$_4$Sb$_{12}$.

This paper is constructed as follows.
We introduce the method of superconducting gap classification based on space group symmetry in Sec.~\ref{sec:gap_classification}.
In Sec.~\ref{sec:line_node}, we show that the condition for unconventional nonsymmorphic line nodes is 2-fold screw symmetry and/or antiferromagnetic (AFM) order with a translation vector along the 2-fold axis.
Then, we classify 59 space groups for the superconducting gap structures.
Next, we show $j_z$-dependent point nodes on $3$- and $6$-fold axes (not on $2$- and $4$-fold axes) in Sec.~\ref{sec:point_node}.
Furthermore, we discuss the presence or absence of such point nodes in hexagonal superconductors UPt$_3$, MoS$_2$, and SrPtAs, and in cubic superconductors UBe$_{13}$ and PrOs$_4$Sb$_{12}$.
Finally, a brief summary and discussion are given in Sec.~\ref{sec:summary}.

\section{Classification theory of superconducting gap}
\label{sec:gap_classification}
Let us briefly introduce the superconducting gap classification based on the space group symmetry for background knowledge of the following Sections.
First, we focus on a magnetic space group $M$, and we restrict $M$ to the high-symmetry $\bm{k}$-point in the BZ.
Then, we define $\gamma^{\bm{k}}(m)$ as a small representation of symmetry operations $m \in {\cal M}^{\bm{k}}$, where ${\cal M}^{\bm{k}} \subset M$ is the little group leaving $\bm{k}$ invariant modulo a reciprocal lattice vector.
$\gamma^{\bm{k}}$ represents the Bloch state with the crystal momentum $\bm{k}$.
In the superconducting state, zero center-of-mass momentum Cooper pairs have to be formed between degenerate states present at $\bm{k}$ and $- \bm{k}$ in the same band when we adopt the weak-coupling BCS theory.
Then, the two states should be connected by some symmetry operations, such as spacial inversion, except for an accidentally degenerate case.
As a result, the representation of Cooper pair wave functions $P^{\bm{k}}$ can be constructed from the representations of the Bloch state $\gamma^{\bm{k}}$.

Next, we calculate the representation of the Cooper pair wave functions $P^{\bm{k}}$.
Let us consider the space group operation $d = \{p_d | \bm{a}_d\} \in M$, where $p_d$ satisfies $p_d \bm{k} \equiv - \bm{k}$ modulo a reciprocal-lattice vector.
Note that the notation $\{p | \bm{a}\}$ is a conventional Seitz space group symbol with a point-group operation $p$ and a translation $\bm{a}$.
The operation $d$ connects two states of the paired electrons.
Although the choice of $d$ is not unique, the magnetic space group of Cooper pair wave functions $\widetilde{\cal M}^{\bm{k}} = {\cal M}^{\bm{k}} + d {\cal M}^{\bm{k}}$ is independent of the choice of $d$.
Therefore, we fix $d$ to a spatial inversion $\{I | \bm{0}\}$ in the following discussion.
Taking into account the antisymmetry of the Cooper pairs and the degeneracy of the two states, we can regard $P^{\bm{k}}$ as an antisymmetrized Kronecker square~\cite{Bradley, Bradley1970}, with zero total momentum, of the induced representation $\gamma^{\bm{k}} \uparrow \widetilde{\cal M}^{\bm{k}}$.
This is obtained in a systematic way by using the double coset decomposition and the corresponding Mackey-Bradley theorem~\cite{Bradley, Bradley1970, Mackey1953},
\begin{subequations}
 \label{eq:Mackey-Bradley}
 \begin{align}
  \chi[P^{\bm{k}}(m)] &= \chi[\gamma^{\bm{k}}(m)] \chi[\gamma^{\bm{k}}(I m I)], \label{eq:Mackey-Bradley_a} \\
  \chi[P^{\bm{k}}(I m)] &= - \chi[\gamma^{\bm{k}}(I m I m)], \label{eq:Mackey-Bradley_b}
 \end{align}
\end{subequations}
where $m \in {\cal M}^{\bm{k}}$, and $\chi$ are characters of the representation.
A proof of the Mackey-Bradley theorem in the context of Cooper pair wave functions is shown in Appendix~\ref{app:Mackey-Bradley}.

Finally, we reduce $P^{\bm{k}}$ into irreducible representations (IRs) of the original crystal point group.
The gap functions should be zero, and thus, the gap nodes appear, if the corresponding IRs do not exist in the results of reductions~\cite{Izyumov1989, Yarzhemsky1992, Yarzhemsky1998}.
Otherwise, the superconducting gap will open in general.
Therefore, the representation of Cooper pair wave functions $P^{\bm{k}}$ tells us the presence or absence of superconducting gap nodes.

Now we comment on the validity of the above method.
Our method of gap classification does not take into account spontaneous symmetry breaking (SSB) in the superconducting state, because the method assumes that all symmetry operations of the normal Bloch state ${\cal M}^{\bm{k}}$ remain in the magnetic space group of Cooper pair wave functions $\widetilde{\cal M}^{\bm{k}} = {\cal M}^{\bm{k}} + I {\cal M}^{\bm{k}}$.
Thus, the method is valid for all the superconducting states in a one-dimensional representation.
However, an application to chiral superconducting states, which spontaneously breaks time-reversal symmetry, is not straightforward.
Later, we discuss implications for this case.
On the other hand, we can treat ferromagnetic superconductors, in which time-reversal symmetry is originally broken in the normal state and ${\cal M}^{\bm{k}}$ does not contain the time-reversal symmetry.

\section{Complete classification of symmetry-protected line nodes}
\label{sec:line_node}
In this section, we review the classification results of symmetry-protected line nodes on mirror- or glide-invariant planes~\cite{Norman1995, Micklitz2009, Kobayashi2016, Yanase2016, Nomoto2016_PRL, Micklitz2017_PRB, Nomoto2017, Micklitz2017_PRL, Sumita2017}, clarifying the condition for the presence or absence of line nodes protected by nonsymmorphic symmetry.
Furthermore, we show an additional constraint: line nodes (gap opening) peculiar to nonsymmorphic systems appear only on the ZF for \textit{primitive or orthorhombic base-centered space groups}.
We provide classification tables (Table~\ref{tab:spacegroup}) of 59 such space groups, which may allow nontrivial superconducting gap structures by nonsymmorphic symmetry.

First, we show constraints on crystal symmetry of the system where the formalism in Sec.~\ref{sec:gap_classification} is applicable.
Since this method requires the symmetry operation connecting $\bm{k}$ and $- \bm{k}$, the system must be invariant under the spatial inversion $I$.
The symmetry-protected line node may appear on the high-symmetry $\bm{k}$-planes when a Fermi surface crosses the plane and the gap function vanishes there.
Thus, we need to consider $\bm{k}$-planes as high-symmetry $\bm{k}$-points in order to discuss line nodes.
Only an identity operation and a mirror reflection (or a glide reflection) are allowed as elements of the unitary little group for the general point on the high-symmetry $\bm{k}$-planes.

\begin{figure}[tbp]
 \centering
 \includegraphics[width=8cm, clip]{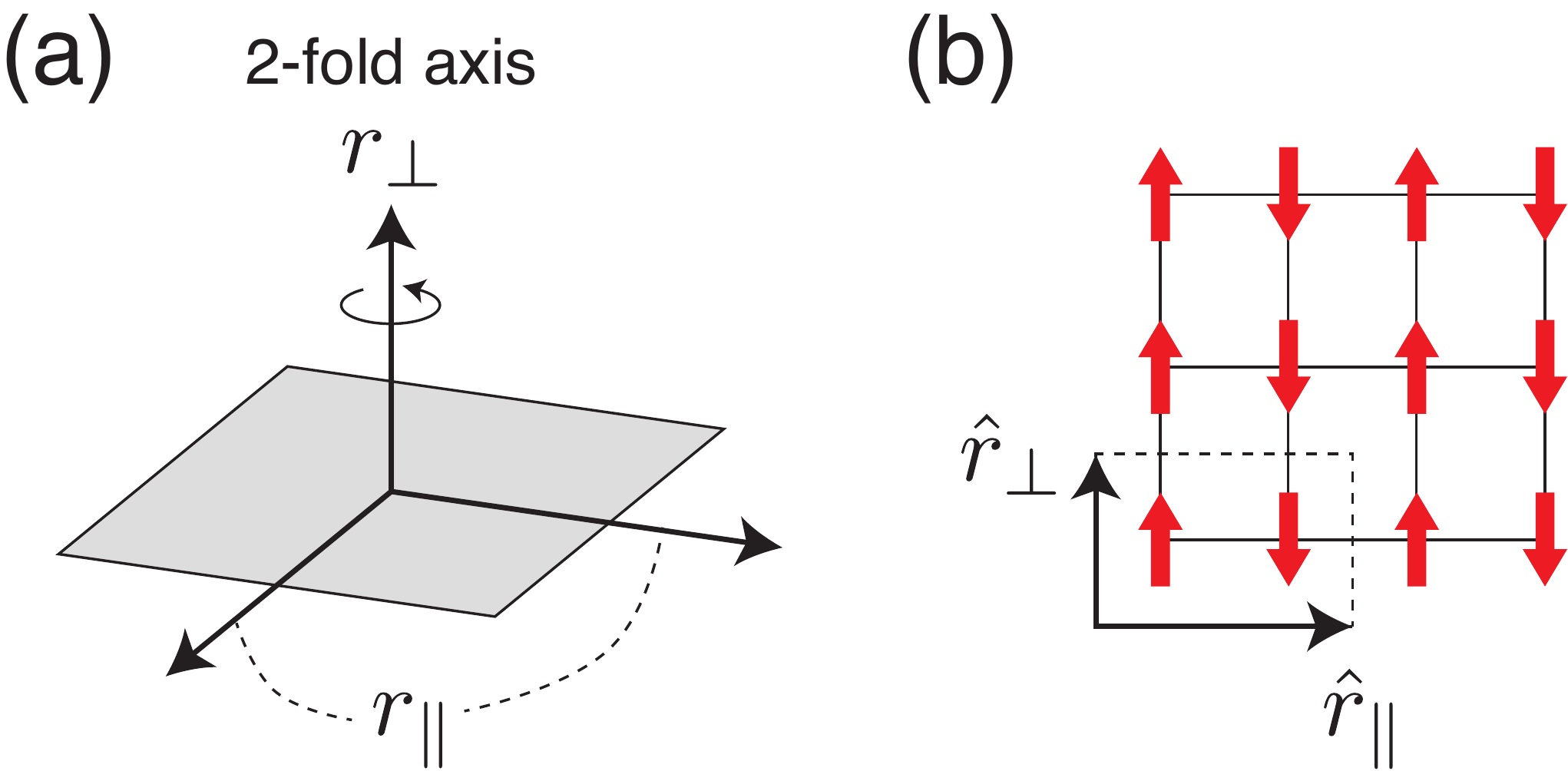}
 \caption{(a) The coordinate $r_\perp$ along the 2-fold axis, and the other coordinates $r_\parallel$ perpendicular to the 2-fold axis. (b) An example of the AFM1 case. The red arrows illustrate magnetic moments on a square lattice, and the dashed line indicates a magnetic unit cell.}
 \label{fig:linenode_system}
\end{figure}

For the above reasons, we assume that the symmetry of the system contains point group $C_{2h}$, which is generated by a spatial inversion and a mirror reflection.
In other words, the space group of the system $G$ has a subgroup $H \subset G$, which is $C_{2h}$ within the point group symmetry.
Taking translations into account, $H$ is classified as follows:
\begin{widetext}
 \begin{equation}
  H = \begin{cases}
       \{E | \bm{0}\} T + \{I | \bm{0}\} T + \{C_{2\perp} | \bm{0}\} T + \{\sigma_\perp | \bm{0}\} T & \text{(RM) Rotation + Mirror,} \\
       \{E | \bm{0}\} T + \{I | \bm{0}\} T + \{C_{2\perp} | \bm{\tau}_\parallel\} T + \{\sigma_\perp | \bm{\tau}_\parallel\} T & \text{(RG) Rotation + Glide,} \\
       \{E | \bm{0}\} T + \{I | \bm{0}\} T + \{C_{2\perp} | \bm{\tau}_\perp\} T + \{\sigma_\perp | \bm{\tau}_\perp\} T & \text{(SM) Screw + Mirror,} \\
       \{E | \bm{0}\} T + \{I | \bm{0}\} T + \{C_{2\perp} | \bm{\tau}_\parallel + \bm{\tau}_\perp\} T + \{\sigma_\perp | \bm{\tau}_\parallel + \bm{\tau}_\perp\} T & \text{(SG) Screw + Glide.}
      \end{cases}
      \label{eq:linenode_spacegroup}
 \end{equation}
\end{widetext}
The translation group $T$ defines a Bravais Lattice, and $\bm{\tau}_i$ are non-primitive translation vectors.
$E$ denotes an identity operation, $C_{2\perp}$ is a $\pi$-rotation around the $r_\perp$ axis, and $\sigma_\perp$ is a mirror reflection about the plane perpendicular to the $r_\perp$ axis.
Note that the direction of the 2-fold axis is represented by a symbol $\perp$, while the other directions orthogonal to the 2-fold axis are represented by a symbol $\parallel$ [Fig.~\ref{fig:linenode_system}(a)].
In Eq.~\eqref{eq:linenode_spacegroup}, the (RM) space group is symmorphic, and the other (RG), (SM), and (SG) space groups are nonsymmorphic.

Next, we discuss the magnetic (anti-unitary) symmetry of the system.
When the system is ferromagnetic (FM), all the time-reversal operation is forbidden.
On the other hand, in the paramagnetic (PM) or AFM state, the system is invariant under the anti-unitary operation $\tilde{\theta}$:
\begin{equation}
 \tilde{\theta} = \begin{cases}
                   \{\theta | \bm{0}\} & \text{(PM),} \\
                   \{\theta | \bm{\tau}_\parallel\} & \text{(AFM1),} \\
                   \{\theta | \bm{\tau}_\perp\} & \text{(AFM2),} \\
                   \{\theta | \bm{\tau}_\parallel + \bm{\tau}_\perp\} & \text{(AFM3),}
                  \end{cases}
                  \label{eq:linenode_antiunitary}
\end{equation}
where $\theta$ is the pure time-reversal operation.
The pure time-reversal operation is allowed in the (PM) state, while the system is invariant under the successive operations of time-reversal and non-primitive translation in the (AFM1)-(AFM3) states.
For example, a magnetic structure of the (AFM1) state is shown in Fig.~\ref{fig:linenode_system}(b): although magnetic moments (red arrows) flip under the time-reversal operation, the magnetic structure recovers after a half-translation $\bm{\tau}_\parallel = \hat{r}_\parallel / 2$.

\begin{table*}[htbp]
 \caption{Classification of superconducting gap on high-symmetry $\bm{k}$-plane. $H$ and $\tilde{\theta}$ specify the space group by Eqs.~\eqref{eq:linenode_spacegroup} and \eqref{eq:linenode_antiunitary}. The representations of Cooper pairs allowed on the high-symmetry $\bm{k}$-planes (BP and ZF) are shown. Materials realizing the space groups are also shown.}
 \label{tab:classification_linenode}
 \begin{center}
  \begin{tabular}{ccccc} \hline\hline
   $H$ & $\tilde{\theta}$ & BP ($k_\perp = 0$) & ZF ($k_\perp = \pi$) & Material \\ \hline
   (RM), (RG) & N/A & $A_u$ & $A_u$ & \\
   (SM), (SG) & N/A & $A_u$ & $B_u$ & UCoGe (FM)~\cite{Nomoto2017}, URhGe~\cite{Nomoto2017} \\
   (RM), (RG) & (PM), (AFM1) & $A_g + 2 A_u + B_u$ & $A_g + 2 A_u + B_u$ & \\
   (RM), (RG) & (AFM2), (AFM3) & $A_g + 2 A_u + B_u$ & $B_g + 3 A_u$ & Sr$_2$IrO$_4$ (vertical)~\cite{Sumita2017}, UPt$_3$ (AFM)~\cite{Micklitz2017_PRL} \\
   (SM), (SG) & (PM), (AFM1) & $A_g + 2 A_u + B_u$ & $A_g + 3 B_u$ & UPt$_3$ (PM)~\cite{Norman1995, Micklitz2009, Kobayashi2016, Yanase2016, Nomoto2016_PRL, Micklitz2017_PRB}, UCoGe (PM)~\cite{Nomoto2017}, CrAs~\cite{Micklitz2017_PRL} \\
   (SM), (SG) & (AFM2), (AFM3) & $A_g + 2 A_u + B_u$ & $B_g + A_u + 2 B_u$ & UPd$_2$Al$_3$~\cite{Fujimoto2006, Nomoto2017, Micklitz2017_PRL}, UNi$_2$Al$_3$~\cite{Micklitz2017_PRL}, Sr$_2$IrO$_4$ (horizontal)~\cite{Sumita2017} \\ \hline\hline
  \end{tabular}
 \end{center}
\end{table*}

By adding the anti-unitary operators \eqref{eq:linenode_antiunitary} to the unitary space group \eqref{eq:linenode_spacegroup}, we can construct the magnetic space group $M = H$ (FM) or $M = H + \tilde{\theta} H$ (PM or AFM).
Based on the magnetic space group, the gap classification introduced in Sec.~\ref{sec:gap_classification} is applied to high-symmetry, namely, mirror- or glide-invariant planes in the BZ: the BP at $k_\perp = 0$ and the ZF at $k_\perp = \pi$.
The obtained results are summarized in Table~\ref{tab:classification_linenode}.
In this table, the representations of superconducting gap functions are classified by the IRs of point group $C_{2h}$.

The classification of a superconducting gap on BPs is consistent with the Sigrist-Ueda method.
On the other hand, the representations allowed on the ZF may differ from those on the BP.
Then, the line nodes (or gap opening) protected by nonsymmorphic symmetry appear on the ZF.
Such a situation is realized when the space group is (SM) or (SG), and/or the pseudo-time-reversal symmetry is (AFM2) or (AFM3).
In other words, when the system preserves the symmetry operation(s) including non-primitive translation $\bm{\tau}_\perp$ perpendicular to the mirror plane, the symmetry ensures nontrivial gap structures beyond the Sigrist-Ueda method.

Results consistent with Table~\ref{tab:classification_linenode} have been recently shown by Micklitz and Norman~\cite{Micklitz2017_PRL}.
Using a Clifford algebra technique, they also confirmed that the line nodes by nonsymmorphic symmetry are protected by a $\mathbb{Z}$ topological number.
A further discussion about the topological stability of the line nodes and resulting surface states, namely Majorana flat bands, is given in another publication~\cite{Kobayashi2017_arXiv}.

\begin{figure*}[tbp]
 \centering
 \includegraphics[width=14cm, clip]{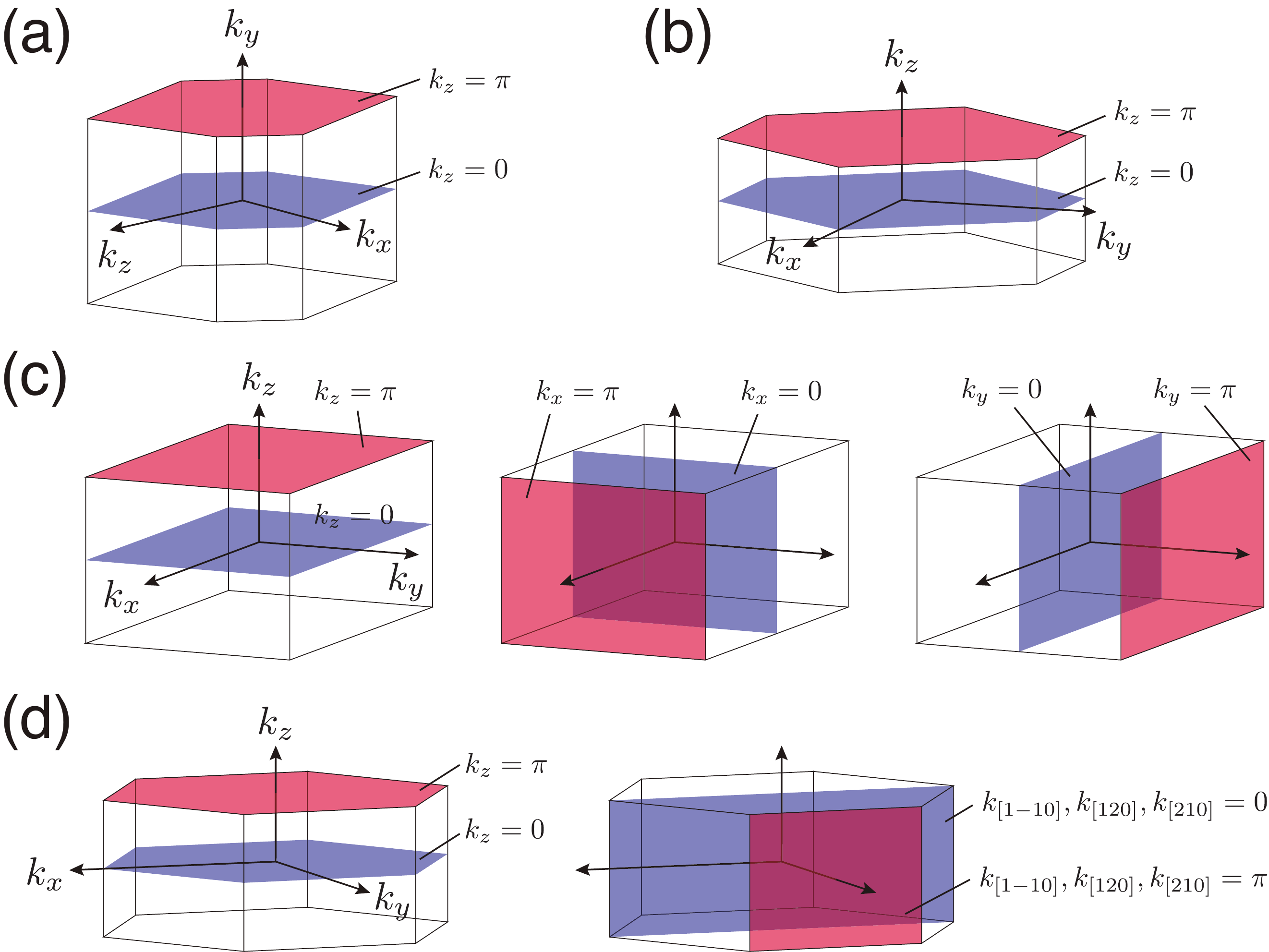}
 \caption{Mirror-invariant BPs and ZFs in the first BZ for (a) monoclinic primitive, (b) orthorhombic base-centered, (c) orthorhombic primitive / tetragonal primitive / cubic primitive, and (d) hexagonal primitive Bravais lattice. The blue and red planes represent BPs and ZFs, respectively.}
 \label{fig:BP_ZF}
\end{figure*}

In the above discussion, we have reviewed the gap classification of line nodes on mirror- or glide-invariant planes.
Here we reconsider constraints on the crystal symmetry of the system.
Space groups containing $C_{2h}$ symmetry are divided into four types: primitive, orthorhombic base-centered, body-centered, and face-centered space groups.
All types of space groups have one or more mirror-invariant BPs ($k_\perp = 0$) in the BZ.
On the other hand, corresponding mirror-invariant ZFs ($k_\perp = \pi$) exist only for primitive or orthorhombic base-centered space groups.
Although some of body-centered or face-centered space groups also have mirror-invariant ZFs, all of them are $k_\perp = 2 \pi$ planes, where the gap classification gives the same result as that on BPs ($k_\perp = 0$ planes).
Examples of mirror-invariant BPs and ZFs in the BZ of a primitive or orthorhombic base-centered Bravais lattice are illustrated in Figs.~\ref{fig:BP_ZF}(a)-(d).

As shown in Table~\ref{tab:classification_linenode}, unconventional line nodes (gap opening) protected by nonsymmorphic symmetry appear on $k_\perp = \pi$ planes.
Therefore, we conclude that nontrivial nonsymmorphic-symmetry-protected line nodes may appear not for a body-centered or face-centered Bravais lattice, but for a \textit{primitive or orthorhombic base-centered Bravais lattice}.
This additional constraint simplifies the classification of space groups with respect to the superconducting gap structure.

\begin{table*}[htbp]
 \caption{Classification of (a) monoclinic, (b) orthorhombic, (c) tetragonal, (d) hexagonal, and (e) cubic space groups. The table for orthorhombic space groups shows primitive and base-centered Bravais lattices, while the other tables show the primitive Bravais lattice. The first and second columns show the number and the name of space groups, respectively. The following column(s) represent the results of classification based on Eq.~\eqref{eq:linenode_spacegroup} when we fix the direction of mirror-invariant BP and ZF [see also Figs.~\ref{fig:BP_ZF}(a)-(d)].}
 \label{tab:spacegroup}
 \begin{center}
  \begin{tabular}{cccp{5mm}cccccp{5mm}cccc} \hline\hline
   \multicolumn{3}{c}{(a)} & & \multicolumn{5}{c}{(b)} & & \multicolumn{4}{c}{(c)} \\ \cline{1-3}\cline{5-10}\cline{11-14}
   No. & Short & $\perp = y$ & & No. & Short & $\perp = x$ & $\perp = y$ & $\perp = z$ & & No. & Short & $\perp = z$ & $\perp = x, y$ \\ \hline
    10 & $P2/m$   & (RM) & & 47 & $Pmmm$ & (RM) & (RM) & (RM) & & 83  & $P4/m$     & (RM) & N/A  \\
    11 & $P2_1/m$ & (SM) & & 48 & $Pnnn$ & (RG) & (RG) & (RG) & & 84  & $P4_2/m$   & (RM) & N/A  \\
    13 & $P2/c$   & (RG) & & 49 & $Pccm$ & (RG) & (RG) & (RM) & & 85  & $P4/n$     & (RG) & N/A  \\
    14 & $P2_1/c$ & (SG) & & 50 & $Pban$ & (RG) & (RG) & (RG) & & 86  & $P4_2/n$   & (RG) & N/A  \\
       &          &      & & 51 & $Pmma$ & (SM) & (RM) & (RG) & & 123 & $P4/mmm$   & (RM) & (RM) \\
       &          &      & & 52 & $Pnna$ & (RG) & (SG) & (RG) & & 124 & $P4/mcc$   & (RM) & (RG) \\
       &          &      & & 53 & $Pmna$ & (RM) & (RG) & (SG) & & 125 & $P4/nbm$   & (RG) & (RG) \\
       &          &      & & 54 & $Pcca$ & (SG) & (RG) & (RG) & & 126 & $P4/nnc$   & (RG) & (RG) \\
       &          &      & & 55 & $Pbam$ & (SG) & (SG) & (RM) & & 127 & $P4/mbm$   & (RM) & (SG) \\
       &          &      & & 56 & $Pccn$ & (SG) & (SG) & (RG) & & 128 & $P4/mnc$   & (RM) & (SG) \\
       &          &      & & 57 & $Pbcm$ & (RG) & (SG) & (SM) & & 129 & $P4/nmm$   & (RG) & (SM) \\
       &          &      & & 58 & $Pnnm$ & (SG) & (SG) & (RM) & & 130 & $P4/ncc$   & (RG) & (SG) \\
       &          &      & & 59 & $Pmmn$ & (SM) & (SM) & (RG) & & 131 & $P4_2/mmc$ & (RM) & (RM) \\
       &          &      & & 60 & $Pbcn$ & (SG) & (RG) & (SG) & & 132 & $P4_2/mcm$ & (RM) & (RG) \\
       &          &      & & 61 & $Pbca$ & (SG) & (SG) & (SG) & & 133 & $P4_2/nbc$ & (RG) & (RG) \\
       &          &      & & 62 & $Pnma$ & (SG) & (SM) & (SG) & & 134 & $P4_2/nnm$ & (RG) & (RG) \\
       &          &      & & 63 & $Cmcm$ & N/A  & N/A  & (SM) & & 135 & $P4_2/mbc$ & (RM) & (SG) \\
       &          &      & & 64 & $Cmca$ & N/A  & N/A  & (SG) & & 136 & $P4_2/mnm$ & (RM) & (SG) \\
       &          &      & & 65 & $Cmmm$ & N/A  & N/A  & (RM) & & 137 & $P4_2/nmc$ & (RG) & (SM) \\
       &          &      & & 66 & $Cccm$ & N/A  & N/A  & (RM) & & 138 & $P4_2/ncm$ & (RG) & (SG) \\
       &          &      & & 67 & $Cmma$ & N/A  & N/A  & (RG) & &     &            &      &      \\
       &          &      & & 68 & $Ccca$ & N/A  & N/A  & (RG) & &     &            &      &      \\ \hline\hline
  \end{tabular}
  \\[5mm]
  \begin{tabular}{ccccp{10mm}ccc} \hline\hline
   \multicolumn{4}{c}{(d)} & & \multicolumn{3}{c}{(e)} \\ \cline{1-4}\cline{6-8}
   No. & Short & $\perp = z$ & $\perp = [1 {-1} 0], [1 2 0], [2 1 0]$ & & No. & Short & $\perp = x, y, z$ \\ \hline
   175 & $P6/m$     & (RM) & N/A  & & 200 & $Pm\bar{3}$  & (RM) \\
   176 & $P6_3/m$   & (SM) & N/A  & & 201 & $Pn\bar{3}$  & (RG) \\
   191 & $P6/mmm$   & (RM) & (RM) & & 205 & $Pa\bar{3}$  & (SG) \\
   192 & $P6/mcc$   & (RM) & (RG) & & 221 & $Pm\bar{3}m$ & (RM) \\
   193 & $P6_3/mcm$ & (SM) & (RM) & & 222 & $Pn\bar{3}n$ & (RG) \\
   194 & $P6_3/mmc$ & (SM) & (RG) & & 223 & $Pm\bar{3}n$ & (RM) \\
       &            &      &      & & 224 & $Pn\bar{3}m$ & (RG) \\ \hline\hline
  \end{tabular}
 \end{center}
\end{table*}

For the above reason, we classify here only primitive and orthorhombic base-centered space groups containing $C_{2h}$ symmetry, which may allow nontrivial gap structures by nonsymmorphic symmetry.
High-symmetry BPs and ZFs [red planes in Figs.~\ref{fig:BP_ZF}(a)-(d)] are classified into (RM), (RG), (SM), or (SG) of Eq.~\eqref{eq:linenode_spacegroup}.
The results are summarized in Tables~\ref{tab:spacegroup}(a)-(e).
Combining Tables~\ref{tab:classification_linenode} and \ref{tab:spacegroup}, we may elucidate the superconducting gap structure on the BPs and ZFs.
Since the symmetry-protected line nodes can appear only on these mirror-invariant $\bm{k}$ planes, the group-theoretical classification of line nodes is completed.

Finally, space groups of the candidate materials for nonsymmorphic line nodes (gap opening) are illustrated in the following.
\begin{itemize}
 \item PM UPt$_3$~\cite{Norman1995, Micklitz2009, Kobayashi2016, Yanase2016, Nomoto2016_PRL, Micklitz2017_PRB}:
       \begin{align*}
        G &= P6_3/mmc \ \text{[Table~\ref{tab:spacegroup}(d)]}, \\
        \tilde{\theta} &= \{\theta | \bm{0}\};
       \end{align*}
 \item PM UCoGe~\cite{Nomoto2017}:
       \begin{align*}
        G &= Pnma \ \text{[Table~\ref{tab:spacegroup}(b)]}, \\
        \tilde{\theta} &= \{\theta | \bm{0}\};
       \end{align*}
 \item FM UCoGe~\cite{Nomoto2017}:
       \begin{equation*}
        G = P2_1/c \ \text{[Table~\ref{tab:spacegroup}(a)]};
       \end{equation*}
 \item AFM UPd$_2$Al$_3$~\cite{Fujimoto2006, Nomoto2017, Micklitz2017_PRL}:
       \begin{align*}
        G &= P2_1/m \ \text{[Table~\ref{tab:spacegroup}(a)]}, \\
        \tilde{\theta} &= \{\theta | \bm{\tau}_\perp\};
       \end{align*}
 \item AFM Sr$_2$IrO$_4$~\cite{Sumita2017}:
       \begin{align*}
        G &= Pcca \ \text{[Table~\ref{tab:spacegroup}(b)]}, \\
        \tilde{\theta} &= \{\theta | \bm{\tau}_\parallel + \bm{\tau}_\perp\}.
       \end{align*}
\end{itemize}

\section{$j_z$-dependent symmetry-protected point nodes}
\label{sec:point_node}
In Sec.~\ref{sec:line_node}, the condition for nontrivial line nodes beyond the Sigrist-Ueda method has been elucidated by using the gap classification on high-symmetry $\bm{k}$ planes.
In this section, we show nontrivial symmetry-protected point nodes using the gap classification on high-symmetry $\bm{k}$ lines, namely the $n$-fold axis ($n = 2$, $3$, $4$, and $6$).
In the following part, nonsymmorphic symmetry does not play any important role~\footnote{The results in Table~\ref{tab:cooper_pair_n-fold_axis} are not changed even when the system has non-primitive translations parallel to the $n$-fold axis, because the phase factor arising from the translations is canceled during calculation of the Mackey-Bradley theorem.}, and thus, we consider symmorphic space groups and the PM state, for simplicity.

The little group on a $n$-fold axis ${\cal M}_n^{\bm{k}}$ is given by
\begin{equation}
 {\cal M}_n^{\bm{k}} = \sum_{m = 0}^{n - 1} \{C_n | \bm{0}\}^m T + \{\theta I | \bm{0}\} \sum_{m = 0}^{n - 1} \{C_n | \bm{0}\}^m T,
\end{equation}
where $C_n$ represents the $n$-fold rotation.
The small representations of ${\cal M}_n^{\bm{k}}$ are obtained by the double-valued IR of corresponding point groups (little co-groups) $C_n = {\cal M}_n^{\bm{k}} / T$ [Tables~\ref{tab:IR_cyclic_point_group}(a)-(d)].
Note that each IR in Table~\ref{tab:IR_cyclic_point_group} is composed of two one-dimensional representations, which are degenerate due to the successive operations of time-reversal and spatial inversion $\{\theta I | \bm{0}\}$.
The subscripts $1 / 2$, $3 / 2$, and $5 / 2$ correspond to the total angular momentum of the Bloch state $j_z = \pm 1 / 2$, $\pm 3 / 2$, and $\pm 5 / 2$, respectively~\cite{Herzberg, Inui-Tanabe-Onodera}.

\begin{table}[htbp]
 \caption{The double-valued IRs of cyclic point groups~\cite{Bradley, Herzberg, Inui-Tanabe-Onodera}.}
 \label{tab:IR_cyclic_point_group}
 \begin{center}
  \begin{tabular}[t]{cccp{10mm}cccc} \hline\hline
   \multicolumn{3}{c}{(a) 2-fold axis} & & \multicolumn{4}{c}{(b) 3-fold axis} \\ \cline{1-3}\cline{5-8}
   $C_2$       & $E$ & $C_2$ & & $C_3$         & $E$   & $C_3$ & $C_3^2$ \\ \hline
   $E_{1 / 2}$ & $2$ & $0$   & & $E_{1 / 2}$   & $2$   & $1$   & $- 1$   \\
               &     &       & & $2 B_{3 / 2}$ & $2$   & $- 2$ & $2$     \\ \hline\hline
  \end{tabular}
  \\[5mm]
  \begin{tabular}[t]{cccccp{5mm}ccccccc} \hline\hline
   \multicolumn{5}{c}{(c) 4-fold axis} & & \multicolumn{7}{c}{(d) 6-fold axis} \\ \cline{1-5}\cline{7-13}
   $C_4$       & $E$ & $C_4$        & $C_4^3$      & $C_4^2$ & & $C_6$       & $E$ & $C_6$        & $C_6^5$      & $C_3$ & $C_3^2$ & $C_2$ \\ \hline
   $E_{1 / 2}$ & $2$ & $\sqrt{2}$   & $- \sqrt{2}$ & $0$     & & $E_{1 / 2}$ & $2$ & $\sqrt{3}$   & $- \sqrt{3}$ & $1$   & $- 1$   & $0$   \\
   $E_{3 / 2}$ & $2$ & $- \sqrt{2}$ & $\sqrt{2}$   & $0$     & & $E_{5 / 2}$ & $2$ & $- \sqrt{3}$ & $\sqrt{3}$   & $1$   & $- 1$   & $0$   \\
               &     &              &              &         & & $E_{3 / 2}$ & $2$ & $0$          & $0$          & $- 2$ & $2$     & $0$   \\ \hline\hline
  \end{tabular}
 \end{center}
\end{table}

From the small representation corresponding to the Bloch wave function, we can calculate $P^{\bm{k}}$, the representation of the Cooper pair wave function, using the Mackey-Bradley theorem [Eq.~\eqref{eq:Mackey-Bradley}].
Since inversion operation $I$ commutes with any symmetry operation in the case of the symmorphic system, Eq.~\eqref{eq:Mackey-Bradley} is simplified,
\begin{subequations}
 \label{eq:Mackey-Bradley_sym}
 \begin{align}
  \chi[P^{\bm{k}}(m)] &= \chi[\gamma^{\bm{k}}(m)]^2, \label{eq:Mackey-Bradley_sym_a} \\
  \chi[P^{\bm{k}}(I m)] &= - \chi[\gamma^{\bm{k}}(m^2)]. \label{eq:Mackey-Bradley_sym_b}
 \end{align}
\end{subequations}
Characters of $P^{\bm{k}}$ obtained by calculating Eq.~\eqref{eq:Mackey-Bradley_sym} are summarized in Tables~\ref{tab:cooper_pair_n-fold_axis}(a)-(d).

\begin{table}[htbp]
 \caption{Characters of representations of Cooper pair wave functions on $n$-fold axis.}
 \label{tab:cooper_pair_n-fold_axis}
 \begin{center}
  \begin{tabular}[t]{cccccp{5mm}ccccc} \hline\hline
   \multicolumn{5}{c}{(a) 2-fold axis} & & \multicolumn{5}{c}{(b) 3-fold axis} \\ \cline{1-5}\cline{7-11}
   $C_{2h}$     & $E$ & $C_2$ & $I$   & $\sigma_h$ & & $S_6$          & $E$ & $C_3, C_3^2$ & $I$   & $I C_3, I C_3^2$ \\ \hline
   $P^{\bm{k}}$ & $4$ & $0$   & $- 2$ & $2$        & & $P^{\bm{k}}_1$ & $4$ & $1$          & $- 2$ & $1$              \\
                &     &       &       &            & & $P^{\bm{k}}_2$ & $4$ & $4$          & $- 2$ & $- 2$            \\ \hline\hline
  \end{tabular}
  \\[5mm]
  \begin{tabular}{ccccccc} \hline\hline
   \multicolumn{7}{c}{(c) 4-fold axis} \\ \hline
   $C_{4h}$     & $E$ & $C_4, C_4^3$ & $C_4^2$ & $I$   & $I C_4, I C_4^3$ & $\sigma_h$ \\ \hline
   $P^{\bm{k}}$ & $4$ & $2$          & $0$     & $- 2$ & $0$              & $2$        \\ \hline\hline
  \end{tabular}
  \\[5mm]
  \begin{tabular}{ccccccccc} \hline\hline
   \multicolumn{9}{c}{(d) 6-fold axis} \\ \hline
   $C_{6h}$       & $E$ & $C_6, C_6^5$ & $C_3, C_3^2$ & $C_2$ & $I$   & $I C_3, I C_3^2$ & $I C_6, I C_6^5$ & $\sigma_h$ \\ \hline
   $P^{\bm{k}}_1$ & $4$ & $3$          & $1$          & $0$   & $- 2$ & $1$              & $- 1$            & $2$        \\
   $P^{\bm{k}}_2$ & $4$ & $0$          & $4$          & $0$   & $- 2$ & $- 2$            & $2$              & $2$        \\ \hline\hline
  \end{tabular}
 \end{center}
\end{table}

From Table~\ref{tab:cooper_pair_n-fold_axis}, we reduce the representation $P^{\bm{k}}$ into IRs of point group $\widetilde{\cal M}_n^{\bm{k}} / T = C_n + I C_n$:
\begin{enumerate}
 \item 2-fold axis ($\widetilde{\cal M}_n^{\bm{k}} / T = C_{2h}$),
       \begin{equation}
        P^{\bm{k}} = A_g + A_u + 2 B_u \qquad (\gamma^{\bm{k}} = E_{1 / 2});
       \end{equation}
 \item 3-fold axis ($\widetilde{\cal M}_n^{\bm{k}} / T = S_6$),
       \begin{subequations}
        \begin{align}
         P^{\bm{k}}_1 &= A_g + A_u + E_u & (\gamma^{\bm{k}} &= E_{1 / 2}), \label{eq:gap_3-fold_a} \\
         P^{\bm{k}}_2 &= A_g + 3 A_u & (\gamma^{\bm{k}} &= 2 B_{3 / 2}); \label{eq:gap_3-fold_b}
        \end{align}
       \end{subequations} 
\item 4-fold axis ($\widetilde{\cal M}_n^{\bm{k}} / T = C_{4h}$),
       \begin{equation}
        P^{\bm{k}} = A_g + A_u + E_u \qquad (\gamma^{\bm{k}} = E_{1 / 2}, E_{3 / 2});
       \end{equation}
 \item 6-fold axis ($\widetilde{\cal M}_n^{\bm{k}} / T = C_{6h}$),
       \begin{subequations}
        \begin{align}
         P^{\bm{k}}_1 &= A_g + A_u + E_{1u} & (\gamma^{\bm{k}} &= E_{1 / 2}, E_{5 / 2}), \label{eq:gap_6-fold_a} \\
         P^{\bm{k}}_2 &= A_g + A_u + 2 B_u & (\gamma^{\bm{k}} &= E_{3 / 2}). \label{eq:gap_6-fold_b}
        \end{align}
       \end{subequations}
\end{enumerate}
In the 2-fold and 4-fold symmetric cases, the representation of the Cooper pair wave function is unique irrespective of the normal Bloch state $\gamma^{\bm{k}}$.
On the other hand, in the 3-fold and 6-fold symmetric cases, two nonequivalent representations of the Cooper pair emerge depending on the Bloch-state angular momentum $j_z$.
For example, on the 3-fold axis, the $E_u$ order parameter is allowed (forbidden) in the case of the $E_{1 / 2}$ ($2 B_{3 / 2}$) Bloch state.
This means that the $E_u$ superconducting gap opens in the energy band of the $j_z = \pm 1 / 2$ state, while point nodes appear for $j_z = \pm 3 / 2$.
Therefore, the presence or absence of point nodes is $j_z$-dependent when the system has 3-fold or 6-fold rotational symmetry.
Such a $j_z$-dependent gap structure is not obtained by the Sigrist-Ueda method.
Thus, the gap structure beyond the Sigrist-Ueda theory may be obtained from Eqs.~\eqref{eq:gap_3-fold_a}, \eqref{eq:gap_3-fold_b}, \eqref{eq:gap_6-fold_a}, and \eqref{eq:gap_6-fold_b}.
In the following subsections, we suggest a material realization of this unusual gap structures in UPt$_3$, and we discuss other candidate superconductors.

\subsection{UPt$_3$ (Space group: $P6_3/mmc$)}
\label{sec:point_node_UPt3}
\begin{figure}[tbp]
 \centering
 \includegraphics[width=6cm, clip]{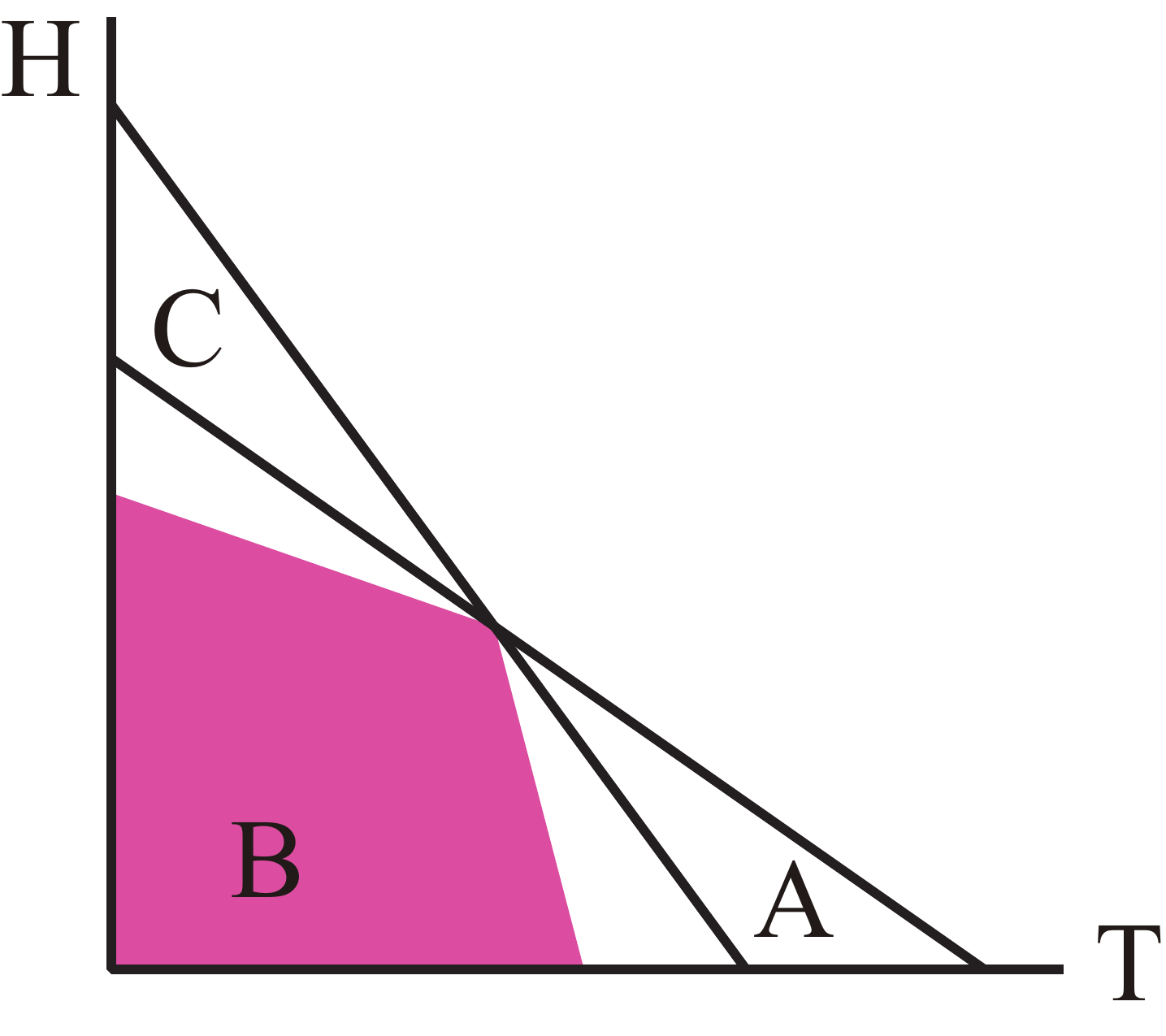}
 \caption{Multiple superconducting phases of UPt$_3$ in the magnetic field-temperature plane~\cite{Sauls1994, Joynt2002}. The shaded region shows the Weyl superconducting phase~\cite{Yanase2016, Yanase2017}.}
 \label{fig:UPt3_phase}
\end{figure}

Superconductivity in UPt$_3$ has been intensively investigated after the discovery of superconductivity in 1980's~\cite{Stewart1984}.
Multiple superconducting phases illustrated in Fig.~\ref{fig:UPt3_phase}~\cite{Fisher1989, Bruls1990, Adenwalla1990, Tou1998} unambiguously exhibit exotic Cooper pairing which is probably categorized into the two-dimensional (2D) IR of point group $D_{6h}$~\cite{Sigrist-Ueda}.
After several theoretical proposals examined by experiments for more than three decades, the $E_{2u}$ representation has been regarded as the most reasonable symmetry of superconducting order parameter~\cite{Sauls1994, Joynt2002}.
In particular, the multiple superconducting phases in the temperature--magnetic-field plane are naturally reproduced by assuming a weak symmetry-breaking term of hexagonal symmetry~\cite{Sauls1994}.
Furthermore, a phase-sensitive measurement~\cite{Strand2010} and the observation of spontaneous time-reversal symmetry breaking~\cite{Luke1993, Schemm2014} in the low-temperature and low-magnetic-field B phase, which was predicted in the $E_{2u}$ state, support the $E_{2u}$ symmetry of superconductivity.

\begin{figure}[tbp]
 \centering
 \includegraphics[width=8cm, clip]{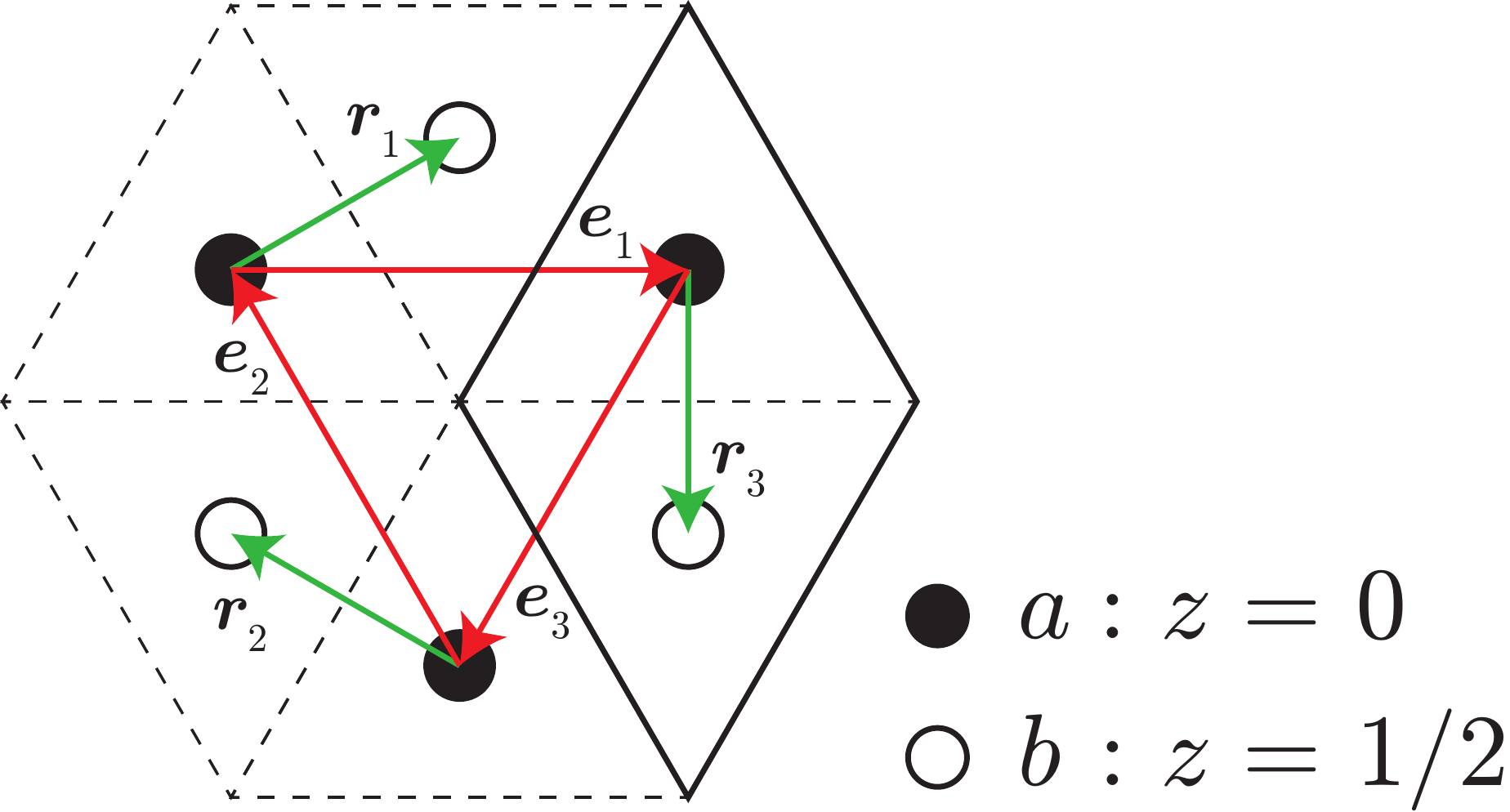}
 \caption{Crystal structure of UPt$_3$. Uranium ions form AB stacked triangular lattice. 2D vectors, $\bm{e}_i$ and $\bm{r}_i$, are shown by arrows. The black solid diamond shows the unit cell.}
 \label{fig:UPt3_structure}
\end{figure}

\begin{figure}[tbp]
 \centering
 \includegraphics[width=6cm, clip]{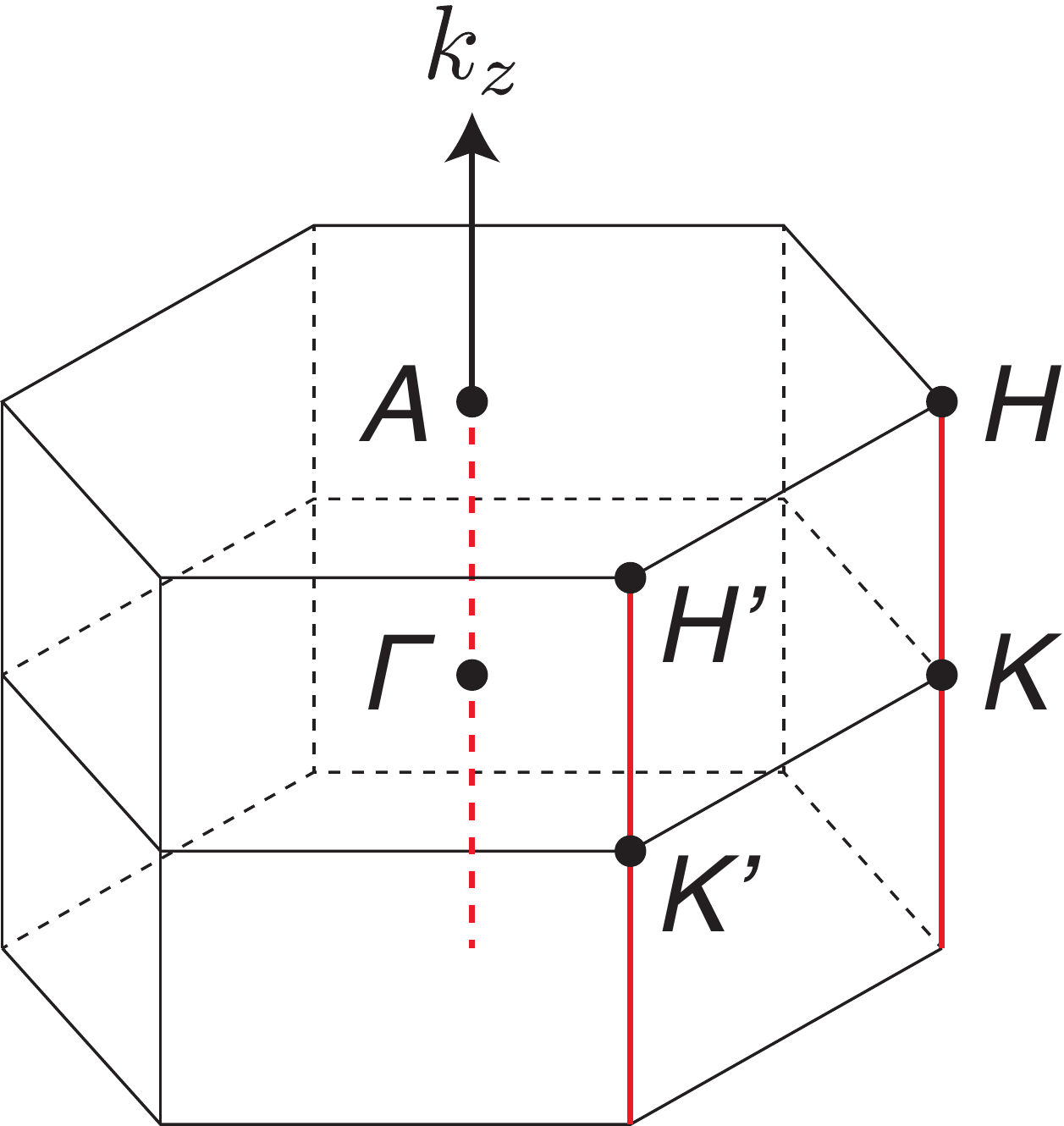}
 \caption{The first BZ of primitive hexagonal lattice. The red lines show 3- and 6-fold rotation symmetric axes.}
 \label{fig:brillouin_zone_hexa}
\end{figure}

The crystal structure of UPt$_3$ is illustrated in Fig.~\ref{fig:UPt3_structure}.
The symmetry of the crystal is represented by nonsymmorphic space group $P6_3/mmc$~\footnote{Symmetry breaking by a weak crystal distortion has been reported~\cite{Walko2001}, although its reliability is under debate. We here assume high symmetry space group $P6_{3}/mmc$.}, which is based on the primitive hexagonal Bravais lattice.
In this space group, the BZ takes the form of Fig.~\ref{fig:brillouin_zone_hexa}.
This BZ has a 3-fold rotation axis on the $K$-$H$ line as well as a 6-fold rotation axis on the $\Gamma$-$A$ line.

Quantum oscillation measurements combined with band-structure calculations~\cite{Joynt2002, Taillefer1988, Norman1988, Kimura1995, McMullan2008, Nomoto2016_PRL} have shown a pair of Fermi surfaces (FSs) centered at the $A$ point ($A$-FSs), three FSs at the $\Gamma$ point ($\Gamma$-FSs), and two FSs at the $K$ point ($K$-FSs) in UPt$_3$.
Although previous studies have clarified gap structures on the $A$-FSs and $\Gamma$-FSs~\cite{Yarzhemsky1992, Norman1995, Yarzhemsky1998, Micklitz2009, Kobayashi2016, Yanase2016, Nomoto2016_PRL, Yanase2017}, those on the $K$-FSs have not been theoretically studied.
From the results of classification theory given in this section, however, it would be interesting to examine the gap structure on the $K$-FSs, since they cross the $K$-$H$ line.
Indeed, we show the intriguing $j_z$-dependent point nodes on the $K$-$H$ line.

\subsubsection{Gap classification}
\label{sec:UPt3_gap_classification}
Now we apply the classification theory introduced in Sec.~\ref{sec:gap_classification} to the space group of UPt$_3$.
In the space group $P6_3/mmc$, the BZ has a 6-fold axis $\Gamma$-$A$, and 3-fold axes $K$-$H$ and $K'$-$H'$ (see Fig.~\ref{fig:brillouin_zone_hexa}).
On the $\Gamma$-$A$ line, the little group has $C_{6v}$ symmetry which results in three small representations $\gamma^{\bm{k}} = E_{1 / 2}$, $E_{3 / 2}$, and $E_{5 / 2}$.
In the superconducting state, we obtain two different representations of the Cooper pair wave function:
\begin{subequations}
 \begin{align}
  P^{\bm{k}}_1 &= A_{1g} + A_{1u} + E_{1u} & (\gamma^{\bm{k}} &= E_{1 / 2}, E_{5 / 2}), \label{eq:UPt3_GAgap_E1/2} \\
  P^{\bm{k}}_2 &= A_{1g} + A_{1u} + B_{1u} + B_{2u} & (\gamma^{\bm{k}} &= E_{3 / 2}), \label{eq:UPt3_GAgap_E3/2}
 \end{align}
\end{subequations}
which have been decomposed into IRs of point group $D_{6h} = C_{6v} + I C_{6v}$.
The same result has been suggested by Yarzhemsky~\cite{Yarzhemsky1992, Yarzhemsky1998}.
From the discussion in this section, UPt$_3$ is considered to possess the $E_{2u}$ superconducting order parameter.
According to Eqs.~\eqref{eq:UPt3_GAgap_E1/2} and \eqref{eq:UPt3_GAgap_E3/2}, the $E_{2u}$ representation is not allowed for any small representations $\gamma^{\bm{k}} = E_{1 / 2}$, $E_{3 / 2}$, and $E_{5 / 2}$.
Therefore, point nodes appear on the $\Gamma$-$A$ line irrespective of the property of the normal Bloch state.
We can obtain the same conclusion with the Sigrist-Ueda method.

On the other hand, the gap structure on the $K$-$H$ ($K'$-$H'$) line is $j_z$-dependent.
The little group has $C_{3v}$ symmetry, which results in two small representations $\gamma^{\bm{k}} = E_{1 / 2}$ and $E_{3 / 2}$.
Corresponding to these two Bloch states, the Cooper pair wave function has two different representations:
\begin{subequations}
 \begin{align}
  P^{\bm{k}}_1 &= A_{1g} + A_{1u} + E_u & (\gamma^{\bm{k}} &= E_{1 / 2}), \label{eq:C3v_gap_E1/2} \\
  P^{\bm{k}}_2 &= A_{1g} + 2 A_{1u} + A_{2u} & (\gamma^{\bm{k}} &= E_{3 / 2}), \label{eq:C3v_gap_E3/2}
 \end{align}
\end{subequations}
which have been decomposed into IRs of point group $D_{3d} = C_{3v} + I C_{3v}$.
Then, $P^{\bm{k}}$ can be induced to the point group $D_{6h}$ with the help of the Frobenius reciprocity theorem~\cite{Bradley}.
The induced representations $P^{\bm{k}} \uparrow D_{6h}$ are summarized in the following equations:
\begin{subequations}
 \begin{align}
  P^{\bm{k}}_1 \uparrow D_{6h} &= A_{1g} + B_{2g} + A_{1u} + B_{2u} + E_{1u} + E_{2u}, \label{eq:UPt3_KHgap_E1/2} \\
  P^{\bm{k}}_2 \uparrow D_{6h} &= A_{1g} + B_{2g} + 2 A_{1u} + A_{2u} + B_{1u} + 2 B_{2u}. \label{eq:UPt3_KHgap_E3/2}
 \end{align}
\end{subequations}
From Eqs.~\eqref{eq:UPt3_KHgap_E1/2} and \eqref{eq:UPt3_KHgap_E3/2}, the $E_{2u}$ superconducting gap opens for $\gamma^{\bm{k}} = E_{1 / 2}$, while point nodes appear for $\gamma^{\bm{k}} = E_{3 / 2}$.
Thus, the gap structure indeed depends on the angular momentum of Bloch states.
In this case, the classification by the Sigrist-Ueda method breaks down since it has taken into account only the pseudo-spin degree of freedom $s = 1 / 2$.
In the following, we demonstrate the nontrivial $j_z$-dependent gap structure by analyzing a microscopic model.

\subsubsection{Model and normal Bloch state}
\label{sec:UPt3_model}
Here we introduce the microscopic model of UPt$_3$, and we clarify the band structure on the $K$-$H$ line.
First, we introduce the Bogoliubov-de Gennes (BdG) Hamiltonian for a two-sublattice model~\cite{Yanase2016, Yanase2017},
\begin{equation}
 {\cal H}_{\text{BdG}} = \frac{1}{2} \sum_{\bm{k}} \bm{C}_{\bm{k}}^\dagger
  \begin{pmatrix}
   \hat{H}_{\text{n}}(\bm{k}) & \hat{\Delta}({\bm{k}}) \\
   \hat{\Delta}(\bm{k})^\dagger & - \hat{H}_{\text{n}}(- \bm{k})^{\text{T}}
  \end{pmatrix}
  \bm{C}_{\bm{k}}, \label{eq:UPt3_model}
\end{equation}
with
\begin{equation}
 \bm{C}_{\bm{k}}^\dagger = (c_{\bm{k} a \uparrow}^\dagger, c_{\bm{k} a \downarrow}^\dagger, c_{\bm{k} b \uparrow}^\dagger, c_{\bm{k} b \downarrow}^\dagger, c_{-\bm{k} a \uparrow}, c_{-\bm{k} a \downarrow}, c_{-\bm{k} b \uparrow}, c_{-\bm{k} b \downarrow}),
\end{equation}
where $\bm{k}$, $m = a, b$, and $s = \uparrow, \downarrow$ are the index of momentum, sublattice, and spin, respectively. 
The BdG Hamiltonian matrix is described by the normal-state Hamiltonian,
\begin{equation}
 \hat{H}_{\text{n}}(\bm{k}) = \begin{pmatrix}
                               \xi(\bm{k}) s_0 + \alpha \bm{g}(\bm{k}) \cdot \bm{s} & a(\bm{k}) s_0 \\
                               a(\bm{k})^* s_0 & \xi(\bm{k}) s_0 - \alpha \bm{g}(\bm{k}) \cdot \bm{s}
                              \end{pmatrix},
                              \label{eq:UPt3_model_normal}
\end{equation}
and the order-parameter part $\hat{\Delta}(\bm{k}) = [\Delta(\bm{k})]_{m s, m' s'}$.
Here $s_i$ represents the Pauli matrix in spin space.
Taking into account the crystal structure of UPt$_3$ illustrated in Fig.~\ref{fig:UPt3_structure}, we adopt an intra-sublattice kinetic energy,
\begin{equation}
 \xi(\bm{k}) = 2 t \sum_{i = 1, 2, 3} \cos{\bm{k}}_\parallel \cdot \bm{e}_i + 2 t_z \cos k_z - \mu, 
\end{equation}
and an inter-sublattice hopping term,
\begin{equation}
 a(\bm{k}) = 2 t' \cos\frac{k_z}{2} \sum_{i = 1, 2, 3} e^{i \bm{k}_\parallel \cdot \bm{r}_i}, \label{eq:UPt3_inter-sublattice}
\end{equation}
with $\bm{k}_\parallel = (k_x, k_y)$.
The basis translation vectors in two dimensions are $\bm{e}_1 = (1, 0)$, $\bm{e}_2 = (- \frac{1}{2}, \frac{\sqrt{3}}{2})$, and $\bm{e}_3 = (- \frac{1}{2}, - \frac{\sqrt{3}}{2})$.
The inter-layer neighboring vectors projected onto the $x$-$y$ plane are given by $\bm{r}_1 = (\frac{1}{2}, \frac{1}{2\sqrt{3}})$, $\bm{r}_2 = (- \frac{1}{2}, \frac{1}{2\sqrt{3}})$, and $\bm{r}_3 = (0, - \frac{1}{\sqrt{3}})$.
These 2D vectors are illustrated in Fig.~\ref{fig:UPt3_structure}.

Although the crystal point group symmetry is centrosymmetric, $D_{6h}$, the local point group symmetry at uranium ions is $D_{3h}$ lacking inversion symmetry.
Then, Kane-Mele antisymmetric spin-orbit coupling (ASOC)~\cite{Kane-Mele2005} with a $g$ vector~\cite{Saito2016},
\begin{equation}
 \bm{g}(\bm{k}) = \hat{z} \sum_{i = 1, 2, 3} \sin\bm{k}_\parallel \cdot \bm{e}_i,
\end{equation}
is allowed by symmetry.
The coupling constant is staggered between the two sublattices, so as to preserve the global $D_{6h}$ point group symmetry~\cite{Kane-Mele2005, Fischer2011, Maruyama2012}.

\begin{figure}[tbp]
 \centering
 \includegraphics[width=8cm, clip]{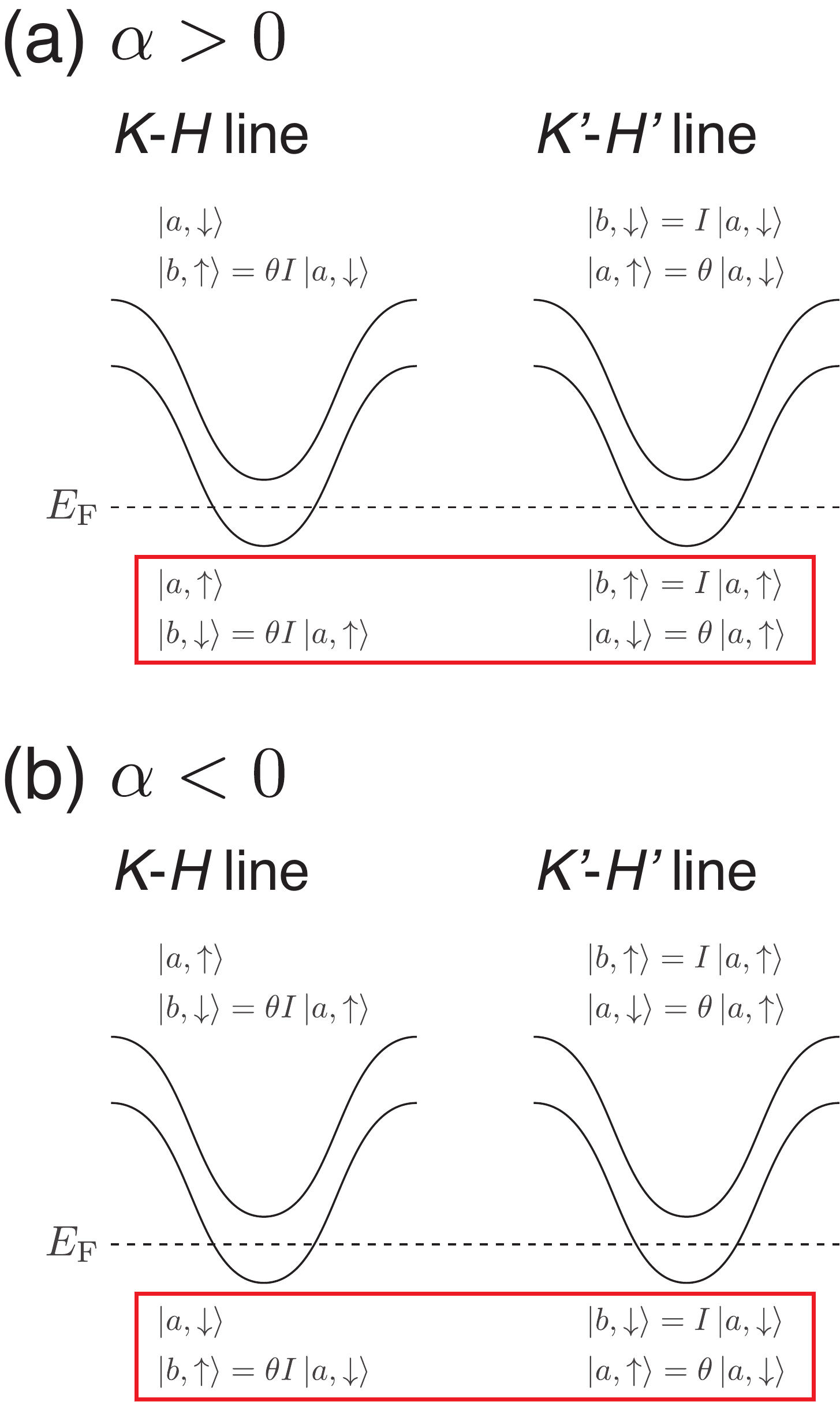}
 \caption{Schematic band structures on the $K$-$H$ and $K'$-$H'$ lines in (a) $\alpha > 0$ case and (b) $\alpha < 0$ case. The wave function of Bloch states crossing the Fermi level is shown in the red frame. The wave function of the upper band is shown above the band.}
 \label{fig:UPt3_degeneracy}
\end{figure}

In order to identify the Bloch state, we calculate the normal energy bands on the $K$-$H$ line from the normal part Hamiltonian [Eq.~\eqref{eq:UPt3_model_normal}].
Although the band originally has 4-fold degeneracy arising from the two sublattices and two spin degrees of freedom, this splits into 2-fold $+$ 2-fold degenerate bands due to the effect of the ASOC term.
The band structures are schematically shown in Figs.~\ref{fig:UPt3_degeneracy}(a) and \ref{fig:UPt3_degeneracy}(b).
When the coupling constant of the ASOC term $\alpha$ is positive, $\ket{a, \uparrow}$ and $\ket{b, \downarrow}$ states cross the Fermi level on the $K$-$H$ line, while $\ket{b, \uparrow}$ and $\ket{a, \downarrow}$ states cross on the $K'$-$H'$ line [Fig.~\ref{fig:UPt3_degeneracy}(a)].
On the other hand, the spin state on the Fermi level changes as shown in Fig.~\ref{fig:UPt3_degeneracy}(b), when the $\alpha$ is negative.
Note that the pure sublattice-based representations construct the basis of energy bands, since the inter-sublattice hopping term Eq.~\eqref{eq:UPt3_inter-sublattice} vanishes on the $K$-$H$ and $K'$-$H'$ lines.
This vanishing has been proved by the symmetry analysis~\cite{Akashi2015, Akashi2017}.

We have investigated the energy band structures of the normal state in the above discussion.
In order to identify superconducting gap structures, therefore, we should solve the following question:
Which representation does the band crossing the Fermi level belong to, $E_{1 / 2}$ or $E_{3 / 2}$~?
The difference between these two representations is the total angular momentum $j_z$ ($= \pm 1 / 2$ or $\pm 3 / 2$) of the Bloch state.
In the next subsection, we show that $j_z$ contains an \textit{effective orbital angular momentum} arising from the permutation of sites, as well as the pure orbital angular momentum and the spin angular momentum.

\subsubsection{Effective orbital angular momentum}
\label{sec:effective_ang_momentum}
Here we investigate the total angular momentum $j_z$ of the Bloch state, and we show that $j_z$ includes an effective orbital angular momentum $\lambda_z$ arising from a Bloch phase of each site, in addition to the pure orbital and spin angular momentum.
Furthermore, we clarify that $j_z$ of the Bloch states crossing the Fermi level depends on the sign of the ASOC $\alpha$.

First, recalling the quantum mechanics, $j_z$ should contains the orbital angular momentum $l_z$ and the spin angular momentum $s_z$.
In our two-sublattice single-orbital model [Eq.~\eqref{eq:UPt3_model_normal}], the orbital degree of freedom is neglected, and then, $l_z = 0$.
Since electrons are spin-$1 / 2$ fermions, the spin angular momentum is $s_z = \pm 1 / 2$.
Therefore, we might consider that $j_z = l_z + s_z = \pm 1 / 2$ in this model.
However, this is not right, as we show below.

\begin{figure}[tbp]
 \centering
 \includegraphics[width=8cm, clip]{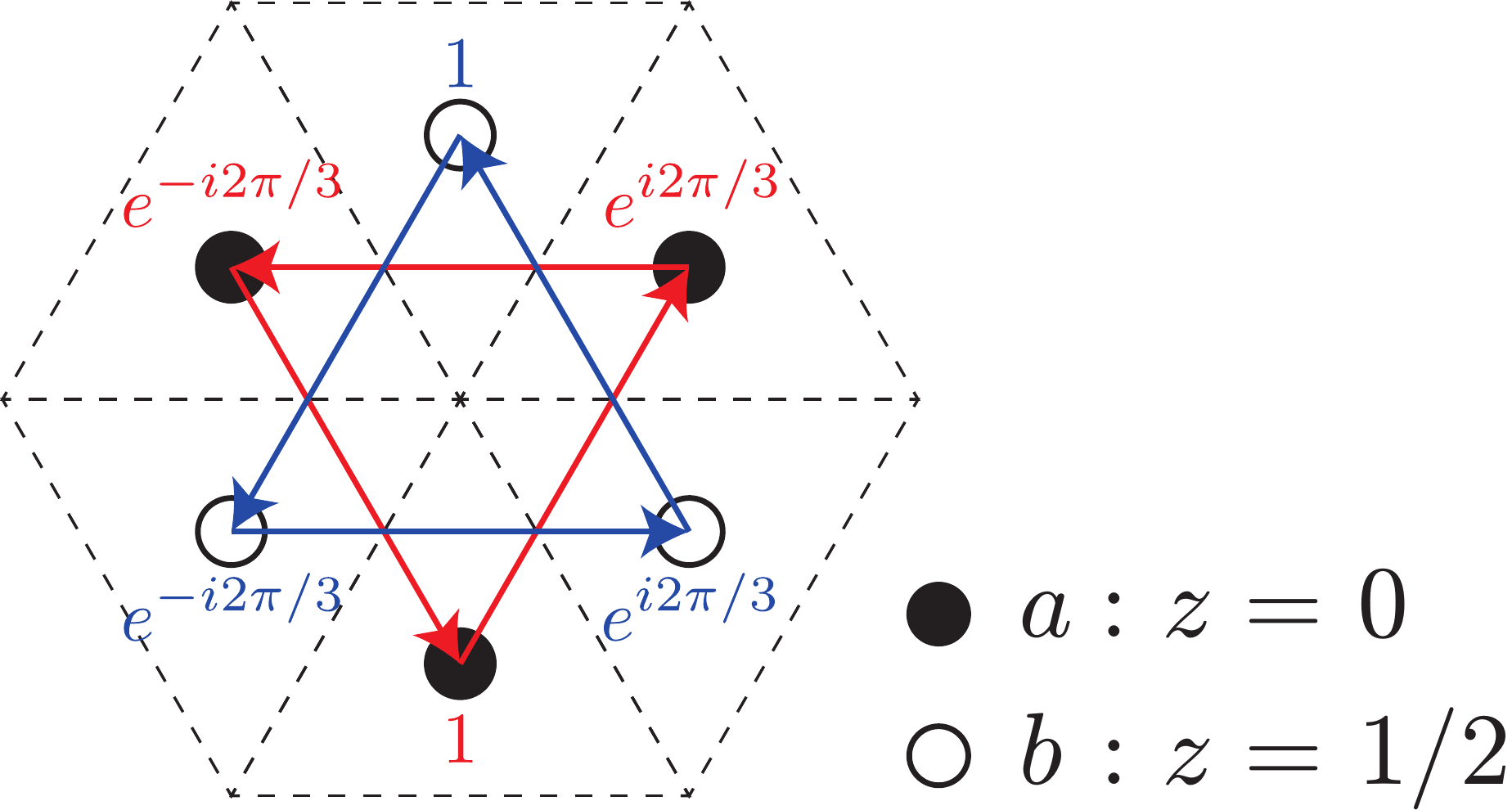}
 \caption{The phase factor $e^{i \bm{k} \cdot \bm{r}}$ on sites for the $K$-point Bloch state. The sublattices $a$ and $b$ obtain different phase values by 3-fold rotation.}
 \label{fig:UPt3_site_phase}
\end{figure}

In order to correctly calculate $j_z$, we have to take into account the effective orbital angular momentum $\lambda_z$ due to the permutation of sites.
The Bloch state has a phase factor (plane-wave part) $e^{i \bm{k} \cdot \bm{r}}$ depending on the site~\cite{Akashi2015, Akashi2017}, which is illustrated in Fig.~\ref{fig:UPt3_site_phase} for the $K$-point Bloch state [$\bm{k} = (4\pi / 3, 0, 0)$].
By operating 3-fold rotation on the $K$-point Bloch state, the $a$ sublattice obtains the phase value $e^{+ i 2\pi / 3}$ (red arrows), while the $b$ sublattice gets $e^{- i 2\pi / 3}$ (blue arrows).
These phase factors, which correspond to $e^{i \lambda_z \theta} \, (\theta = 2\pi / 3)$, indicate that the sublattices $a$ and $b$ possess an effective orbital angular momentum $\lambda_z = + 1$ and $- 1$, respectively.
The Bloch state at the $K'$ point has a complex conjugate phase factor to that on the $K$-point, which results in $\lambda_z = - 1$ ($+ 1$) for the $a$ ($b$) sublattice.
For a more general argument, the effective orbital angular momentum can be calculated by analyzing the space-group transformation of the Bloch state (see Appendix~\ref{app:effective_ang_momentum}).

Using the above discussion, we calculate the total angular momentum of the Bloch state by $j_z = l_z + s_z + \lambda_z$.
For example, in the $\ket{a, \uparrow}$ state on the $K$-$H$ line [see Fig.~\ref{fig:UPt3_degeneracy}(a)], $l_z = 0$, $s_z = + 1 / 2$, $\lambda_z = + 1$, so that we obtain $j_z = + 3 / 2$.
The total angular momenta of all states are summarized in Table~\ref{tab:UPt3_ang_momentum}.
From Figs.~\ref{fig:UPt3_degeneracy}(a) and \ref{fig:UPt3_degeneracy}(b), and Table~\ref{tab:UPt3_ang_momentum}, we identify the representations of the Bloch states crossing the Fermi level as follows.
\begin{enumerate}
 \item $\alpha > 0$: $E_{3 / 2}$ representation, because
       \[
        \ket{a, \uparrow} = \ket{j_z = + \tfrac{3}{2}}, \quad \ket{b, \downarrow} = \ket{j_z = - \tfrac{3}{2}},
       \]
       on the $K$-$H$ line, and
       \[
        \ket{b, \uparrow} = \ket{j_z = + \tfrac{3}{2}}, \quad \ket{a, \downarrow} = \ket{j_z = - \tfrac{3}{2}},
       \]
       on the $K'$-$H'$ line.
 \item $\alpha < 0$: $E_{1 / 2}$ representation, because
       \[
        \ket{a, \downarrow} = \ket{j_z = + \tfrac{1}{2}}, \quad \ket{b, \uparrow} = \ket{j_z = - \tfrac{1}{2}},
       \]
       on the $K$-$H$ line, and
       \[
        \ket{b, \downarrow} = \ket{j_z = + \tfrac{1}{2}}, \quad \ket{a, \uparrow} = \ket{j_z = - \tfrac{1}{2}},
       \]
       on the $K'$-$H'$ line.
\end{enumerate}
Assuming the $E_{2u}$ superconducting order parameter, therefore, the gap classification theory indicates that point nodes emerge on the $K$-$H$ ($K'$-$H'$) line when $\alpha > 0$, while the gap opens otherwise [see Eqs.~\eqref{eq:UPt3_KHgap_E1/2} and \eqref{eq:UPt3_KHgap_E3/2}].
In the next subsection, we demonstrate such unusual gap structures by a numerical analysis of the microscopic model.

\begin{table}[htbp]
 \caption{The total angular momentum of the Bloch state.}
 \label{tab:UPt3_ang_momentum}
 \begin{center}
  \begin{tabular}{cccccp{5mm}cccc} \hline\hline
   & \multicolumn{4}{c}{$K$-$H$ line} & & \multicolumn{4}{c}{$K'$-$H'$ line} \\ \cline{2-5}\cline{7-10}
   & $l_z$ & $s_z$ & $\lambda_z$ & $j_z$ & & $l_z$ & $s_z$ & $\lambda_z$ & $j_z$ \\ \hline
   $\ket{a, \uparrow}$   & $0$ & $+ 1 / 2$ & $+ 1$ & $+ 3 / 2$ & & $0$ & $+ 1 / 2$ & $- 1$ & $- 1 / 2$ \\
   $\ket{a, \downarrow}$ & $0$ & $- 1 / 2$ & $+ 1$ & $+ 1 / 2$ & & $0$ & $- 1 / 2$ & $- 1$ & $- 3 / 2$ \\
   $\ket{b, \uparrow}$   & $0$ & $+ 1 / 2$ & $- 1$ & $- 1 / 2$ & & $0$ & $+ 1 / 2$ & $+ 1$ & $+ 3 / 2$ \\
   $\ket{b, \downarrow}$ & $0$ & $- 1 / 2$ & $- 1$ & $- 3 / 2$ & & $0$ & $- 1 / 2$ & $+ 1$ & $+ 1 / 2$ \\ \hline\hline
  \end{tabular}
 \end{center}
\end{table}

\subsubsection{Gap structures depending on Bloch-state angular momentum}
\label{sec:UPt3_gap_structure}
Now we demonstrate unconventional $j_z$-dependent gap structures using the numerical calculation of the microscopic model.
To investigate the superconducting gap structures, we consider the two-component order parameters in the $E_{2u}$ IR of point group $D_{6h}$:
\begin{equation}
 \hat{\Delta}(\bm{k}) = \eta_1 \hat{\Gamma}_1^{E_{2u}} + \eta_2 \hat{\Gamma}_2^{E_{2u}}. \label{eq:UPt3_order}
\end{equation}
The two-component order parameters are parametrized as
\begin{equation}
 (\eta_1, \eta_2) = \Delta (1, i\eta) / \sqrt{1 + \eta^2}, \label{eq:UPt3_eta}
\end{equation}
with a real variable $\eta$.
The basis functions $\hat{\Gamma}_1^{E_{2u}}$ and $\hat{\Gamma}_2^{E_{2u}}$ are admixtures of some harmonics.
Adopting the neighboring Cooper pairs in the crystal lattice of uranium ions, we obtain the basis functions
\begin{align}
 \hat{\Gamma}_1^{E_{2u}} &= \Bigl[ \delta_1 \{p^{\text{(intra)}}_x(\bm{k}) s_x - p^{\text{(intra)}}_y(\bm{k}) s_y\} \sigma_0 \notag \\
 & + \delta_2 \{p^{\text{(inter)}}_x(\bm{k}) s_x - p^{\text{(inter)}}_y(\bm{k}) s_y\} \sigma_+ / 2 \notag \\
 & + \delta_2 \{p^{\text{(inter)}}_x(\bm{k})^* s_x - p^{\text{(inter)}}_y(\bm{k})^* s_y\} \sigma_- / 2 \notag \\
 & + f_{(x^2 - y^2)z}(\bm{k}) s_z \sigma_x - d_{yz}(\bm{k}) s_z \sigma_y \Bigr] i s_y, \label{eq:UPt3_order_1} \displaybreak[2] \\
 \hat{\Gamma}_2^{E_{2u}} &= \Bigl[ \delta_1 \{p^{\text{(intra)}}_y(\bm{k}) s_x + p^{\text{(intra)}}_x(\bm{k}) s_y\} \sigma_0 \notag \\
 & + \delta_2 \{p^{\text{(inter)}}_y(\bm{k}) s_x + p^{\text{(inter)}}_x(\bm{k}) s_y\} \sigma_+ / 2 \notag \\
 & + \delta_2 \{p^{\text{(inter)}}_y(\bm{k})^* s_x + p^{\text{(inter)}}_x(\bm{k})^* s_y\} \sigma_- / 2 \notag \\
 & + f_{xyz}(\bm{k}) s_z \sigma_x - d_{xz}(\bm{k}) s_z \sigma_y \Bigr] i s_y, \label{eq:UPt3_order_2}
\end{align}
which are composed of the intra-sublattice $p$-wave, inter-sublattice $p$-wave, and inter-sublattice $d + f$-wave components given by
\begin{align}
 p^{\text{(intra)}}_x(\bm{k}) &= \sum_{i} e_i^x \sin\bm{k}_\parallel \cdot \bm{e}_i, \\
 p^{\text{(intra)}}_y(\bm{k}) &= \sum_{i} e_i^y \sin\bm{k}_\parallel \cdot \bm{e}_i, \displaybreak[2] \\
 p^{\text{(inter)}}_x(\bm{k}) &= - i \sqrt{3} \cos\frac{k_z}{2} \sum_{i} r_i^x e^{i \bm{k}_\parallel \cdot \bm{r}_i}, \\
 p^{\text{(inter)}}_y(\bm{k}) &= - i \sqrt{3} \cos\frac{k_z}{2} \sum_{i} r_i^y e^{i \bm{k}_\parallel \cdot \bm{r}_i}, \displaybreak[2] \\
 d_{xz}(\bm{k}) &= - \sqrt{3} \sin\frac{k_z}{2} \Im \sum_{i} r_i^x e^{i \bm{k}_\parallel \cdot \bm{r}_i}, \\
 d_{yz}(\bm{k}) &= - \sqrt{3} \sin\frac{k_z}{2} \Im \sum_{i} r_i^y e^{i \bm{k}_\parallel \cdot \bm{r}_i}, \displaybreak[2] \\
 f_{xyz}(\bm{k}) &= - \sqrt{3} \sin\frac{k_z}{2} \Re \sum_{i} r_i^x e^{i \bm{k}_\parallel \cdot \bm{r}_i}, \\
 f_{(x^2 - y^2)z}(\bm{k}) &= - \sqrt{3} \sin\frac{k_z}{2} \Re \sum_{i} r_i^y e^{i \bm{k}_\parallel \cdot \bm{r}_i}.
\end{align}
Pauli matrices in the spin and sublattice space are denoted by $s_i$ and $\sigma_i$, respectively.
$\sigma_+$ and $\sigma_-$ are defined by $\sigma_{\pm} = \sigma_x \pm i \sigma_y$.

A similar model was introduced to investigate the topological superconductivity in UPt$_3$~\cite{Yanase2016, Yanase2017}, and it has recently been studied to show the polar Kerr effect~\cite{WangZ2017} and the odd-frequency Cooper pairs~\cite{Triola2018}.
In these previous studies, the inter-sublattice $p$-wave component was neglected.
Here we take into account the inter-sublattice $p$-wave component and show that it actually plays an essential role for the $j_z$-dependent point node, because the other components vanish on the $K$-$H$ line.
Here we assume here that the $d + f$-wave component is dominant among all the order parameters since the purely $f$-wave state reproduces the multiple superconducting phase diagram illustrated in Fig.~\ref{fig:UPt3_phase}~\cite{Fisher1989, Bruls1990, Adenwalla1990, Sauls1994, Joynt2002}.
On the other hand, an admixture of $p$-wave components allowed by symmetry changes the gap structure.
Thus, we take into account small intra- and inter-sublattice $p$-wave components with $0 < |\delta_1| \ll 1$ and $0 < |\delta_2| \ll 1$, respectively.

Now we briefly review the multiple superconducting phases illustrated in Fig.~\ref{fig:UPt3_phase}~\cite{Fisher1989, Bruls1990, Adenwalla1990, Sauls1994, Joynt2002}.
The A, B, and C phases are characterized by the ratio of two-component order parameters $\eta = \eta_2 / i \eta_1$ summarized in Table~\ref{tab:range_eta}.
A pure imaginary ratio of $\eta_1$ and $\eta_2$ in the B phase implies the chiral superconducting state, which maximally gains the condensation energy.
A recent theoretical study based on our two-sublattice model~\cite{WangZ2017} has shown the polar Kerr effect consistent with the experiment~\cite{Schemm2014}.
Owing to the $p$-wave components, the B phase is a nonunitary state.
It has been considered that the A and C phases are stabilized by weak symmetry breaking of hexagonal structure, possibly induced by weak antiferromagnetic order~\cite{Sauls1994, Joynt2002, Aeppli1988, Hayden1992}.
We assume here that the A phase is the $\Gamma_2$ state ($\eta = \infty$), while the C phase is the $\Gamma_1$ state ($\eta = 0$), and we assume non-negative $\eta \geq 0$ without loss of generality.

\begin{table}[htbp]
 \caption{Range of the parameter $\eta$ in the A, B, and C phases of UPt$_3$.}
 \label{tab:range_eta}
 \begin{center}
  \begin{tabular}{cc} \hline\hline
   A phase & $|\eta| = \infty$ \\
   B phase & $0 \leq |\eta| \leq \infty$ \\
   C phase & $|\eta| = 0$ \\ \hline\hline
  \end{tabular}
 \end{center}
\end{table}

\begin{figure}[tbp]
 \centering
 \includegraphics[width=8cm, clip]{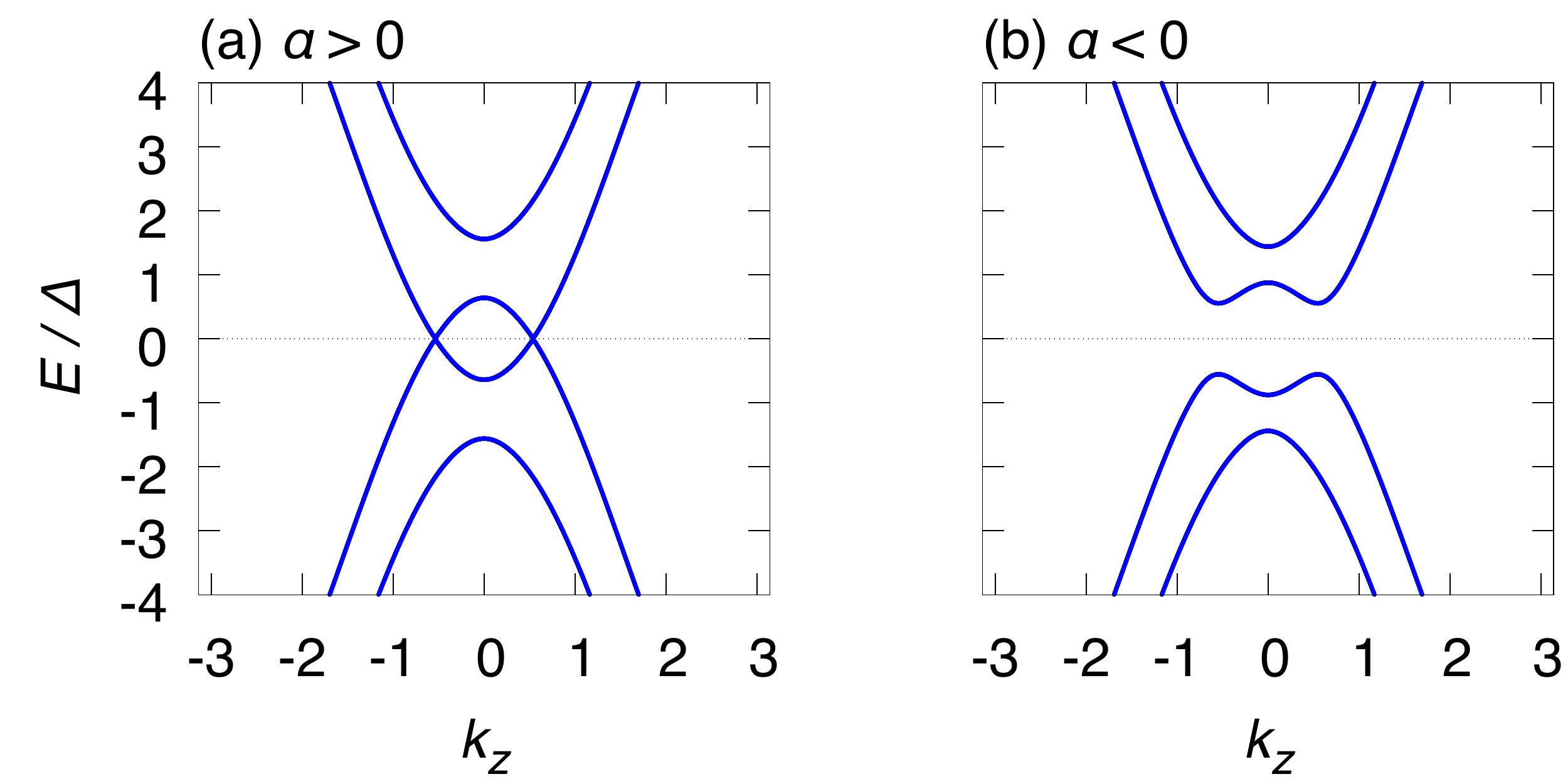}
 \caption{The quasiparticle energy dispersion on the $K$-$H$ line in (a) $\alpha = 0.2 > 0$ case and (b) $\alpha = - 0.2 < 0$ case. We assume the superconducting phase preserving time-reversal symmetry, namely A phase ($\eta = \infty$) or C phase ($\eta = 0$). The other parameters $(t, t_z, t', \mu, \Delta, \delta_1, \delta_2) = (1, - 1, 0.4, - 5.2, 0.5, 0.04, 0.2)$ are assumed so that the $K$-FSs of UPt$_3$ are reproduced.}
 \label{fig:UPt3_eigenKH}
\end{figure}

Analyzing the BdG Hamiltonian Eq.~\eqref{eq:UPt3_model} including the two-component order parameters Eqs.~\eqref{eq:UPt3_order}-\eqref{eq:UPt3_order_2}, we investigate the superconducting gap structures on the $K$-$H$ line.
Figures~\ref{fig:UPt3_eigenKH}(a) and \ref{fig:UPt3_eigenKH}(b) represent the calculated quasiparticle energy dispersion in the A phase, which show that point nodes emerge in the positive $\alpha$ case, while the gap opens in the negative $\alpha$ case.
Qualitatively the same results are obtained in the C phase.
All the results are consistent with the gap classification theory based on the space group (Sec.~\ref{sec:UPt3_gap_classification}).
Thus, it is confirmed that the gap structures depending on the Bloch-state angular momentum $j_z$ are realized on the $K$-FSs of UPt$_3$.

Here we discuss the effects of SSB in the superconducting phase.
As mentioned at the end of Sec.~\ref{sec:gap_classification}, our method of gap classification does not take into account SSB of the ordered state.
Therefore, the results of gap classification [Eqs.~\eqref{eq:UPt3_KHgap_E1/2} and \eqref{eq:UPt3_KHgap_E3/2}] cannot be applied to the SSB phase in a straightforward way.
The two-dimensional $E_{2u}$ state discussed above corresponds to the case.
Thus, we comment on two types of SSB in the superconducting phase: crystal symmetry breaking and time-reversal symmetry breaking.
First, in the $\eta \neq 1$ case, the superconducting order parameter spontaneously breaks 3-fold crystal rotation symmetry.
However, the effects of such SSB due to the superconducting order parameter are considered negligibly small in the weak-coupling region, $\Delta / E_F \ll 1$.
Thus, even though a point node on the $K$-$H$ line is gapped by 3-fold rotation symmetry breaking, the gap should be rather small.
Therefore, the $j_z$-dependent point nodes or gap opening should be experimentally distinguishable, irrespective of the presence or absence of the SSB.

Second, we discuss SSB of time-reversal symmetry in the B phase.
Since the B phase is nonunitary, the 2-fold-degenerate energy band splits by spontaneous time-reversal symmetry breaking in the superconducting phase.
For $\eta = 1$, indeed, the order parameters Eqs.~\eqref{eq:UPt3_order}-\eqref{eq:UPt3_order_2} produce point nodes on the $K$-$H$ line even when the Bloch state belongs to the $E_{1 / 2}$ representation.
Although those point nodes are not protected by the crystal symmetry, they are topologically protected (Weyl node)~\cite{Ishizuka2017}.
Unless the parameter takes the special value $\eta = 1$, our group-theoretical classification is also consistent with the gap structure in the B phase.

In this subsection, we have revealed the $j_z$-dependent point nodes or gap opening corresponding to the sign of the ASOC term in the effective model Eq.~\eqref{eq:UPt3_model}.
The remaining question is which representation is realized in UPt$_3$.
According to our first-principles band-structure calculation, the Bloch state on the $K$-$H$ line belongs to the $E_{1 / 2}$ representation~\cite{Ishizuka2017}.
Combining this first-principles calculation and the gap classification theory, we conclude that the superconducting gap opens on the $K$-$H$ line in UPt$_3$ except for the topologically-protected point nodes emerging in the B phase.

\subsection{MoS$_2$ and SrPtAs (Space group: $P6_3/mmc$)}
\label{sec:point_node_MoS2-SrPtAs}
Now we discuss gap structures in hexagonal superconductors (I) MoS$_2$ and (II) SrPtAs.
These compounds have the same space group symmetry as UPt$_3$.
First, we introduce the backgrounds of these materials below.

(I) MoS$_2$ is a member of the group-VI transition-metal dichalcogenides \textit{MX}$_2$ ($M = \text{Mo}, \text{W}$; $X = \text{S}, \text{Se}, \text{Te}$).
Superconductivity of MoS$_2$ has been observed in the ion-gated atomically thin 2D system~\cite{Ye2012, Lu2015, Shi2015, Saito2016, Costanzo2016}, and in the bulk intercalated system~\cite{Woollam1977, Zhang2016}.
In these electron-doped systems, Mo-$d_{z^2}$ orbitals contribute to the spin-split lowest conduction bands, which form FSs around the $K$ point~\cite{Liu2015}.
However, the FSs do not cross the $K$-$H$ line, because the lowest conduction bands are almost dispersionless along this line due to the absence of the nearest inter-layer hopping~\cite{Akashi2015}.
On the other hand, $d_{x^2 - y^2} \pm id_{xy}$ orbitals of Mo ions contribute to the spin-split top valence bands~\cite{Liu2015}, which have sizable dispersion on the $K$-$H$ line in the 2H stacking structure~\cite{Akashi2015}.
Thus, these top valence bands may form FSs crossing the $K$-$H$ line in a hole-doped MoS$_2$.

(II) SrPtAs is a pnictide superconductor with a hexagonal lattice rather than the square lattice in iron pnictides~\cite{Ishida2009}.
First-principle studies using local density approximation show 2D FSs enclosing the $\Gamma$-$A$ line, a 2D FS enclosing the $K$-$H$ line, and a three-dimensional FS crossing the $K$-$H$ line~\cite{Shein2011, Youn2012, Youn2012_arXiv}.
A muon spin-rotation/relaxation measurement suggests time-reversal symmetry breaking and a nodeless pairing gap~\cite{Biswas2013}.
However, recent $^{195}$Pt-NMR and $^{75}$As-NQR measurements support a spin-singlet $s$-wave superconducting state with an isotropic gap~\cite{Matano2014}.
Since these experimental results look incompatible, the pairing symmetry of SrPtAs is still under debate.

In these materials characterized by the hexagonal space group $P6_3/mmc$, the superconducting gap on the $K$-$H$ line is classified by Eqs.~\eqref{eq:UPt3_KHgap_E1/2} and \eqref{eq:UPt3_KHgap_E3/2}.
The result of the gap classification in each material is discussed below.

(I) Although the symmetry of superconductivity in MoS$_2$ has not been determined, a recent theoretical study~\cite{Nakamura2017} suggests the conventional BCS state ($A_{1g}$) in the paramagnetic regime, and the pair-density-wave (PDW) state ($B_{2u}$) under the external magnetic field.
Since both Eqs.~\eqref{eq:UPt3_KHgap_E1/2} and \eqref{eq:UPt3_KHgap_E3/2} contain $A_{1g}$ and $B_{2u}$ representations, the superconducting gap opens on the $K$-$H$ line irrespective of the Bloch-state angular momentum.
The $A_{1g}$ symmetry is supported by recent first-principle calculations, which take into account electron-phonon interactions~\cite{Ge2013, Rosner2014, Das2015}.
On the other hand, topological superconductivity in electron-doped~\cite{Yuan2014} and hole-doped~\cite{Hsu2017} monolayer MoS$_2$ has been theoretically proposed, where the pairing symmetry is classified into $A_1$ or $E$ representation of the point group $C_{3v}$.
Assuming that the pairing symmetry in bulk MoS$_2$ is the same as that in monolayer MoS$_2$, these representations are induced to $D_{6h}$ as
\begin{align}
 A_1 \uparrow D_{6h} &= A_{1g} + B_{2g} + A_{2u} + B_{1u}, \\
 E \uparrow D_{6h} &= E_{1g} + E_{2g} + E_{1u} + E_{2u}.
\end{align}
Therefore, the presence or absence of point nodes on the $K$-$H$ line depends on the choice of the basis function of the $A_1$ or $E$ representation, which is summarized in Table~\ref{tab:gap_MoS2}.
The superconducting gap structure depends on the effective angular momentum $j_z$ when the pairing symmetry is $A_{2u}$, $B_{1u}$, $E_{1u}$, or $E_{2u}$.

\begin{table}[htbp]
 \caption{Gap structure on the $K$-$H$ line in bulk MoS$_2$ where the pairing symmetry belongs to $A_1$ or $E$ representation of $C_{3v}$. ``PS'' in the first and second columns represents pairing symmetry.}
 \label{tab:gap_MoS2}
 \begin{tabular}{cccc} \hline\hline
  PS in $C_{3v}$ & PS in $D_{6h}$ & Bloch state & Gap structure \\ \hline
  $A_1$ & $A_{1g}$ & $E_{1 / 2}, E_{3 / 2}$ & Gap \\
  & $B_{2g}$ & $E_{1 / 2}, E_{3 / 2}$ & Gap \\
  & $A_{2u}$ & $E_{1 / 2}$ & Point node \\
  & & $E_{3 / 2}$ & Gap \\
  & $B_{1u}$ & $E_{1 / 2}$ & Point node \\
  & & $E_{3 / 2}$ & Gap \\ \hline
  $E$ & $E_{1g}$ & $E_{1 / 2}, E_{3 / 2}$ & Point node \\
  & $E_{2g}$ & $E_{1 / 2}, E_{3 / 2}$ & Point node \\
  & $E_{1u}$ & $E_{1 / 2}$ & Gap \\
  & & $E_{3 / 2}$ & Point node \\
  & $E_{2u}$ & $E_{1 / 2}$ & Gap \\
  & & $E_{3 / 2}$ & Point node \\ \hline\hline
 \end{tabular}
\end{table}

(II) In SrPtAs, on the other hand, the $E_{2g}$ state with a chiral $d$-wave pairing~\cite{Fischer2014} and the $B_{1u}$ state with an $f$-wave pairing~\cite{Goryo2012, WangW2014, Bruckner2014, Fischer2015} have been proposed besides the fully-gapped $A_{1g}$ order parameter suggested by the nuclear magnetic resonance (NMR) measurement~\cite{Matano2014}.
Since $E_{2g}$ representation is not allowed in both Eqs.~\eqref{eq:UPt3_KHgap_E1/2} and \eqref{eq:UPt3_KHgap_E3/2}, the chiral $d$-wave state hosts point nodes on the $K$-$H$ line, which is incompatible with the nodeless gap structure.
The $B_{1u}$ state is consistent with the nodeless gap structure, if the Bloch state on the $K$-$H$ line belongs to $E_{3 / 2}$, although this one-dimensional representation is incompatible with broken time-reversal symmetry.
The $E_{1u}$ and $E_{2u}$ superconducting states for the $E_{1 / 2}$ Bloch state are consistent with both nodeless gap and broken time-reversal symmetry.
However, these odd-parity superconducting states are incompatible with NMR Knight shift measurement, which indicates the decrease of spin susceptibility below $T_{\text{c}}$~\cite{Matano2014}.

We have elucidated the superconducting gap structures on the 3- or 6-fold axes of the hexagonal materials.
On the other hand, cubic systems also have 3-fold axes as illustrated in Fig.~\ref{fig:cubic_BZ}.
In the following part we investigate the existence of $j_z$-dependent point nodes in the cubic superconductors, UBe$_{13}$ and PrOs$_4$Sb$_{12}$.

\subsection{UBe$_{13}$ (Space group: $Fm\bar{3}c$)}
Here we discuss the gap structures in a cubic heavy-fermion superconductor UBe$_{13}$.
Although superconductivity in UBe$_{13}$ was discovered in 1983~\cite{Ott1983}, the nature and the symmetry of superconductivity are still under debate.
A point-nodal $p$-wave~\cite{Ott1984} and line-nodal~\cite{MacLaughlin1984} superconductivity have been proposed, while recent angle-resolved heat-capacity measurements have suggested a fully opened superconducting gap~\cite{Shimizu2015}.
Furthermore, another mystery about UBe$_{13}$ is the emergence of a second phase transition in the superconducting state when a small amount of U atoms are replaced by Th~\cite{Smith1985, Ott1985}; $\mu$SR~\cite{Heffner1990} and thermal-expansion~\cite{Kromer1998} experiments have reported the existence of four superconducting phases.

\begin{figure}[tbp]
 \centering
 \includegraphics[width=8cm, clip]{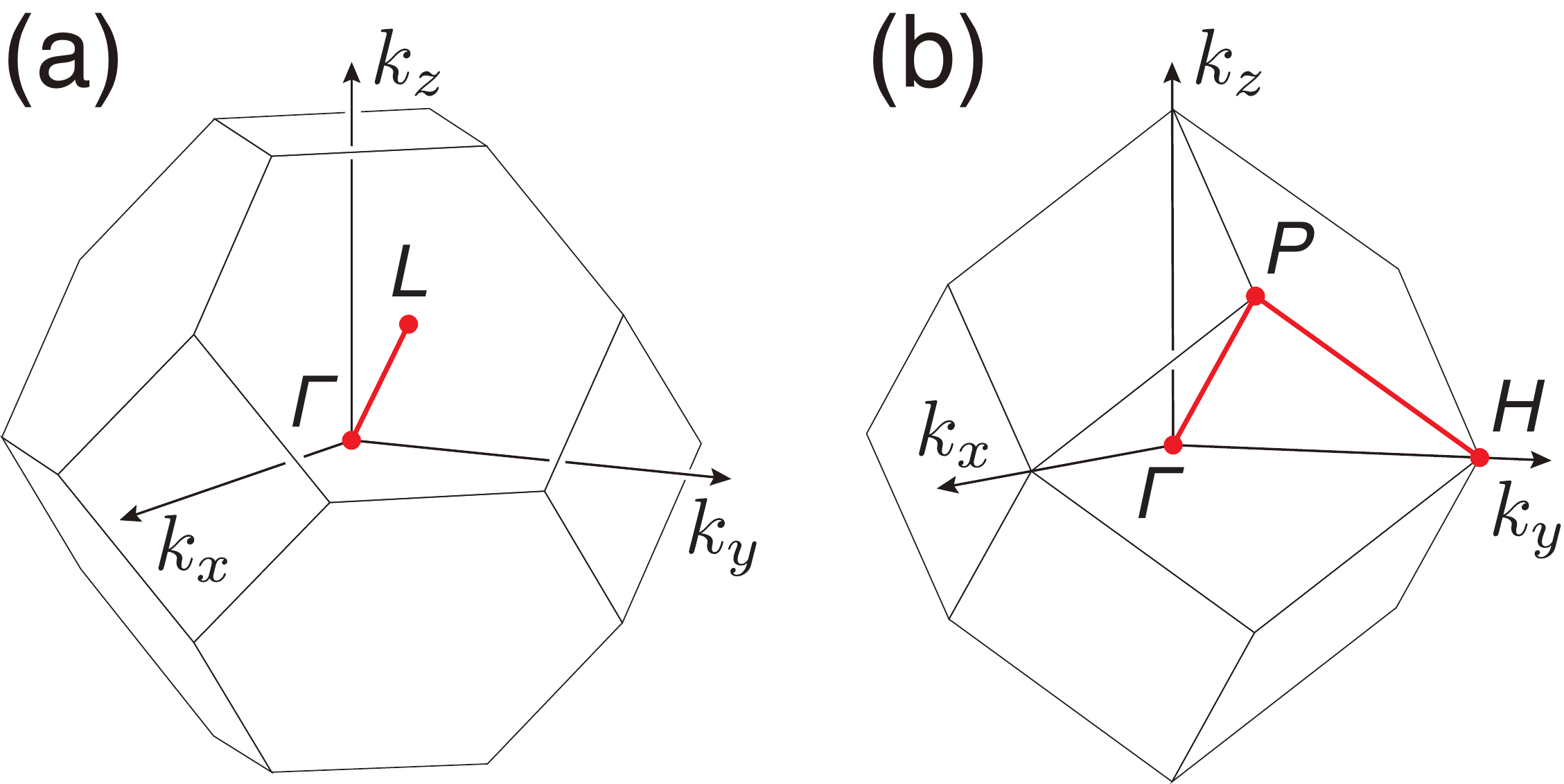}
 \caption{The first BZ of (a) face-centered cubic lattice and (b) body-centered cubic lattice. The red lines show 3-fold rotation symmetric axes.}
 \label{fig:cubic_BZ}
\end{figure}

The space group of UBe$_{13}$ is face-centered cubic $Fm\bar{3}c$, where the BZ has a 3-fold rotation axis $\Gamma$-$L$ [see Fig.~\ref{fig:cubic_BZ}(a)].
Although first-principle calculations show only a tiny FS crossing the $\Gamma$-$L$ line~\cite{Takegahara2000, Maehira2002}, such a FS structure has not been confirmed by experiments.
Thus, we carry out the gap classification on the $\Gamma$-$L$ line, assuming the existence of FSs in the $[1 1 1]$ direction.
The little group on the $\Gamma$-$L$ line has $C_{3v}$ symmetry, which results in the two distinct representations of the Cooper pair wave function given by Eqs.~\eqref{eq:C3v_gap_E1/2} and \eqref{eq:C3v_gap_E3/2}, corresponding to two small representations $\gamma^{\bm{k}} = E_{1 / 2}$ and $E_{3 / 2}$.
Inducing $P^{\bm{k}}$ to the original crystal point group $O_h$, we obtain the induced representation $P^{\bm{k}} \uparrow O_h$ summarized in the following equations:
\begin{subequations}
 \begin{align}
  P^{\bm{k}}_1 \uparrow O_h &= A_{1g} + T_{2g} + A_{1u} + E_u + T_{1u} + 2 T_{2u} \notag \\
  & \hspace*{40mm} (\gamma^{\bm{k}} = E_{1 / 2}), \label{eq:UBe13_GLgap_E1/2} \\
  P^{\bm{k}}_2 \uparrow O_h &= A_{1g} + T_{2g} + 2 A_{1u} + A_{2u} + T_{1u} + 2 T_{2u} \notag \\
  & \hspace*{40mm} (\gamma^{\bm{k}} = E_{3 / 2}). \label{eq:UBe13_GLgap_E3/2}
 \end{align}
\end{subequations}

For UBe$_{13}$, the $E_u$ state~\cite{Shimizu2017, Machida2018} and the accidental $A_{1u} + A_{2u}$ mixed state~\cite{Shimizu2017} have been proposed, consistent with the double transition in U$_{1 - x}$Th$_x$Be$_{13}$.
In the former $E_u$ state, $j_z$-dependent point nodes emerge: the superconducting gap opens for $\gamma^{\bm{k}} = E_{1 / 2}$, while point nodes appear for $\gamma^{\bm{k}} = E_{3 / 2}$ [see Eqs.~\eqref{eq:UBe13_GLgap_E1/2} and \eqref{eq:UBe13_GLgap_E3/2}].
Previous studies based on the $E_u$ scenario simply assumed the $E_{1 / 2}$ Bloch state and obtained the full gap superconducting state in the time-reversal symmetric A and C phases~\cite{Shimizu2017}.
Their results are not valid when the FS crossing the $\Gamma$-$L$ line is formed by the $E_{3 / 2}$ Bloch state.
In the accidentally mixed state, on the other hand, the $A_{1u}$ component makes the gap open irrespective of the angular momentum of the Bloch state.
The $A_{2u}$ component gives rise to the gap only for the $E_{3 / 2}$ Bloch state.
In both $E_u$ and $A_{1u} + A_{2u}$ scenarios, it is necessary to identify the angular momentum of Bloch states in order to relate the symmetry and gap structure of superconductivity.
The experimental data should be carefully interpreted by taking into account this fact.

\subsection{PrOs$_4$Sb$_{12}$ (Space group: $Im\bar{3}$)}
Next, we consider the gap structures in PrOs$_4$Sb$_{12}$.
PrOs$_4$Sb$_{12}$ is a heavy-fermion superconductor with the filled skutterudite structure \textit{RT}$_4$\textit{X}$_{12}$ ($R = \text{rare earth or U}$; $T = \text{Fe}, \text{Ru}, \text{Os}$; $X = \text{P}, \text{As}, \text{Sb}$).
Many studies have reported the manifestation of unconventional superconductivity in PrOs$_4$Sb$_{12}$~\cite{Maple2006, Aoki2007}.
For example, multiple superconducting phases have been suggested by specific-heat~\cite{Maple2002, Vollmer2003} and thermal transport~\cite{Izawa2003} measurements.
However, the superconducting pairing symmetry in PrOs$_4$Sb$_{12}$ remains unclear even now: a point-nodal superconductivity has been suggested by some nuclear quadrupole resonance (NQR)~\cite{Katayama2007}, penetration depth~\cite{Chia2003}, and specific heat~\cite{Maple2002, Vollmer2003} studies, while other thermal conductivity~\cite{Seyfarth2006}, $\mu$SR~\cite{MacLaughlin2002}, and NQR~\cite{Kotegawa2003} measurements have proposed a fully gapped Fermi surface.
Furthermore, several experiments have observed time-reversal symmetry breaking in the low-temperature and low-magnetic-field superconducting phase (B phase)~\cite{Aoki2003, Levenson2016_arXiv}.

Here we carry out a group-theoretical analysis for the gap structure of PrOs$_4$Sb$_{12}$.
PrOs$_4$Sb$_{12}$ has a body-centered cubic space group $Im\bar{3}$, where the BZ has 3-fold rotation axes $\Gamma$-$P$ and $P$-$H$ [see Fig.~\ref{fig:cubic_BZ}(b)].
The FS topology of PrOs$_4$Sb$_{12}$ has been confirmed by the combination of dHvA experiment and first-principles calculation~\cite{Sugawara2002}.
The determined FS consists of three parts, two of which cross the $\Gamma$-$P$ line and the other does the $P$-$H$ line.
Therefore, the gap structure on these lines is worth considering.
The little group on the $\Gamma$-$P$ and $P$-$H$ lines has $C_3$ symmetry, which results in two small representations $\gamma^{\bm{k}} = E_{1 / 2}$ and $2 B_{3 / 2}$, given by Table~\ref{tab:IR_cyclic_point_group}(b).
Corresponding to these two Bloch states, the Cooper pair wave function has two nonequivalent representations as shown in Eqs.~\eqref{eq:gap_3-fold_a} and \eqref{eq:gap_3-fold_b}.
Thus, the induced representation $P^{\bm{k}} \uparrow T_h$ is obtained in the following equations:
\begin{subequations}
 \begin{align}
  P^{\bm{k}}_1 \uparrow T_h &= A_g + T_g + A_u + E_u + 3 T_u & (\gamma^{\bm{k}} = E_{1 / 2}), \label{eq:PrOs4Sb12_GPgap_1/2} \\
  P^{\bm{k}}_2 \uparrow T_h &= A_g + T_g + 3 A_u + 3 T_u & (\gamma^{\bm{k}} = 2 B_{3 / 2}). \label{eq:PrOs4Sb12_GPgap_3/2}
 \end{align}
\end{subequations}

Theoretical studies have suggested various possibilities of the pairing symmetry in PrOs$_4$Sb$_{12}$~\cite{Sergienko2004, Maki2003, Maki2004, Goryo2003, Miyake2003, Ichioka2003}.
For example, the three-dimensional $T_g$ and $T_u$ states~\cite{Sergienko2004}, the mixed $A_g + E_g$ state with a $s + g$-wave pairing~\cite{Maki2003}, and that with a $s + id$-wave pairing~\cite{Goryo2003} have been proposed.
In these cases, the superconducting gap opens on the $\Gamma$-$P$ and $P$-$H$ lines irrespective of the angular momentum of the Bloch state.
However, the $j_z$-dependent point nodes may emerge, if the order parameter belongs to the $E_u$ representation [see Eqs.~\eqref{eq:PrOs4Sb12_GPgap_1/2} and \eqref{eq:PrOs4Sb12_GPgap_3/2}].

\section{Summary and discussion}
\label{sec:summary}
In this paper, we investigated the unconventional superconducting gap structures beyond the results of the Sigrist-Ueda method~\cite{Sigrist-Ueda}.
The group theoretical classification of a gap function enables us to deal with the nonsymmorphic space group symmetry and representations of Bloch wave functions, which are neglected in the Sigrist-Ueda method for classification of an order parameter based on the point group.
Using this method, the nontrivial gap structures have been elucidated as follows.

When the system has symmetry including non-primitive translation parallel to a 2-fold axis, the Cooper pair wave functions on the BP and the ZF, which are perpendicular to the 2-fold axis, have different representations as a consequence of the nonsymmorphic symmetry.
In this case, therefore, line nodes (or a gap opening) protected by nonsymmorphic symmetry may emerge on the BZ boundary.
Indeed, such nontrivial gap structures have been suggested in real materials: UPt$_3$~\cite{Norman1995, Micklitz2009, Kobayashi2016, Yanase2016, Nomoto2016_PRL, Micklitz2017_PRB}, UCoGe~\cite{Nomoto2017}, UPd$_2$Al$_3$~\cite{Fujimoto2006, Nomoto2017, Micklitz2017_PRL}, and Sr$_2$IrO$_4$~\cite{Sumita2017}.
We classified the gap structure of all the centrosymmetric space groups.
From the list of space groups, we may understand the symmetry-protected line node for each crystal, magnetic, and superconducting symmetries.

Furthermore, we clarified the existence of \textit{$j_z$-dependent point nodes (gap opening)} on the 3- or 6-fold axis in the BZ.
The classification of a gap function by Sigrist and Ueda breaks down on these high symmetry axes because the representations of Cooper pairs depend on the angular momentum of the Bloch wave function $j_z$.
Then, the relation between the point nodes and the symmetry of Cooper pairs depends on $j_z$.
We proposed such a $j_z$-dependent point node in a heavy fermion superconductor UPt$_3$ by the group-theoretical analysis and the model calculation.
In UPt$_3$, the angular momentum $j_z$ of Bloch wave functions on the $K$-$H$ line depends on the sign of the ASOC term, since $j_z$ contains the \textit{effective orbital angular momentum} $\lambda_z$ arising from the permutation of sites.
As a result, the sign of ASOC determines whether point nodes emerge on the $K$-$H$ line or not.
This is a rare case in which physical properties depend on the sign of the ASOC term.
Based on the results of classification, we also discussed the gap structure and pairing symmetry in MoS$_2$, SrPtAs, UBe$_{13}$, and PrOs$_4$Sb$_{12}$.
Since the space groups of these superconductors contain 3- or 6-fold rotation symmetry, the gap structure depends on $j_z$.
Thus, we need to appropriately take into account the Bloch wave function for interpretations of experimental data.
Some unconventional superconducting states proposed for these compounds were discussed in light of precise classification theory.

Recently, as represented by $j = 3 / 2$ fermions in half-Heusler superconductors, the angular momentum of electrons in condensed matters (i.e. multipole degrees of freedom) is attracting much attention~\cite{Chadov2010, Lin2010, Xiao2010, Al-Sawai2010, Brydon2016, Kim2016_arXiv, Venderbos2018, Nomoto2016_PRB, Agterberg2017, Timm2017, Bzdusek2017, Savary2017, Boettcher2018}.
Our study sheds light on a new aspect of angular momentum physics in superconductors.
$j_z$-dependent point nodes may emerge especially in odd-parity hexagonal or cubic superconductors.

\begin{acknowledgments}
 The authors are grateful to T.~Nomoto, S.~Kobayashi, M.~Sato, R.~Akashi, J.~Ishizuka, and A.~Daido for fruitful discussions.
 This work was supported by Grant-in Aid for Scientific Research on Innovative Areas ``J-Physics'' (15H05884) and ``Topological Materials Science'' (16H00991) from JSPS of Japan, and by JSPS KAKENHI Grants No. 15K05164, No. 15H05745, and No. 17J09908.
\end{acknowledgments}

\appendix
\section{Proof of Mackey-Bradley theorem in the context of Cooper pair wave function}
\label{app:Mackey-Bradley}
We prove here the Mackey-Bradley theorem~\cite{Bradley, Bradley1970, Mackey1953} described by Eq.~\eqref{eq:Mackey-Bradley} by considering the symmetry transformation of the Cooper pair wave function.
First, we introduce a creation operator of a Bloch state in a band-based representation denoted by $c_{\Gamma \alpha}^\dagger(\bm{k})$, where $\Gamma$ and $\alpha$ are the IR of the little group ${\cal M}^{\bm{k}}$ and its basis, respectively.
This operator is transformed as
\begin{equation}
 m c_{\Gamma \alpha}^\dagger(\bm{k}) m^{- 1} = \sum_{\alpha'} c_{\Gamma \alpha'}^\dagger(\bm{k}) \gamma^{\bm{k}, \Gamma}_{\alpha' \alpha}(m), \label{eq:Bloch_spacegroup_trsf}
\end{equation}
by a space group operation $m \in {\cal M}^{\bm{k}}$, where $\gamma^{\bm{k}, \Gamma}$ is the representation matrix of the small representation $\Gamma$.
Then, we define the Cooper pair wave function
\begin{equation}
 \Delta^{\Gamma}_{\alpha \beta}(\bm{k}) = c_{\Gamma \alpha}^\dagger(\bm{k}) \cdot I c_{\Gamma \beta}^\dagger(\bm{k}) I,
\end{equation}
which is assumed to form a pair between two Bloch states belonging to the same IR $\Gamma$.
Using Eq.~\eqref{eq:Bloch_spacegroup_trsf}, the Cooper pair wave function is transformed by $m \in {\cal M}^{\bm{k}}$ as follows:
\begin{align}
 & m \Delta^{\Gamma}_{\alpha \beta}(\bm{k}) m^{- 1} \notag \\
 &= m c_{\Gamma \alpha}^\dagger(\bm{k}) m^{- 1} \cdot m I c_{\Gamma \beta}^\dagger(\bm{k}) I m^{- 1} \notag \displaybreak[2] \\
 &= m c_{\Gamma \alpha}^\dagger(\bm{k}) m^{- 1} \cdot I (I m I) c_{\Gamma \beta}^\dagger(\bm{k}) (I m I)^{- 1} I \notag \\
 &= \sum_{\alpha'} c_{\Gamma \alpha'}^\dagger(\bm{k}) \gamma^{\bm{k}, \Gamma}_{\alpha' \alpha}(m) \cdot I \biggl\{ \sum_{\beta'} c_{\Gamma \beta'}^\dagger(\bm{k}) \gamma^{\bm{k}, \Gamma}_{\beta' \beta}(I m I) \biggr\} I \notag \displaybreak[2] \\
 &= \sum_{\alpha' \beta'} \{ c_{\Gamma \alpha'}^\dagger(\bm{k}) \cdot I c_{\Gamma \beta'}^\dagger(\bm{k}) I \} \gamma^{\bm{k}, \Gamma}_{\alpha' \alpha}(m) \gamma^{\bm{k}, \Gamma}_{\beta' \beta}(I m I) \notag \\
 &= \sum_{\alpha' \beta'} \Delta^{\Gamma}_{\alpha' \beta'}(\bm{k}) \gamma^{\bm{k}, \Gamma}_{\alpha' \alpha}(m) \gamma^{\bm{k}, \Gamma}_{\beta' \beta}(I m I).
\end{align}
On the other hand, $\Delta^{\Gamma}_{\alpha \beta}(\bm{k})$ is transformed by $I m \in I {\cal M}^{\bm{k}}$ as
\begin{align}
 & I m \Delta^{\Gamma}_{\alpha \beta}(\bm{k}) m^{- 1} I \notag \\
 &= \sum_{\alpha' \beta'} I \{ c_{\Gamma \alpha'}^\dagger(\bm{k}) \cdot I c_{\Gamma \beta'}^\dagger(\bm{k}) I \} I \gamma^{\bm{k}, \Gamma}_{\alpha' \alpha}(m) \gamma^{\bm{k}, \Gamma}_{\beta' \beta}(I m I) \notag \displaybreak[2] \\
 &= \sum_{\alpha' \beta'} \{ I c_{\Gamma \alpha'}^\dagger(\bm{k}) I \cdot c_{\Gamma \beta'}^\dagger(\bm{k}) \} \gamma^{\bm{k}, \Gamma}_{\alpha' \alpha}(m) \gamma^{\bm{k}, \Gamma}_{\beta' \beta}(I m I) \notag \\
 &= - \sum_{\alpha' \beta'} \{ c_{\Gamma \beta'}^\dagger(\bm{k}) \cdot I c_{\Gamma \alpha'}^\dagger(\bm{k}) I \} \gamma^{\bm{k}, \Gamma}_{\alpha' \alpha}(m) \gamma^{\bm{k}, \Gamma}_{\beta' \beta}(I m I) \notag \displaybreak[2] \\
 &= - \sum_{\alpha' \beta'} \{ c_{\Gamma \alpha'}^\dagger(\bm{k}) \cdot I c_{\Gamma \beta'}^\dagger(\bm{k}) I \} \gamma^{\bm{k}, \Gamma}_{\beta' \alpha}(m) \gamma^{\bm{k}, \Gamma}_{\alpha' \beta}(I m I) \notag \\
 &= - \sum_{\alpha' \beta'} \Delta^{\Gamma}_{\alpha' \beta'}(\bm{k}) \gamma^{\bm{k}, \Gamma}_{\beta' \alpha}(m) \gamma^{\bm{k}, \Gamma}_{\alpha' \beta}(I m I),
\end{align}
where we use the anticommutation relation of fermions.

From the above calculations, we obtain the representation of the Cooper pair wave function $P^{\bm{k}, \Gamma}$:
\begin{subequations}
 \begin{align}
  P^{\bm{k}, \Gamma}_{\alpha \beta, \alpha' \beta'}(m) &= \gamma^{\bm{k}, \Gamma}_{\alpha' \alpha}(m) \gamma^{\bm{k}, \Gamma}_{\beta' \beta}(I m I), \\
  P^{\bm{k}, \Gamma}_{\alpha \beta, \alpha' \beta'}(I m) &= - \gamma^{\bm{k}, \Gamma}_{\beta' \alpha}(m) \gamma^{\bm{k}, \Gamma}_{\alpha' \beta}(I m I).
 \end{align}
\end{subequations}
Therefore, the character of $P^{\bm{k}, \Gamma}$ is given by
\begin{subequations}
 \begin{align}
  \chi[P^{\bm{k}, \Gamma}(m)] &= \sum_{\alpha \beta} \gamma^{\bm{k}, \Gamma}_{\alpha \alpha}(m) \gamma^{\bm{k}, \Gamma}_{\beta \beta}(I m I) \notag \\
  &= \chi[\gamma^{\bm{k}, \Gamma}(m)] \chi[\gamma^{\bm{k}, \Gamma}(I m I)], \\
  \chi[P^{\bm{k}, \Gamma}(I m)] &= - \sum_{\alpha \beta} \gamma^{\bm{k}, \Gamma}_{\beta \alpha}(m) \gamma^{\bm{k}, \Gamma}_{\alpha \beta}(I m I) \notag \\
  &= - \chi[\gamma^{\bm{k}, \Gamma}(m I m I)].
 \end{align}
\end{subequations}
These equations are Mackey-Bradley theorem described in Eqs.~\eqref{eq:Mackey-Bradley_a} and \eqref{eq:Mackey-Bradley_b}.

\section{Effective orbital angular momentum due to site permutation}
\label{app:effective_ang_momentum}
In Sec.~\ref{sec:effective_ang_momentum}, we introduced the effective orbital angular momentum $\lambda_z$ due to the permutation of uranium sites.
Here, we provide a general formulation for this angular momentum considering the transformation of the Bloch state wave function.
First, we introduce a creation operator of a Bloch state in a sublattice-based representation denoted by $c_{m \zeta}^\dagger(\bm{k})$, where $m$ and $\zeta = l_z + s_z$ are the indices of the sublattice and angular momentum, respectively.
Fourier transformation of the Bloch state is defined as
\begin{equation}
 c_{m \zeta}^\dagger(\bm{k}) = \sum_{\bm{R}} e^{- i \bm{k} \cdot \bm{R}} c_\zeta^\dagger(\bm{R} + \bm{r}_m),
\end{equation}
where $\bm{R}$ represents the position for the unit cell (lattice vector) and $\bm{r}_m$ is the relative position of the $m$ sublattice in a unit cell.
Using this equation, the creation operator is transformed by a space group operation $g = \{p | \bm{a}\}$ as
\begin{align}
 & g c_{m \zeta}^\dagger(\bm{k}) g^{- 1} \notag \\
 &= \sum_{\bm{R}} e^{- i \bm{k} \cdot \bm{R}} g c_\zeta^\dagger(\bm{R} + \bm{r}_m) g^{- 1} \notag \\
 &= \sum_{\bm{R}} e^{- i \bm{k} \cdot \bm{R}} \sum_{\zeta'} c_{\zeta'}^\dagger(p (\bm{R} + \bm{r}_m) + \bm{a}) D^{(\tilde{j})}_{\zeta' \zeta}(p). \notag
\intertext{Defining $\bm{R}' + \bm{r}_{p m} \equiv p (\bm{R} + \bm{r}_m) + \bm{a}$, we have}
 &= \sum_{\bm{R}'} e^{- i \bm{k} \cdot [p^{- 1} (\bm{R}' + \bm{r}_{p m} - \bm{a} - p \bm{r}_m)]} \notag \\
 & \qquad \times \sum_{\zeta'} c_{\zeta'}^\dagger(\bm{R}' + \bm{r}_{p m}) D^{(\tilde{j})}_{\zeta' \zeta}(p) \notag \displaybreak[2] \\
 &= e^{i p \bm{k} \cdot \bm{a}} \sum_{\zeta'} \left( \sum_{\bm{R}'} e^{- i p \bm{k} \cdot \bm{R}'} c_{\zeta'}^\dagger(\bm{R}' + \bm{r}_{p m}) \right) \notag \\
 & \qquad \times e^{- i p \bm{k} \cdot (\bm{r}_{p m} - p \bm{r}_m)} D^{(\tilde{j})}_{\zeta' \zeta}(p) \notag \displaybreak[2] \\
 &= e^{i p \bm{k} \cdot \bm{a}} \sum_{\zeta'} c_{p m, \zeta'}^\dagger(p \bm{k}) e^{- i p \bm{k} \cdot (\bm{r}_{p m} - p \bm{r}_m)} D^{(\tilde{j})}_{\zeta' \zeta}(p) \notag \\
 &= e^{i p \bm{k} \cdot \bm{a}} \sum_{m'} \sum_{\zeta'} c_{m' \zeta'}^\dagger(p \bm{k}) D^{\text{(perm)}}_{m' m}(p, \bm{k}) D^{(\tilde{j})}_{\zeta' \zeta}(p), \label{eq:trsf_Bloch}
\end{align}
where $D^{(\tilde{j})}(p)$ is a representation matrix of $p$ in $\tilde{j} = l + s$ space.
From Eq.~\eqref{eq:trsf_Bloch}, we define a representation matrix indicating the permutation of sites as
\begin{equation}
 D^{\text{(perm)}}_{m' m}(p, \bm{k}) = e^{- i p \bm{k} \cdot (\bm{r}_{p m} - p \bm{r}_m)} \delta_{m', p m}.
\end{equation}
The phase factor in this matrix corresponds to the effective orbital angular momentum $\lambda_z$.
For example, a 3-fold rotation in UPt$_3$ is represented by
\begin{equation}
 D^{\text{(perm)}}(C_3, \bm{k}) =
  \bordermatrix{
  & a & b \cr
  a & e^{+ i 2\pi / 3} & 0 \cr
  b & 0 & e^{- i 2\pi / 3}
  },
\end{equation}
on the $K$ point $\bm{k} = (4\pi / 3, 0, 0)$.
This phase factor gives the effective angular momentum $\lambda_z = \pm 1$ as we demonstrated in Sec.~\ref{sec:effective_ang_momentum}.


\begin{thebibliography}{134}%
\makeatletter
\providecommand \@ifxundefined [1]{%
 \@ifx{#1\undefined}
}%
\providecommand \@ifnum [1]{%
 \ifnum #1\expandafter \@firstoftwo
 \else \expandafter \@secondoftwo
 \fi
}%
\providecommand \@ifx [1]{%
 \ifx #1\expandafter \@firstoftwo
 \else \expandafter \@secondoftwo
 \fi
}%
\providecommand \natexlab [1]{#1}%
\providecommand \enquote  [1]{``#1''}%
\providecommand \bibnamefont  [1]{#1}%
\providecommand \bibfnamefont [1]{#1}%
\providecommand \citenamefont [1]{#1}%
\providecommand \href@noop [0]{\@secondoftwo}%
\providecommand \href [0]{\begingroup \@sanitize@url \@href}%
\providecommand \@href[1]{\@@startlink{#1}\@@href}%
\providecommand \@@href[1]{\endgroup#1\@@endlink}%
\providecommand \@sanitize@url [0]{\catcode `\\12\catcode `\$12\catcode
  `\&12\catcode `\#12\catcode `\^12\catcode `\_12\catcode `\%12\relax}%
\providecommand \@@startlink[1]{}%
\providecommand \@@endlink[0]{}%
\providecommand \url  [0]{\begingroup\@sanitize@url \@url }%
\providecommand \@url [1]{\endgroup\@href {#1}{\urlprefix }}%
\providecommand \urlprefix  [0]{URL }%
\providecommand \Eprint [0]{\href }%
\providecommand \doibase [0]{http://dx.doi.org/}%
\providecommand \selectlanguage [0]{\@gobble}%
\providecommand \bibinfo  [0]{\@secondoftwo}%
\providecommand \bibfield  [0]{\@secondoftwo}%
\providecommand \translation [1]{[#1]}%
\providecommand \BibitemOpen [0]{}%
\providecommand \bibitemStop [0]{}%
\providecommand \bibitemNoStop [0]{.\EOS\space}%
\providecommand \EOS [0]{\spacefactor3000\relax}%
\providecommand \BibitemShut  [1]{\csname bibitem#1\endcsname}%
\let\auto@bib@innerbib\@empty
\bibitem [{\citenamefont {Matsuda}\ \emph {et~al.}(2006)\citenamefont
  {Matsuda}, \citenamefont {Izawa},\ and\ \citenamefont
  {Vekhter}}]{Matsuda2006}%
  \BibitemOpen
  \bibfield  {author} {\bibinfo {author} {\bibfnamefont {Y.}~\bibnamefont
  {Matsuda}}, \bibinfo {author} {\bibfnamefont {K.}~\bibnamefont {Izawa}}, \
  and\ \bibinfo {author} {\bibfnamefont {I.}~\bibnamefont {Vekhter}},\ }\href
  {http://stacks.iop.org/0953-8984/18/i=44/a=R01} {\bibfield  {journal}
  {\bibinfo  {journal} {J. Phys.: Conden. Matter}\ }\textbf {\bibinfo {volume}
  {18}},\ \bibinfo {pages} {R705} (\bibinfo {year} {2006})}\BibitemShut
  {NoStop}%
\bibitem [{\citenamefont {Sakakibara}\ \emph {et~al.}(2016)\citenamefont
  {Sakakibara}, \citenamefont {Kittaka},\ and\ \citenamefont
  {Machida}}]{Sakakibara2016}%
  \BibitemOpen
  \bibfield  {author} {\bibinfo {author} {\bibfnamefont {T.}~\bibnamefont
  {Sakakibara}}, \bibinfo {author} {\bibfnamefont {S.}~\bibnamefont {Kittaka}},
  \ and\ \bibinfo {author} {\bibfnamefont {K.}~\bibnamefont {Machida}},\ }\href
  {http://stacks.iop.org/0034-4885/79/i=9/a=094002} {\bibfield  {journal}
  {\bibinfo  {journal} {Rep. Prog. Phys.}\ }\textbf {\bibinfo {volume} {79}},\
  \bibinfo {pages} {094002} (\bibinfo {year} {2016})}\BibitemShut {NoStop}%
\bibitem [{\citenamefont {Volovik}\ and\ \citenamefont
  {Gor'kov}(1984)}]{Volovik1984}%
  \BibitemOpen
  \bibfield  {author} {\bibinfo {author} {\bibfnamefont {G.~E.}\ \bibnamefont
  {Volovik}}\ and\ \bibinfo {author} {\bibfnamefont {L.~P.}\ \bibnamefont
  {Gor'kov}},\ }\href@noop {} {\bibfield  {journal} {\bibinfo  {journal}
  {Pis'ma Zh. Eksp. Teor. Fiz.}\ }\textbf {\bibinfo {volume} {39}},\ \bibinfo
  {pages} {550} (\bibinfo {year} {1984})}\BibitemShut {NoStop}%
\bibitem [{\citenamefont {Volovik}\ and\ \citenamefont
  {Gor'kov}(1985)}]{Volovik1985}%
  \BibitemOpen
  \bibfield  {author} {\bibinfo {author} {\bibfnamefont {G.~E.}\ \bibnamefont
  {Volovik}}\ and\ \bibinfo {author} {\bibfnamefont {L.~P.}\ \bibnamefont
  {Gor'kov}},\ }\href@noop {} {\bibfield  {journal} {\bibinfo  {journal} {Zh.
  Eksp. Teor. Fiz.}\ }\textbf {\bibinfo {volume} {88}},\ \bibinfo {pages}
  {1412} (\bibinfo {year} {1985})}\BibitemShut {NoStop}%
\bibitem [{\citenamefont {Anderson}(1984)}]{Anderson1984}%
  \BibitemOpen
  \bibfield  {author} {\bibinfo {author} {\bibfnamefont {P.~W.}\ \bibnamefont
  {Anderson}},\ }\href {\doibase 10.1103/PhysRevB.30.4000} {\bibfield
  {journal} {\bibinfo  {journal} {Phys. Rev. B}\ }\textbf {\bibinfo {volume}
  {30}},\ \bibinfo {pages} {4000} (\bibinfo {year} {1984})}\BibitemShut
  {NoStop}%
\bibitem [{\citenamefont {Sigrist}\ and\ \citenamefont
  {Ueda}(1991)}]{Sigrist-Ueda}%
  \BibitemOpen
  \bibfield  {author} {\bibinfo {author} {\bibfnamefont {M.}~\bibnamefont
  {Sigrist}}\ and\ \bibinfo {author} {\bibfnamefont {K.}~\bibnamefont {Ueda}},\
  }\href {\doibase 10.1103/RevModPhys.63.239} {\bibfield  {journal} {\bibinfo
  {journal} {Rev. Mod. Phys.}\ }\textbf {\bibinfo {volume} {63}},\ \bibinfo
  {pages} {239} (\bibinfo {year} {1991})}\BibitemShut {NoStop}%
\bibitem [{\citenamefont {Yarzhemsky}\ and\ \citenamefont
  {Murav'ev}(1992)}]{Yarzhemsky1992}%
  \BibitemOpen
  \bibfield  {author} {\bibinfo {author} {\bibfnamefont {V.~G.}\ \bibnamefont
  {Yarzhemsky}}\ and\ \bibinfo {author} {\bibfnamefont {E.~N.}\ \bibnamefont
  {Murav'ev}},\ }\href {http://stacks.iop.org/0953-8984/4/i=13/a=015}
  {\bibfield  {journal} {\bibinfo  {journal} {J. Phys.: Condens. Matter}\
  }\textbf {\bibinfo {volume} {4}},\ \bibinfo {pages} {3525} (\bibinfo {year}
  {1992})}\BibitemShut {NoStop}%
\bibitem [{\citenamefont {Norman}(1995)}]{Norman1995}%
  \BibitemOpen
  \bibfield  {author} {\bibinfo {author} {\bibfnamefont {M.~R.}\ \bibnamefont
  {Norman}},\ }\href {\doibase 10.1103/PhysRevB.52.15093} {\bibfield  {journal}
  {\bibinfo  {journal} {Phys. Rev. B}\ }\textbf {\bibinfo {volume} {52}},\
  \bibinfo {pages} {15093} (\bibinfo {year} {1995})}\BibitemShut {NoStop}%
\bibitem [{\citenamefont {Yarzhemsky}(1998)}]{Yarzhemsky1998}%
  \BibitemOpen
  \bibfield  {author} {\bibinfo {author} {\bibfnamefont {V.~G.}\ \bibnamefont
  {Yarzhemsky}},\ }\href {\doibase
  10.1002/(SICI)1521-3951(199809)209:1<101::AID-PSSB101>3.0.CO;2-N} {\bibfield
  {journal} {\bibinfo  {journal} {Phys. Status Solidi B}\ }\textbf {\bibinfo
  {volume} {209}},\ \bibinfo {pages} {101} (\bibinfo {year}
  {1998})}\BibitemShut {NoStop}%
\bibitem [{\citenamefont {Yarzhemsky}(2000)}]{Yarzhemsky2000}%
  \BibitemOpen
  \bibfield  {author} {\bibinfo {author} {\bibfnamefont {V.~G.}\ \bibnamefont
  {Yarzhemsky}},\ }\href {\doibase
  10.1002/1097-461X(2000)80:2<133::AID-QUA9>3.0.CO;2-B} {\bibfield  {journal}
  {\bibinfo  {journal} {Int. J. Quant. Chem.}\ }\textbf {\bibinfo {volume}
  {80}},\ \bibinfo {pages} {133} (\bibinfo {year} {2000})}\BibitemShut
  {NoStop}%
\bibitem [{\citenamefont {Yarzhemsky}(2003)}]{Yarzhemsky2003}%
  \BibitemOpen
  \bibfield  {author} {\bibinfo {author} {\bibfnamefont {V.~G.}\ \bibnamefont
  {Yarzhemsky}},\ }\href {\doibase 10.1063/1.1612402} {\bibfield  {journal}
  {\bibinfo  {journal} {AIP Conf. Proc.}\ }\textbf {\bibinfo {volume} {678}},\
  \bibinfo {pages} {343} (\bibinfo {year} {2003})}\BibitemShut {NoStop}%
\bibitem [{\citenamefont {Yarzhemsky}(2008)}]{Yarzhemsky2008}%
  \BibitemOpen
  \bibfield  {author} {\bibinfo {author} {\bibfnamefont {V.~G.}\ \bibnamefont
  {Yarzhemsky}},\ }\href@noop {} {\bibfield  {journal} {\bibinfo  {journal} {J.
  Opt. Adv. Mater.}\ }\textbf {\bibinfo {volume} {10}},\ \bibinfo {pages}
  {1759} (\bibinfo {year} {2008})}\BibitemShut {NoStop}%
\bibitem [{\citenamefont {Micklitz}\ and\ \citenamefont
  {Norman}(2009)}]{Micklitz2009}%
  \BibitemOpen
  \bibfield  {author} {\bibinfo {author} {\bibfnamefont {T.}~\bibnamefont
  {Micklitz}}\ and\ \bibinfo {author} {\bibfnamefont {M.~R.}\ \bibnamefont
  {Norman}},\ }\href {\doibase 10.1103/PhysRevB.80.100506} {\bibfield
  {journal} {\bibinfo  {journal} {Phys. Rev. B}\ }\textbf {\bibinfo {volume}
  {80}},\ \bibinfo {pages} {100506} (\bibinfo {year} {2009})}\BibitemShut
  {NoStop}%
\bibitem [{\citenamefont {Kobayashi}\ \emph {et~al.}(2016)\citenamefont
  {Kobayashi}, \citenamefont {Yanase},\ and\ \citenamefont
  {Sato}}]{Kobayashi2016}%
  \BibitemOpen
  \bibfield  {author} {\bibinfo {author} {\bibfnamefont {S.}~\bibnamefont
  {Kobayashi}}, \bibinfo {author} {\bibfnamefont {Y.}~\bibnamefont {Yanase}}, \
  and\ \bibinfo {author} {\bibfnamefont {M.}~\bibnamefont {Sato}},\ }\href
  {\doibase 10.1103/PhysRevB.94.134512} {\bibfield  {journal} {\bibinfo
  {journal} {Phys. Rev. B}\ }\textbf {\bibinfo {volume} {94}},\ \bibinfo
  {pages} {134512} (\bibinfo {year} {2016})}\BibitemShut {NoStop}%
\bibitem [{\citenamefont {Yanase}(2016)}]{Yanase2016}%
  \BibitemOpen
  \bibfield  {author} {\bibinfo {author} {\bibfnamefont {Y.}~\bibnamefont
  {Yanase}},\ }\href {\doibase 10.1103/PhysRevB.94.174502} {\bibfield
  {journal} {\bibinfo  {journal} {Phys. Rev. B}\ }\textbf {\bibinfo {volume}
  {94}},\ \bibinfo {pages} {174502} (\bibinfo {year} {2016})}\BibitemShut
  {NoStop}%
\bibitem [{\citenamefont {Nomoto}\ and\ \citenamefont
  {Ikeda}(2016)}]{Nomoto2016_PRL}%
  \BibitemOpen
  \bibfield  {author} {\bibinfo {author} {\bibfnamefont {T.}~\bibnamefont
  {Nomoto}}\ and\ \bibinfo {author} {\bibfnamefont {H.}~\bibnamefont {Ikeda}},\
  }\href {\doibase 10.1103/PhysRevLett.117.217002} {\bibfield  {journal}
  {\bibinfo  {journal} {Phys. Rev. Lett.}\ }\textbf {\bibinfo {volume} {117}},\
  \bibinfo {pages} {217002} (\bibinfo {year} {2016})}\BibitemShut {NoStop}%
\bibitem [{\citenamefont {Micklitz}\ and\ \citenamefont
  {Norman}(2017{\natexlab{a}})}]{Micklitz2017_PRB}%
  \BibitemOpen
  \bibfield  {author} {\bibinfo {author} {\bibfnamefont {T.}~\bibnamefont
  {Micklitz}}\ and\ \bibinfo {author} {\bibfnamefont {M.~R.}\ \bibnamefont
  {Norman}},\ }\href {\doibase 10.1103/PhysRevB.95.024508} {\bibfield
  {journal} {\bibinfo  {journal} {Phys. Rev. B}\ }\textbf {\bibinfo {volume}
  {95}},\ \bibinfo {pages} {024508} (\bibinfo {year}
  {2017}{\natexlab{a}})}\BibitemShut {NoStop}%
\bibitem [{\citenamefont {Nomoto}\ and\ \citenamefont
  {Ikeda}(2017)}]{Nomoto2017}%
  \BibitemOpen
  \bibfield  {author} {\bibinfo {author} {\bibfnamefont {T.}~\bibnamefont
  {Nomoto}}\ and\ \bibinfo {author} {\bibfnamefont {H.}~\bibnamefont {Ikeda}},\
  }\href {\doibase 10.7566/JPSJ.86.023703} {\bibfield  {journal} {\bibinfo
  {journal} {J. Phys. Soc. Jpn.}\ }\textbf {\bibinfo {volume} {86}},\ \bibinfo
  {pages} {023703} (\bibinfo {year} {2017})}\BibitemShut {NoStop}%
\bibitem [{\citenamefont {Micklitz}\ and\ \citenamefont
  {Norman}(2017{\natexlab{b}})}]{Micklitz2017_PRL}%
  \BibitemOpen
  \bibfield  {author} {\bibinfo {author} {\bibfnamefont {T.}~\bibnamefont
  {Micklitz}}\ and\ \bibinfo {author} {\bibfnamefont {M.~R.}\ \bibnamefont
  {Norman}},\ }\href {\doibase 10.1103/PhysRevLett.118.207001} {\bibfield
  {journal} {\bibinfo  {journal} {Phys. Rev. Lett.}\ }\textbf {\bibinfo
  {volume} {118}},\ \bibinfo {pages} {207001} (\bibinfo {year}
  {2017}{\natexlab{b}})}\BibitemShut {NoStop}%
\bibitem [{\citenamefont {Sumita}\ \emph {et~al.}(2017)\citenamefont {Sumita},
  \citenamefont {Nomoto},\ and\ \citenamefont {Yanase}}]{Sumita2017}%
  \BibitemOpen
  \bibfield  {author} {\bibinfo {author} {\bibfnamefont {S.}~\bibnamefont
  {Sumita}}, \bibinfo {author} {\bibfnamefont {T.}~\bibnamefont {Nomoto}}, \
  and\ \bibinfo {author} {\bibfnamefont {Y.}~\bibnamefont {Yanase}},\ }\href
  {\doibase 10.1103/PhysRevLett.119.027001} {\bibfield  {journal} {\bibinfo
  {journal} {Phys. Rev. Lett.}\ }\textbf {\bibinfo {volume} {119}},\ \bibinfo
  {pages} {027001} (\bibinfo {year} {2017})}\BibitemShut {NoStop}%
\bibitem [{\citenamefont {Blount}(1985)}]{Blount1985}%
  \BibitemOpen
  \bibfield  {author} {\bibinfo {author} {\bibfnamefont {E.~I.}\ \bibnamefont
  {Blount}},\ }\href {\doibase 10.1103/PhysRevB.32.2935} {\bibfield  {journal}
  {\bibinfo  {journal} {Phys. Rev. B}\ }\textbf {\bibinfo {volume} {32}},\
  \bibinfo {pages} {2935} (\bibinfo {year} {1985})}\BibitemShut {NoStop}%
\bibitem [{\citenamefont {Daido}\ and\ \citenamefont
  {Yanase}(2016)}]{Daido2016}%
  \BibitemOpen
  \bibfield  {author} {\bibinfo {author} {\bibfnamefont {A.}~\bibnamefont
  {Daido}}\ and\ \bibinfo {author} {\bibfnamefont {Y.}~\bibnamefont {Yanase}},\
  }\href {\doibase 10.1103/PhysRevB.94.054519} {\bibfield  {journal} {\bibinfo
  {journal} {Phys. Rev. B}\ }\textbf {\bibinfo {volume} {94}},\ \bibinfo
  {pages} {054519} (\bibinfo {year} {2016})}\BibitemShut {NoStop}%
\bibitem [{\citenamefont {Yanase}\ and\ \citenamefont
  {Shiozaki}(2017)}]{Yanase2017}%
  \BibitemOpen
  \bibfield  {author} {\bibinfo {author} {\bibfnamefont {Y.}~\bibnamefont
  {Yanase}}\ and\ \bibinfo {author} {\bibfnamefont {K.}~\bibnamefont
  {Shiozaki}},\ }\href {\doibase 10.1103/PhysRevB.95.224514} {\bibfield
  {journal} {\bibinfo  {journal} {Phys. Rev. B}\ }\textbf {\bibinfo {volume}
  {95}},\ \bibinfo {pages} {224514} (\bibinfo {year} {2017})}\BibitemShut
  {NoStop}%
\bibitem [{\citenamefont {Kozii}\ \emph {et~al.}(2016)\citenamefont {Kozii},
  \citenamefont {Venderbos},\ and\ \citenamefont {Fu}}]{Kozii2016}%
  \BibitemOpen
  \bibfield  {author} {\bibinfo {author} {\bibfnamefont {V.}~\bibnamefont
  {Kozii}}, \bibinfo {author} {\bibfnamefont {J.~W.~F.}\ \bibnamefont
  {Venderbos}}, \ and\ \bibinfo {author} {\bibfnamefont {L.}~\bibnamefont
  {Fu}},\ }\href@noop {} {\bibfield  {journal} {\bibinfo  {journal} {Sci.
  Adv.}\ }\textbf {\bibinfo {volume} {2}},\ \bibinfo {pages} {e1601835}
  (\bibinfo {year} {2016})}\BibitemShut {NoStop}%
\bibitem [{\citenamefont {Venderbos}\ \emph {et~al.}(2016)\citenamefont
  {Venderbos}, \citenamefont {Kozii},\ and\ \citenamefont
  {Fu}}]{Venderbos2016}%
  \BibitemOpen
  \bibfield  {author} {\bibinfo {author} {\bibfnamefont {J.~W.~F.}\
  \bibnamefont {Venderbos}}, \bibinfo {author} {\bibfnamefont {V.}~\bibnamefont
  {Kozii}}, \ and\ \bibinfo {author} {\bibfnamefont {L.}~\bibnamefont {Fu}},\
  }\href {\doibase 10.1103/PhysRevB.94.180504} {\bibfield  {journal} {\bibinfo
  {journal} {Phys. Rev. B}\ }\textbf {\bibinfo {volume} {94}},\ \bibinfo
  {pages} {180504} (\bibinfo {year} {2016})}\BibitemShut {NoStop}%
\bibitem [{\citenamefont {Sau}\ and\ \citenamefont {Tewari}(2012)}]{Sau2012}%
  \BibitemOpen
  \bibfield  {author} {\bibinfo {author} {\bibfnamefont {J.~D.}\ \bibnamefont
  {Sau}}\ and\ \bibinfo {author} {\bibfnamefont {S.}~\bibnamefont {Tewari}},\
  }\href {\doibase 10.1103/PhysRevB.86.104509} {\bibfield  {journal} {\bibinfo
  {journal} {Phys. Rev. B}\ }\textbf {\bibinfo {volume} {86}},\ \bibinfo
  {pages} {104509} (\bibinfo {year} {2012})}\BibitemShut {NoStop}%
\bibitem [{\citenamefont {Goswami}\ and\ \citenamefont
  {Nevidomskyy}(2015)}]{Goswami2015}%
  \BibitemOpen
  \bibfield  {author} {\bibinfo {author} {\bibfnamefont {P.}~\bibnamefont
  {Goswami}}\ and\ \bibinfo {author} {\bibfnamefont {A.~H.}\ \bibnamefont
  {Nevidomskyy}},\ }\href {\doibase 10.1103/PhysRevB.92.214504} {\bibfield
  {journal} {\bibinfo  {journal} {Phys. Rev. B}\ }\textbf {\bibinfo {volume}
  {92}},\ \bibinfo {pages} {214504} (\bibinfo {year} {2015})}\BibitemShut
  {NoStop}%
\bibitem [{\citenamefont {Fischer}\ \emph {et~al.}(2014)\citenamefont
  {Fischer}, \citenamefont {Neupert}, \citenamefont {Platt}, \citenamefont
  {Schnyder}, \citenamefont {Hanke}, \citenamefont {Goryo}, \citenamefont
  {Thomale},\ and\ \citenamefont {Sigrist}}]{Fischer2014}%
  \BibitemOpen
  \bibfield  {author} {\bibinfo {author} {\bibfnamefont {M.~H.}\ \bibnamefont
  {Fischer}}, \bibinfo {author} {\bibfnamefont {T.}~\bibnamefont {Neupert}},
  \bibinfo {author} {\bibfnamefont {C.}~\bibnamefont {Platt}}, \bibinfo
  {author} {\bibfnamefont {A.~P.}\ \bibnamefont {Schnyder}}, \bibinfo {author}
  {\bibfnamefont {W.}~\bibnamefont {Hanke}}, \bibinfo {author} {\bibfnamefont
  {J.}~\bibnamefont {Goryo}}, \bibinfo {author} {\bibfnamefont
  {R.}~\bibnamefont {Thomale}}, \ and\ \bibinfo {author} {\bibfnamefont
  {M.}~\bibnamefont {Sigrist}},\ }\href {\doibase 10.1103/PhysRevB.89.020509}
  {\bibfield  {journal} {\bibinfo  {journal} {Phys. Rev. B}\ }\textbf {\bibinfo
  {volume} {89}},\ \bibinfo {pages} {020509} (\bibinfo {year}
  {2014})}\BibitemShut {NoStop}%
\bibitem [{\citenamefont {Meng}\ and\ \citenamefont
  {Balents}(2012)}]{Meng2012}%
  \BibitemOpen
  \bibfield  {author} {\bibinfo {author} {\bibfnamefont {T.}~\bibnamefont
  {Meng}}\ and\ \bibinfo {author} {\bibfnamefont {L.}~\bibnamefont {Balents}},\
  }\href {\doibase 10.1103/PhysRevB.86.054504} {\bibfield  {journal} {\bibinfo
  {journal} {Phys. Rev. B}\ }\textbf {\bibinfo {volume} {86}},\ \bibinfo
  {pages} {054504} (\bibinfo {year} {2012})}\BibitemShut {NoStop}%
\bibitem [{\citenamefont {Yang}\ \emph {et~al.}(2014)\citenamefont {Yang},
  \citenamefont {Pan},\ and\ \citenamefont {Zhang}}]{Yang2014}%
  \BibitemOpen
  \bibfield  {author} {\bibinfo {author} {\bibfnamefont {S.~A.}\ \bibnamefont
  {Yang}}, \bibinfo {author} {\bibfnamefont {H.}~\bibnamefont {Pan}}, \ and\
  \bibinfo {author} {\bibfnamefont {F.}~\bibnamefont {Zhang}},\ }\href
  {\doibase 10.1103/PhysRevLett.113.046401} {\bibfield  {journal} {\bibinfo
  {journal} {Phys. Rev. Lett.}\ }\textbf {\bibinfo {volume} {113}},\ \bibinfo
  {pages} {046401} (\bibinfo {year} {2014})}\BibitemShut {NoStop}%
\bibitem [{\citenamefont {Volovik}(2017)}]{Volovik2017}%
  \BibitemOpen
  \bibfield  {author} {\bibinfo {author} {\bibfnamefont {G.~E.}\ \bibnamefont
  {Volovik}},\ }\href {\doibase 10.1134/S0021364017040063} {\bibfield
  {journal} {\bibinfo  {journal} {JETP Letters}\ }\textbf {\bibinfo {volume}
  {105}},\ \bibinfo {pages} {273} (\bibinfo {year} {2017})}\BibitemShut
  {NoStop}%
\bibitem [{\citenamefont {Bradley}\ and\ \citenamefont
  {Cracknell}(1972)}]{Bradley}%
  \BibitemOpen
  \bibfield  {author} {\bibinfo {author} {\bibfnamefont {C.~J.}\ \bibnamefont
  {Bradley}}\ and\ \bibinfo {author} {\bibfnamefont {A.~P.}\ \bibnamefont
  {Cracknell}},\ }\href@noop {} {\emph {\bibinfo {title} {The Mathematical
  Theory of Symmetry in Solids}}}\ (\bibinfo  {publisher} {Oxford University
  Press},\ \bibinfo {address} {Oxford},\ \bibinfo {year} {1972})\BibitemShut
  {NoStop}%
\bibitem [{\citenamefont {Bradley}\ and\ \citenamefont
  {Davies}(1970)}]{Bradley1970}%
  \BibitemOpen
  \bibfield  {author} {\bibinfo {author} {\bibfnamefont {C.~J.}\ \bibnamefont
  {Bradley}}\ and\ \bibinfo {author} {\bibfnamefont {B.~L.}\ \bibnamefont
  {Davies}},\ }\href@noop {} {\bibfield  {journal} {\bibinfo  {journal} {J.
  Math. Phys.}\ }\textbf {\bibinfo {volume} {11}},\ \bibinfo {pages} {1536}
  (\bibinfo {year} {1970})}\BibitemShut {NoStop}%
\bibitem [{\citenamefont {Mackey}(1953)}]{Mackey1953}%
  \BibitemOpen
  \bibfield  {author} {\bibinfo {author} {\bibfnamefont {G.~W.}\ \bibnamefont
  {Mackey}},\ }\href@noop {} {\bibfield  {journal} {\bibinfo  {journal} {Am. J.
  Math.}\ }\textbf {\bibinfo {volume} {75}},\ \bibinfo {pages} {387} (\bibinfo
  {year} {1953})}\BibitemShut {NoStop}%
\bibitem [{\citenamefont {Izyumov}\ \emph {et~al.}(1989)\citenamefont
  {Izyumov}, \citenamefont {Laptev},\ and\ \citenamefont
  {Syromyatnikov}}]{Izyumov1989}%
  \BibitemOpen
  \bibfield  {author} {\bibinfo {author} {\bibfnamefont {Y.}~\bibnamefont
  {Izyumov}}, \bibinfo {author} {\bibfnamefont {V.}~\bibnamefont {Laptev}}, \
  and\ \bibinfo {author} {\bibfnamefont {V.}~\bibnamefont {Syromyatnikov}},\
  }\href {\doibase 10.1142/S0217979289000890} {\bibfield  {journal} {\bibinfo
  {journal} {Int. J. Mod. Phys. B}\ }\textbf {\bibinfo {volume} {3}},\ \bibinfo
  {pages} {1377} (\bibinfo {year} {1989})}\BibitemShut {NoStop}%
\bibitem [{\citenamefont {Fujimoto}(2006)}]{Fujimoto2006}%
  \BibitemOpen
  \bibfield  {author} {\bibinfo {author} {\bibfnamefont {S.}~\bibnamefont
  {Fujimoto}},\ }\href {\doibase 10.1143/JPSJ.75.083704} {\bibfield  {journal}
  {\bibinfo  {journal} {J. Phys. Soc. Jpn.}\ }\textbf {\bibinfo {volume}
  {75}},\ \bibinfo {pages} {083704} (\bibinfo {year} {2006})}\BibitemShut
  {NoStop}%
\bibitem [{\citenamefont {Kobayashi}\ \emph {et~al.}()\citenamefont
  {Kobayashi}, \citenamefont {Sumita}, \citenamefont {Yanase},\ and\
  \citenamefont {Sato}}]{Kobayashi2017_arXiv}%
  \BibitemOpen
  \bibfield  {author} {\bibinfo {author} {\bibfnamefont {S.}~\bibnamefont
  {Kobayashi}}, \bibinfo {author} {\bibfnamefont {S.}~\bibnamefont {Sumita}},
  \bibinfo {author} {\bibfnamefont {Y.}~\bibnamefont {Yanase}}, \ and\ \bibinfo
  {author} {\bibfnamefont {M.}~\bibnamefont {Sato}},\ }\Eprint
  {http://arxiv.org/abs/1711.06421} {arXiv:1711.06421} \BibitemShut {NoStop}%
\bibitem [{Note1()}]{Note1}%
  \BibitemOpen
  \bibinfo {note} {The results in Table~\ref {tab:cooper_pair_n-fold_axis} are
  not changed even when the system has non-primitive translations parallel to
  the $n$-fold axis, because the phase factor arising from the translations is
  cancelled during calculation of the Mackey-Bradley theorem.}\BibitemShut
  {Stop}%
\bibitem [{\citenamefont {Herzberg}(1966)}]{Herzberg}%
  \BibitemOpen
  \bibfield  {author} {\bibinfo {author} {\bibfnamefont {G.}~\bibnamefont
  {Herzberg}},\ }\href@noop {} {\emph {\bibinfo {title} {Molecular Spectra and
  Molecular Structure}}},\ Vol.~\bibinfo {volume} {3}\ (\bibinfo  {publisher}
  {Van Nostrand},\ \bibinfo {address} {New York},\ \bibinfo {year}
  {1966})\BibitemShut {NoStop}%
\bibitem [{\citenamefont {Inui}\ \emph {et~al.}(1990)\citenamefont {Inui},
  \citenamefont {Tanabe},\ and\ \citenamefont {Onodera}}]{Inui-Tanabe-Onodera}%
  \BibitemOpen
  \bibfield  {author} {\bibinfo {author} {\bibfnamefont {T.}~\bibnamefont
  {Inui}}, \bibinfo {author} {\bibfnamefont {Y.}~\bibnamefont {Tanabe}}, \ and\
  \bibinfo {author} {\bibfnamefont {Y.}~\bibnamefont {Onodera}},\ }\href@noop
  {} {\emph {\bibinfo {title} {Group theory and its applications in
  physics}}},\ \bibinfo {series} {Springer Series in Solid-State Sciences},
  Vol.~\bibinfo {volume} {78}\ (\bibinfo  {publisher} {Springer-Verlag Berlin
  Heidelberg},\ \bibinfo {address} {Berlin, Heidelberg},\ \bibinfo {year}
  {1990})\BibitemShut {NoStop}%
\bibitem [{\citenamefont {Stewart}\ \emph {et~al.}(1984)\citenamefont
  {Stewart}, \citenamefont {Fisk}, \citenamefont {Willis},\ and\ \citenamefont
  {Smith}}]{Stewart1984}%
  \BibitemOpen
  \bibfield  {author} {\bibinfo {author} {\bibfnamefont {G.~R.}\ \bibnamefont
  {Stewart}}, \bibinfo {author} {\bibfnamefont {Z.}~\bibnamefont {Fisk}},
  \bibinfo {author} {\bibfnamefont {J.~O.}\ \bibnamefont {Willis}}, \ and\
  \bibinfo {author} {\bibfnamefont {J.~L.}\ \bibnamefont {Smith}},\ }\href
  {\doibase 10.1103/PhysRevLett.52.679} {\bibfield  {journal} {\bibinfo
  {journal} {Phys. Rev. Lett.}\ }\textbf {\bibinfo {volume} {52}},\ \bibinfo
  {pages} {679} (\bibinfo {year} {1984})}\BibitemShut {NoStop}%
\bibitem [{\citenamefont {Fisher}\ \emph {et~al.}(1989)\citenamefont {Fisher},
  \citenamefont {Kim}, \citenamefont {Woodfield}, \citenamefont {Phillips},
  \citenamefont {Taillefer}, \citenamefont {Hasselbach}, \citenamefont
  {Flouquet}, \citenamefont {Giorgi},\ and\ \citenamefont
  {Smith}}]{Fisher1989}%
  \BibitemOpen
  \bibfield  {author} {\bibinfo {author} {\bibfnamefont {R.~A.}\ \bibnamefont
  {Fisher}}, \bibinfo {author} {\bibfnamefont {S.}~\bibnamefont {Kim}},
  \bibinfo {author} {\bibfnamefont {B.~F.}\ \bibnamefont {Woodfield}}, \bibinfo
  {author} {\bibfnamefont {N.~E.}\ \bibnamefont {Phillips}}, \bibinfo {author}
  {\bibfnamefont {L.}~\bibnamefont {Taillefer}}, \bibinfo {author}
  {\bibfnamefont {K.}~\bibnamefont {Hasselbach}}, \bibinfo {author}
  {\bibfnamefont {J.}~\bibnamefont {Flouquet}}, \bibinfo {author}
  {\bibfnamefont {A.~L.}\ \bibnamefont {Giorgi}}, \ and\ \bibinfo {author}
  {\bibfnamefont {J.~L.}\ \bibnamefont {Smith}},\ }\href {\doibase
  10.1103/PhysRevLett.62.1411} {\bibfield  {journal} {\bibinfo  {journal}
  {Phys. Rev. Lett.}\ }\textbf {\bibinfo {volume} {62}},\ \bibinfo {pages}
  {1411} (\bibinfo {year} {1989})}\BibitemShut {NoStop}%
\bibitem [{\citenamefont {Bruls}\ \emph {et~al.}(1990)\citenamefont {Bruls},
  \citenamefont {Weber}, \citenamefont {Wolf}, \citenamefont {Thalmeier},
  \citenamefont {L\"uthi}, \citenamefont {de~Visser},\ and\ \citenamefont
  {Menovsky}}]{Bruls1990}%
  \BibitemOpen
  \bibfield  {author} {\bibinfo {author} {\bibfnamefont {G.}~\bibnamefont
  {Bruls}}, \bibinfo {author} {\bibfnamefont {D.}~\bibnamefont {Weber}},
  \bibinfo {author} {\bibfnamefont {B.}~\bibnamefont {Wolf}}, \bibinfo {author}
  {\bibfnamefont {P.}~\bibnamefont {Thalmeier}}, \bibinfo {author}
  {\bibfnamefont {B.}~\bibnamefont {L\"uthi}}, \bibinfo {author} {\bibfnamefont
  {A.}~\bibnamefont {de~Visser}}, \ and\ \bibinfo {author} {\bibfnamefont
  {A.}~\bibnamefont {Menovsky}},\ }\href {\doibase 10.1103/PhysRevLett.65.2294}
  {\bibfield  {journal} {\bibinfo  {journal} {Phys. Rev. Lett.}\ }\textbf
  {\bibinfo {volume} {65}},\ \bibinfo {pages} {2294} (\bibinfo {year}
  {1990})}\BibitemShut {NoStop}%
\bibitem [{\citenamefont {Adenwalla}\ \emph {et~al.}(1990)\citenamefont
  {Adenwalla}, \citenamefont {Lin}, \citenamefont {Ran}, \citenamefont {Zhao},
  \citenamefont {Ketterson}, \citenamefont {Sauls}, \citenamefont {Taillefer},
  \citenamefont {Hinks}, \citenamefont {Levy},\ and\ \citenamefont
  {Sarma}}]{Adenwalla1990}%
  \BibitemOpen
  \bibfield  {author} {\bibinfo {author} {\bibfnamefont {S.}~\bibnamefont
  {Adenwalla}}, \bibinfo {author} {\bibfnamefont {S.~W.}\ \bibnamefont {Lin}},
  \bibinfo {author} {\bibfnamefont {Q.~Z.}\ \bibnamefont {Ran}}, \bibinfo
  {author} {\bibfnamefont {Z.}~\bibnamefont {Zhao}}, \bibinfo {author}
  {\bibfnamefont {J.~B.}\ \bibnamefont {Ketterson}}, \bibinfo {author}
  {\bibfnamefont {J.~A.}\ \bibnamefont {Sauls}}, \bibinfo {author}
  {\bibfnamefont {L.}~\bibnamefont {Taillefer}}, \bibinfo {author}
  {\bibfnamefont {D.~G.}\ \bibnamefont {Hinks}}, \bibinfo {author}
  {\bibfnamefont {M.}~\bibnamefont {Levy}}, \ and\ \bibinfo {author}
  {\bibfnamefont {B.~K.}\ \bibnamefont {Sarma}},\ }\href {\doibase
  10.1103/PhysRevLett.65.2298} {\bibfield  {journal} {\bibinfo  {journal}
  {Phys. Rev. Lett.}\ }\textbf {\bibinfo {volume} {65}},\ \bibinfo {pages}
  {2298} (\bibinfo {year} {1990})}\BibitemShut {NoStop}%
\bibitem [{\citenamefont {Tou}\ \emph {et~al.}(1998)\citenamefont {Tou},
  \citenamefont {Kitaoka}, \citenamefont {Ishida}, \citenamefont {Asayama},
  \citenamefont {Kimura}, \citenamefont {\={O}nuki}, \citenamefont {Yamamoto},
  \citenamefont {Haga},\ and\ \citenamefont {Maezawa}}]{Tou1998}%
  \BibitemOpen
  \bibfield  {author} {\bibinfo {author} {\bibfnamefont {H.}~\bibnamefont
  {Tou}}, \bibinfo {author} {\bibfnamefont {Y.}~\bibnamefont {Kitaoka}},
  \bibinfo {author} {\bibfnamefont {K.}~\bibnamefont {Ishida}}, \bibinfo
  {author} {\bibfnamefont {K.}~\bibnamefont {Asayama}}, \bibinfo {author}
  {\bibfnamefont {N.}~\bibnamefont {Kimura}}, \bibinfo {author} {\bibfnamefont
  {Y.}~\bibnamefont {\={O}nuki}}, \bibinfo {author} {\bibfnamefont
  {E.}~\bibnamefont {Yamamoto}}, \bibinfo {author} {\bibfnamefont
  {Y.}~\bibnamefont {Haga}}, \ and\ \bibinfo {author} {\bibfnamefont
  {K.}~\bibnamefont {Maezawa}},\ }\href {\doibase 10.1103/PhysRevLett.80.3129}
  {\bibfield  {journal} {\bibinfo  {journal} {Phys. Rev. Lett.}\ }\textbf
  {\bibinfo {volume} {80}},\ \bibinfo {pages} {3129} (\bibinfo {year}
  {1998})}\BibitemShut {NoStop}%
\bibitem [{\citenamefont {Sauls}(1994)}]{Sauls1994}%
  \BibitemOpen
  \bibfield  {author} {\bibinfo {author} {\bibfnamefont {J.~A.}\ \bibnamefont
  {Sauls}},\ }\href {\doibase 10.1080/00018739400101475} {\bibfield  {journal}
  {\bibinfo  {journal} {Advances in Physics}\ }\textbf {\bibinfo {volume}
  {43}},\ \bibinfo {pages} {113} (\bibinfo {year} {1994})}\BibitemShut
  {NoStop}%
\bibitem [{\citenamefont {Joynt}\ and\ \citenamefont
  {Taillefer}(2002)}]{Joynt2002}%
  \BibitemOpen
  \bibfield  {author} {\bibinfo {author} {\bibfnamefont {R.}~\bibnamefont
  {Joynt}}\ and\ \bibinfo {author} {\bibfnamefont {L.}~\bibnamefont
  {Taillefer}},\ }\href {\doibase 10.1103/RevModPhys.74.235} {\bibfield
  {journal} {\bibinfo  {journal} {Rev. Mod. Phys.}\ }\textbf {\bibinfo {volume}
  {74}},\ \bibinfo {pages} {235} (\bibinfo {year} {2002})}\BibitemShut
  {NoStop}%
\bibitem [{\citenamefont {Strand}\ \emph {et~al.}(2010)\citenamefont {Strand},
  \citenamefont {Bahr}, \citenamefont {Van~Harlingen}, \citenamefont {Davis},
  \citenamefont {Gannon},\ and\ \citenamefont {Halperin}}]{Strand2010}%
  \BibitemOpen
  \bibfield  {author} {\bibinfo {author} {\bibfnamefont {J.~D.}\ \bibnamefont
  {Strand}}, \bibinfo {author} {\bibfnamefont {D.~J.}\ \bibnamefont {Bahr}},
  \bibinfo {author} {\bibfnamefont {D.~J.}\ \bibnamefont {Van~Harlingen}},
  \bibinfo {author} {\bibfnamefont {J.~P.}\ \bibnamefont {Davis}}, \bibinfo
  {author} {\bibfnamefont {W.~J.}\ \bibnamefont {Gannon}}, \ and\ \bibinfo
  {author} {\bibfnamefont {W.~P.}\ \bibnamefont {Halperin}},\ }\href {\doibase
  10.1126/science.1187943} {\bibfield  {journal} {\bibinfo  {journal}
  {Science}\ }\textbf {\bibinfo {volume} {328}},\ \bibinfo {pages} {1368}
  (\bibinfo {year} {2010})}\BibitemShut {NoStop}%
\bibitem [{\citenamefont {Luke}\ \emph {et~al.}(1993)\citenamefont {Luke},
  \citenamefont {Keren}, \citenamefont {Le}, \citenamefont {Wu}, \citenamefont
  {Uemura}, \citenamefont {Bonn}, \citenamefont {Taillefer},\ and\
  \citenamefont {Garrett}}]{Luke1993}%
  \BibitemOpen
  \bibfield  {author} {\bibinfo {author} {\bibfnamefont {G.~M.}\ \bibnamefont
  {Luke}}, \bibinfo {author} {\bibfnamefont {A.}~\bibnamefont {Keren}},
  \bibinfo {author} {\bibfnamefont {L.~P.}\ \bibnamefont {Le}}, \bibinfo
  {author} {\bibfnamefont {W.~D.}\ \bibnamefont {Wu}}, \bibinfo {author}
  {\bibfnamefont {Y.~J.}\ \bibnamefont {Uemura}}, \bibinfo {author}
  {\bibfnamefont {D.~A.}\ \bibnamefont {Bonn}}, \bibinfo {author}
  {\bibfnamefont {L.}~\bibnamefont {Taillefer}}, \ and\ \bibinfo {author}
  {\bibfnamefont {J.~D.}\ \bibnamefont {Garrett}},\ }\href {\doibase
  10.1103/PhysRevLett.71.1466} {\bibfield  {journal} {\bibinfo  {journal}
  {Phys. Rev. Lett.}\ }\textbf {\bibinfo {volume} {71}},\ \bibinfo {pages}
  {1466} (\bibinfo {year} {1993})}\BibitemShut {NoStop}%
\bibitem [{\citenamefont {Schemm}\ \emph {et~al.}(2014)\citenamefont {Schemm},
  \citenamefont {Gannon}, \citenamefont {Wishne}, \citenamefont {Halperin},\
  and\ \citenamefont {Kapitulnik}}]{Schemm2014}%
  \BibitemOpen
  \bibfield  {author} {\bibinfo {author} {\bibfnamefont {E.~R.}\ \bibnamefont
  {Schemm}}, \bibinfo {author} {\bibfnamefont {W.~J.}\ \bibnamefont {Gannon}},
  \bibinfo {author} {\bibfnamefont {C.~M.}\ \bibnamefont {Wishne}}, \bibinfo
  {author} {\bibfnamefont {W.~P.}\ \bibnamefont {Halperin}}, \ and\ \bibinfo
  {author} {\bibfnamefont {A.}~\bibnamefont {Kapitulnik}},\ }\href {\doibase
  10.1126/science.1248552} {\bibfield  {journal} {\bibinfo  {journal}
  {Science}\ }\textbf {\bibinfo {volume} {345}},\ \bibinfo {pages} {190}
  (\bibinfo {year} {2014})}\BibitemShut {NoStop}%
\bibitem [{Note2()}]{Note2}%
  \BibitemOpen
  \bibinfo {note} {Symmetry breaking by a weak crystal distortion has been
  reported~\cite {Walko2001}, although its reliability is under debate. We here
  assume high symmetry space group $P6_{3}/mmc$.}\BibitemShut {Stop}%
\bibitem [{\citenamefont {Walko}\ \emph {et~al.}(2001)\citenamefont {Walko},
  \citenamefont {Hong}, \citenamefont {Chandrasekhar~Rao}, \citenamefont
  {Wawrzak}, \citenamefont {Seidman}, \citenamefont {Halperin},\ and\
  \citenamefont {Bedzyk}}]{Walko2001}%
  \BibitemOpen
  \bibfield  {author} {\bibinfo {author} {\bibfnamefont {D.~A.}\ \bibnamefont
  {Walko}}, \bibinfo {author} {\bibfnamefont {J.-I.}\ \bibnamefont {Hong}},
  \bibinfo {author} {\bibfnamefont {T.~V.}\ \bibnamefont {Chandrasekhar~Rao}},
  \bibinfo {author} {\bibfnamefont {Z.}~\bibnamefont {Wawrzak}}, \bibinfo
  {author} {\bibfnamefont {D.~N.}\ \bibnamefont {Seidman}}, \bibinfo {author}
  {\bibfnamefont {W.~P.}\ \bibnamefont {Halperin}}, \ and\ \bibinfo {author}
  {\bibfnamefont {M.~J.}\ \bibnamefont {Bedzyk}},\ }\href {\doibase
  10.1103/PhysRevB.63.054522} {\bibfield  {journal} {\bibinfo  {journal} {Phys.
  Rev. B}\ }\textbf {\bibinfo {volume} {63}},\ \bibinfo {pages} {054522}
  (\bibinfo {year} {2001})}\BibitemShut {NoStop}%
\bibitem [{\citenamefont {Taillefer}\ and\ \citenamefont
  {Lonzarich}(1988)}]{Taillefer1988}%
  \BibitemOpen
  \bibfield  {author} {\bibinfo {author} {\bibfnamefont {L.}~\bibnamefont
  {Taillefer}}\ and\ \bibinfo {author} {\bibfnamefont {G.~G.}\ \bibnamefont
  {Lonzarich}},\ }\href {\doibase 10.1103/PhysRevLett.60.1570} {\bibfield
  {journal} {\bibinfo  {journal} {Phys. Rev. Lett.}\ }\textbf {\bibinfo
  {volume} {60}},\ \bibinfo {pages} {1570} (\bibinfo {year}
  {1988})}\BibitemShut {NoStop}%
\bibitem [{\citenamefont {Norman}\ \emph {et~al.}(1988)\citenamefont {Norman},
  \citenamefont {Albers}, \citenamefont {Boring},\ and\ \citenamefont
  {Christensen}}]{Norman1988}%
  \BibitemOpen
  \bibfield  {author} {\bibinfo {author} {\bibfnamefont {M.~R.}\ \bibnamefont
  {Norman}}, \bibinfo {author} {\bibfnamefont {R.~C.}\ \bibnamefont {Albers}},
  \bibinfo {author} {\bibfnamefont {A.~M.}\ \bibnamefont {Boring}}, \ and\
  \bibinfo {author} {\bibfnamefont {N.~E.}\ \bibnamefont {Christensen}},\
  }\href {\doibase http://dx.doi.org/10.1016/0038-1098(88)91109-X} {\bibfield
  {journal} {\bibinfo  {journal} {Solid State Communications}\ }\textbf
  {\bibinfo {volume} {68}},\ \bibinfo {pages} {245} (\bibinfo {year}
  {1988})}\BibitemShut {NoStop}%
\bibitem [{\citenamefont {Kimura}\ \emph {et~al.}(1995)\citenamefont {Kimura},
  \citenamefont {Settai}, \citenamefont {\={O}nuki}, \citenamefont {Toshima},
  \citenamefont {Yamamoto}, \citenamefont {Maezawa}, \citenamefont {Aoki},\
  and\ \citenamefont {Harima}}]{Kimura1995}%
  \BibitemOpen
  \bibfield  {author} {\bibinfo {author} {\bibfnamefont {N.}~\bibnamefont
  {Kimura}}, \bibinfo {author} {\bibfnamefont {R.}~\bibnamefont {Settai}},
  \bibinfo {author} {\bibfnamefont {Y.}~\bibnamefont {\={O}nuki}}, \bibinfo
  {author} {\bibfnamefont {H.}~\bibnamefont {Toshima}}, \bibinfo {author}
  {\bibfnamefont {E.}~\bibnamefont {Yamamoto}}, \bibinfo {author}
  {\bibfnamefont {K.}~\bibnamefont {Maezawa}}, \bibinfo {author} {\bibfnamefont
  {H.}~\bibnamefont {Aoki}}, \ and\ \bibinfo {author} {\bibfnamefont
  {H.}~\bibnamefont {Harima}},\ }\href {\doibase 10.1143/JPSJ.64.3881}
  {\bibfield  {journal} {\bibinfo  {journal} {J. Phys. Soc. Jpn.}\ }\textbf
  {\bibinfo {volume} {64}},\ \bibinfo {pages} {3881} (\bibinfo {year}
  {1995})}\BibitemShut {NoStop}%
\bibitem [{\citenamefont {McMullan}\ \emph {et~al.}(2008)\citenamefont
  {McMullan}, \citenamefont {Rourke}, \citenamefont {Norman}, \citenamefont
  {Huxley}, \citenamefont {Doiron-Leyraud}, \citenamefont {Flouquet},
  \citenamefont {Lonzarich}, \citenamefont {McCollam},\ and\ \citenamefont
  {Julian}}]{McMullan2008}%
  \BibitemOpen
  \bibfield  {author} {\bibinfo {author} {\bibfnamefont {G.~J.}\ \bibnamefont
  {McMullan}}, \bibinfo {author} {\bibfnamefont {P.~M.~C.}\ \bibnamefont
  {Rourke}}, \bibinfo {author} {\bibfnamefont {M.~R.}\ \bibnamefont {Norman}},
  \bibinfo {author} {\bibfnamefont {A.~D.}\ \bibnamefont {Huxley}}, \bibinfo
  {author} {\bibfnamefont {N.}~\bibnamefont {Doiron-Leyraud}}, \bibinfo
  {author} {\bibfnamefont {J.}~\bibnamefont {Flouquet}}, \bibinfo {author}
  {\bibfnamefont {G.~G.}\ \bibnamefont {Lonzarich}}, \bibinfo {author}
  {\bibfnamefont {A.}~\bibnamefont {McCollam}}, \ and\ \bibinfo {author}
  {\bibfnamefont {S.~R.}\ \bibnamefont {Julian}},\ }\href
  {http://stacks.iop.org/1367-2630/10/i=5/a=053029} {\bibfield  {journal}
  {\bibinfo  {journal} {New Journal of Physics}\ }\textbf {\bibinfo {volume}
  {10}},\ \bibinfo {pages} {053029} (\bibinfo {year} {2008})}\BibitemShut
  {NoStop}%
\bibitem [{\citenamefont {Kane}\ and\ \citenamefont
  {Mele}(2005)}]{Kane-Mele2005}%
  \BibitemOpen
  \bibfield  {author} {\bibinfo {author} {\bibfnamefont {C.~L.}\ \bibnamefont
  {Kane}}\ and\ \bibinfo {author} {\bibfnamefont {E.~J.}\ \bibnamefont
  {Mele}},\ }\href {\doibase 10.1103/PhysRevLett.95.226801} {\bibfield
  {journal} {\bibinfo  {journal} {Phys. Rev. Lett.}\ }\textbf {\bibinfo
  {volume} {95}},\ \bibinfo {pages} {226801} (\bibinfo {year}
  {2005})}\BibitemShut {NoStop}%
\bibitem [{\citenamefont {Saito}\ \emph {et~al.}(2016)\citenamefont {Saito},
  \citenamefont {Nakamura}, \citenamefont {Bahramy}, \citenamefont {Kohama},
  \citenamefont {Ye}, \citenamefont {Kasahara}, \citenamefont {Nakagawa},
  \citenamefont {Onga}, \citenamefont {Tokunaga}, \citenamefont {Nojima},
  \citenamefont {Yanase},\ and\ \citenamefont {Iwasa}}]{Saito2016}%
  \BibitemOpen
  \bibfield  {author} {\bibinfo {author} {\bibfnamefont {Y.}~\bibnamefont
  {Saito}}, \bibinfo {author} {\bibfnamefont {Y.}~\bibnamefont {Nakamura}},
  \bibinfo {author} {\bibfnamefont {M.~S.}\ \bibnamefont {Bahramy}}, \bibinfo
  {author} {\bibfnamefont {Y.}~\bibnamefont {Kohama}}, \bibinfo {author}
  {\bibfnamefont {J.}~\bibnamefont {Ye}}, \bibinfo {author} {\bibfnamefont
  {Y.}~\bibnamefont {Kasahara}}, \bibinfo {author} {\bibfnamefont
  {Y.}~\bibnamefont {Nakagawa}}, \bibinfo {author} {\bibfnamefont
  {M.}~\bibnamefont {Onga}}, \bibinfo {author} {\bibfnamefont {M.}~\bibnamefont
  {Tokunaga}}, \bibinfo {author} {\bibfnamefont {T.}~\bibnamefont {Nojima}},
  \bibinfo {author} {\bibfnamefont {Y.}~\bibnamefont {Yanase}}, \ and\ \bibinfo
  {author} {\bibfnamefont {Y.}~\bibnamefont {Iwasa}},\ }\href@noop {}
  {\bibfield  {journal} {\bibinfo  {journal} {Nat. Phys.}\ }\textbf {\bibinfo
  {volume} {12}},\ \bibinfo {pages} {144} (\bibinfo {year} {2016})}\BibitemShut
  {NoStop}%
\bibitem [{\citenamefont {Fischer}\ \emph {et~al.}(2011)\citenamefont
  {Fischer}, \citenamefont {Loder},\ and\ \citenamefont
  {Sigrist}}]{Fischer2011}%
  \BibitemOpen
  \bibfield  {author} {\bibinfo {author} {\bibfnamefont {M.~H.}\ \bibnamefont
  {Fischer}}, \bibinfo {author} {\bibfnamefont {F.}~\bibnamefont {Loder}}, \
  and\ \bibinfo {author} {\bibfnamefont {M.}~\bibnamefont {Sigrist}},\ }\href
  {\doibase 10.1103/PhysRevB.84.184533} {\bibfield  {journal} {\bibinfo
  {journal} {Phys. Rev. B}\ }\textbf {\bibinfo {volume} {84}},\ \bibinfo
  {pages} {184533} (\bibinfo {year} {2011})}\BibitemShut {NoStop}%
\bibitem [{\citenamefont {Maruyama}\ \emph {et~al.}(2012)\citenamefont
  {Maruyama}, \citenamefont {Sigrist},\ and\ \citenamefont
  {Yanase}}]{Maruyama2012}%
  \BibitemOpen
  \bibfield  {author} {\bibinfo {author} {\bibfnamefont {D.}~\bibnamefont
  {Maruyama}}, \bibinfo {author} {\bibfnamefont {M.}~\bibnamefont {Sigrist}}, \
  and\ \bibinfo {author} {\bibfnamefont {Y.}~\bibnamefont {Yanase}},\ }\href
  {\doibase 10.1143/JPSJ.81.034702} {\bibfield  {journal} {\bibinfo  {journal}
  {J. Phys. Soc. Jpn.}\ }\textbf {\bibinfo {volume} {81}},\ \bibinfo {pages}
  {034702} (\bibinfo {year} {2012})}\BibitemShut {NoStop}%
\bibitem [{\citenamefont {Akashi}\ \emph {et~al.}(2015)\citenamefont {Akashi},
  \citenamefont {Ochi}, \citenamefont {Bord\'acs}, \citenamefont {Suzuki},
  \citenamefont {Tokura}, \citenamefont {Iwasa},\ and\ \citenamefont
  {Arita}}]{Akashi2015}%
  \BibitemOpen
  \bibfield  {author} {\bibinfo {author} {\bibfnamefont {R.}~\bibnamefont
  {Akashi}}, \bibinfo {author} {\bibfnamefont {M.}~\bibnamefont {Ochi}},
  \bibinfo {author} {\bibfnamefont {S.}~\bibnamefont {Bord\'acs}}, \bibinfo
  {author} {\bibfnamefont {R.}~\bibnamefont {Suzuki}}, \bibinfo {author}
  {\bibfnamefont {Y.}~\bibnamefont {Tokura}}, \bibinfo {author} {\bibfnamefont
  {Y.}~\bibnamefont {Iwasa}}, \ and\ \bibinfo {author} {\bibfnamefont
  {R.}~\bibnamefont {Arita}},\ }\href {\doibase
  10.1103/PhysRevApplied.4.014002} {\bibfield  {journal} {\bibinfo  {journal}
  {Phys. Rev. Applied}\ }\textbf {\bibinfo {volume} {4}},\ \bibinfo {pages}
  {014002} (\bibinfo {year} {2015})}\BibitemShut {NoStop}%
\bibitem [{\citenamefont {Akashi}\ \emph {et~al.}(2017)\citenamefont {Akashi},
  \citenamefont {Iida}, \citenamefont {Yamamoto},\ and\ \citenamefont
  {Yoshizawa}}]{Akashi2017}%
  \BibitemOpen
  \bibfield  {author} {\bibinfo {author} {\bibfnamefont {R.}~\bibnamefont
  {Akashi}}, \bibinfo {author} {\bibfnamefont {Y.}~\bibnamefont {Iida}},
  \bibinfo {author} {\bibfnamefont {K.}~\bibnamefont {Yamamoto}}, \ and\
  \bibinfo {author} {\bibfnamefont {K.}~\bibnamefont {Yoshizawa}},\ }\href
  {\doibase 10.1103/PhysRevB.95.245401} {\bibfield  {journal} {\bibinfo
  {journal} {Phys. Rev. B}\ }\textbf {\bibinfo {volume} {95}},\ \bibinfo
  {pages} {245401} (\bibinfo {year} {2017})}\BibitemShut {NoStop}%
\bibitem [{\citenamefont {Wang}\ \emph {et~al.}(2017)\citenamefont {Wang},
  \citenamefont {Berlinsky}, \citenamefont {Zwicknagl},\ and\ \citenamefont
  {Kallin}}]{WangZ2017}%
  \BibitemOpen
  \bibfield  {author} {\bibinfo {author} {\bibfnamefont {Z.}~\bibnamefont
  {Wang}}, \bibinfo {author} {\bibfnamefont {J.}~\bibnamefont {Berlinsky}},
  \bibinfo {author} {\bibfnamefont {G.}~\bibnamefont {Zwicknagl}}, \ and\
  \bibinfo {author} {\bibfnamefont {C.}~\bibnamefont {Kallin}},\ }\href
  {\doibase 10.1103/PhysRevB.96.174511} {\bibfield  {journal} {\bibinfo
  {journal} {Phys. Rev. B}\ }\textbf {\bibinfo {volume} {96}},\ \bibinfo
  {pages} {174511} (\bibinfo {year} {2017})}\BibitemShut {NoStop}%
\bibitem [{\citenamefont {Triola}\ and\ \citenamefont
  {Black-Schaffer}(2018)}]{Triola2018}%
  \BibitemOpen
  \bibfield  {author} {\bibinfo {author} {\bibfnamefont {C.}~\bibnamefont
  {Triola}}\ and\ \bibinfo {author} {\bibfnamefont {A.~M.}\ \bibnamefont
  {Black-Schaffer}},\ }\href {\doibase 10.1103/PhysRevB.97.064505} {\bibfield
  {journal} {\bibinfo  {journal} {Phys. Rev. B}\ }\textbf {\bibinfo {volume}
  {97}},\ \bibinfo {pages} {064505} (\bibinfo {year} {2018})}\BibitemShut
  {NoStop}%
\bibitem [{\citenamefont {Aeppli}\ \emph {et~al.}(1988)\citenamefont {Aeppli},
  \citenamefont {Bucher}, \citenamefont {Broholm}, \citenamefont {Kjems},
  \citenamefont {Baumann},\ and\ \citenamefont {Hufnagl}}]{Aeppli1988}%
  \BibitemOpen
  \bibfield  {author} {\bibinfo {author} {\bibfnamefont {G.}~\bibnamefont
  {Aeppli}}, \bibinfo {author} {\bibfnamefont {E.}~\bibnamefont {Bucher}},
  \bibinfo {author} {\bibfnamefont {C.}~\bibnamefont {Broholm}}, \bibinfo
  {author} {\bibfnamefont {J.~K.}\ \bibnamefont {Kjems}}, \bibinfo {author}
  {\bibfnamefont {J.}~\bibnamefont {Baumann}}, \ and\ \bibinfo {author}
  {\bibfnamefont {J.}~\bibnamefont {Hufnagl}},\ }\href {\doibase
  10.1103/PhysRevLett.60.615} {\bibfield  {journal} {\bibinfo  {journal} {Phys.
  Rev. Lett.}\ }\textbf {\bibinfo {volume} {60}},\ \bibinfo {pages} {615}
  (\bibinfo {year} {1988})}\BibitemShut {NoStop}%
\bibitem [{\citenamefont {Hayden}\ \emph {et~al.}(1992)\citenamefont {Hayden},
  \citenamefont {Taillefer}, \citenamefont {Vettier},\ and\ \citenamefont
  {Flouquet}}]{Hayden1992}%
  \BibitemOpen
  \bibfield  {author} {\bibinfo {author} {\bibfnamefont {S.~M.}\ \bibnamefont
  {Hayden}}, \bibinfo {author} {\bibfnamefont {L.}~\bibnamefont {Taillefer}},
  \bibinfo {author} {\bibfnamefont {C.}~\bibnamefont {Vettier}}, \ and\
  \bibinfo {author} {\bibfnamefont {J.}~\bibnamefont {Flouquet}},\ }\href
  {\doibase 10.1103/PhysRevB.46.8675} {\bibfield  {journal} {\bibinfo
  {journal} {Phys. Rev. B}\ }\textbf {\bibinfo {volume} {46}},\ \bibinfo
  {pages} {8675} (\bibinfo {year} {1992})}\BibitemShut {NoStop}%
\bibitem [{\citenamefont {Ishizuka}\ \emph {et~al.}()\citenamefont {Ishizuka},
  \citenamefont {Sumita},\ and\ \citenamefont {Yanase}}]{Ishizuka2017}%
  \BibitemOpen
  \bibfield  {author} {\bibinfo {author} {\bibfnamefont {J.}~\bibnamefont
  {Ishizuka}}, \bibinfo {author} {\bibfnamefont {S.}~\bibnamefont {Sumita}}, \
  and\ \bibinfo {author} {\bibfnamefont {Y.}~\bibnamefont {Yanase}},\ }\bibinfo
  {note} {in preparation.}\BibitemShut {Stop}%
\bibitem [{\citenamefont {Ye}\ \emph {et~al.}(2012)\citenamefont {Ye},
  \citenamefont {Zhang}, \citenamefont {Akashi}, \citenamefont {Bahramy},
  \citenamefont {Arita},\ and\ \citenamefont {Iwasa}}]{Ye2012}%
  \BibitemOpen
  \bibfield  {author} {\bibinfo {author} {\bibfnamefont {J.~T.}\ \bibnamefont
  {Ye}}, \bibinfo {author} {\bibfnamefont {Y.~J.}\ \bibnamefont {Zhang}},
  \bibinfo {author} {\bibfnamefont {R.}~\bibnamefont {Akashi}}, \bibinfo
  {author} {\bibfnamefont {M.~S.}\ \bibnamefont {Bahramy}}, \bibinfo {author}
  {\bibfnamefont {R.}~\bibnamefont {Arita}}, \ and\ \bibinfo {author}
  {\bibfnamefont {Y.}~\bibnamefont {Iwasa}},\ }\href {\doibase
  10.1126/science.1228006} {\bibfield  {journal} {\bibinfo  {journal}
  {Science}\ }\textbf {\bibinfo {volume} {338}},\ \bibinfo {pages} {1193}
  (\bibinfo {year} {2012})}\BibitemShut {NoStop}%
\bibitem [{\citenamefont {Lu}\ \emph {et~al.}(2015)\citenamefont {Lu},
  \citenamefont {Zheliuk}, \citenamefont {Leermakers}, \citenamefont {Yuan},
  \citenamefont {Zeitler}, \citenamefont {Law},\ and\ \citenamefont
  {Ye}}]{Lu2015}%
  \BibitemOpen
  \bibfield  {author} {\bibinfo {author} {\bibfnamefont {J.~M.}\ \bibnamefont
  {Lu}}, \bibinfo {author} {\bibfnamefont {O.}~\bibnamefont {Zheliuk}},
  \bibinfo {author} {\bibfnamefont {I.}~\bibnamefont {Leermakers}}, \bibinfo
  {author} {\bibfnamefont {N.~F.~Q.}\ \bibnamefont {Yuan}}, \bibinfo {author}
  {\bibfnamefont {U.}~\bibnamefont {Zeitler}}, \bibinfo {author} {\bibfnamefont
  {K.~T.}\ \bibnamefont {Law}}, \ and\ \bibinfo {author} {\bibfnamefont
  {J.~T.}\ \bibnamefont {Ye}},\ }\href {\doibase 10.1126/science.aab2277}
  {\bibfield  {journal} {\bibinfo  {journal} {Science}\ }\textbf {\bibinfo
  {volume} {350}},\ \bibinfo {pages} {1353} (\bibinfo {year}
  {2015})}\BibitemShut {NoStop}%
\bibitem [{\citenamefont {Shi}\ \emph {et~al.}(2015)\citenamefont {Shi},
  \citenamefont {Ye}, \citenamefont {Zhang}, \citenamefont {Suzuki},
  \citenamefont {Yoshida}, \citenamefont {Miyazaki}, \citenamefont {Inoue},
  \citenamefont {Saito},\ and\ \citenamefont {Iwasa}}]{Shi2015}%
  \BibitemOpen
  \bibfield  {author} {\bibinfo {author} {\bibfnamefont {W.}~\bibnamefont
  {Shi}}, \bibinfo {author} {\bibfnamefont {J.}~\bibnamefont {Ye}}, \bibinfo
  {author} {\bibfnamefont {Y.}~\bibnamefont {Zhang}}, \bibinfo {author}
  {\bibfnamefont {R.}~\bibnamefont {Suzuki}}, \bibinfo {author} {\bibfnamefont
  {M.}~\bibnamefont {Yoshida}}, \bibinfo {author} {\bibfnamefont
  {J.}~\bibnamefont {Miyazaki}}, \bibinfo {author} {\bibfnamefont
  {N.}~\bibnamefont {Inoue}}, \bibinfo {author} {\bibfnamefont
  {Y.}~\bibnamefont {Saito}}, \ and\ \bibinfo {author} {\bibfnamefont
  {Y.}~\bibnamefont {Iwasa}},\ }\href@noop {} {\bibfield  {journal} {\bibinfo
  {journal} {Sci. Rep.}\ }\textbf {\bibinfo {volume} {5}},\ \bibinfo {pages}
  {12534} (\bibinfo {year} {2015})}\BibitemShut {NoStop}%
\bibitem [{\citenamefont {Costanzo}\ \emph {et~al.}(2016)\citenamefont
  {Costanzo}, \citenamefont {Jo}, \citenamefont {Berger},\ and\ \citenamefont
  {Morpurgo}}]{Costanzo2016}%
  \BibitemOpen
  \bibfield  {author} {\bibinfo {author} {\bibfnamefont {D.}~\bibnamefont
  {Costanzo}}, \bibinfo {author} {\bibfnamefont {S.}~\bibnamefont {Jo}},
  \bibinfo {author} {\bibfnamefont {H.}~\bibnamefont {Berger}}, \ and\ \bibinfo
  {author} {\bibfnamefont {A.~F.}\ \bibnamefont {Morpurgo}},\ }\href@noop {}
  {\bibfield  {journal} {\bibinfo  {journal} {Nat. Nanotechnol.}\ }\textbf
  {\bibinfo {volume} {11}},\ \bibinfo {pages} {339} (\bibinfo {year}
  {2016})}\BibitemShut {NoStop}%
\bibitem [{\citenamefont {Woollam}\ and\ \citenamefont
  {Somoano}(1977)}]{Woollam1977}%
  \BibitemOpen
  \bibfield  {author} {\bibinfo {author} {\bibfnamefont {J.~A.}\ \bibnamefont
  {Woollam}}\ and\ \bibinfo {author} {\bibfnamefont {R.~B.}\ \bibnamefont
  {Somoano}},\ }\href {\doibase 10.1016/0025-5416(77)90048-9} {\bibfield
  {journal} {\bibinfo  {journal} {Materials Science and Engineering}\ }\textbf
  {\bibinfo {volume} {31}},\ \bibinfo {pages} {289} (\bibinfo {year}
  {1977})}\BibitemShut {NoStop}%
\bibitem [{\citenamefont {Zhang}\ \emph {et~al.}(2016)\citenamefont {Zhang},
  \citenamefont {Tsai}, \citenamefont {Chapman}, \citenamefont {Khestanova},
  \citenamefont {Waters},\ and\ \citenamefont {Grigorieva}}]{Zhang2016}%
  \BibitemOpen
  \bibfield  {author} {\bibinfo {author} {\bibfnamefont {R.}~\bibnamefont
  {Zhang}}, \bibinfo {author} {\bibfnamefont {I.-L.}\ \bibnamefont {Tsai}},
  \bibinfo {author} {\bibfnamefont {J.}~\bibnamefont {Chapman}}, \bibinfo
  {author} {\bibfnamefont {E.}~\bibnamefont {Khestanova}}, \bibinfo {author}
  {\bibfnamefont {J.}~\bibnamefont {Waters}}, \ and\ \bibinfo {author}
  {\bibfnamefont {I.~V.}\ \bibnamefont {Grigorieva}},\ }\href {\doibase
  10.1021/acs.nanolett.5b04361} {\bibfield  {journal} {\bibinfo  {journal}
  {Nano Lett.}\ }\textbf {\bibinfo {volume} {16}},\ \bibinfo {pages} {629}
  (\bibinfo {year} {2016})}\BibitemShut {NoStop}%
\bibitem [{\citenamefont {Liu}\ \emph {et~al.}(2015)\citenamefont {Liu},
  \citenamefont {Xiao}, \citenamefont {Yao}, \citenamefont {Xu},\ and\
  \citenamefont {Yao}}]{Liu2015}%
  \BibitemOpen
  \bibfield  {author} {\bibinfo {author} {\bibfnamefont {G.-B.}\ \bibnamefont
  {Liu}}, \bibinfo {author} {\bibfnamefont {D.}~\bibnamefont {Xiao}}, \bibinfo
  {author} {\bibfnamefont {Y.}~\bibnamefont {Yao}}, \bibinfo {author}
  {\bibfnamefont {X.}~\bibnamefont {Xu}}, \ and\ \bibinfo {author}
  {\bibfnamefont {W.}~\bibnamefont {Yao}},\ }\href {\doibase
  10.1039/C4CS00301B} {\bibfield  {journal} {\bibinfo  {journal} {Chem. Soc.
  Rev.}\ }\textbf {\bibinfo {volume} {44}},\ \bibinfo {pages} {2643} (\bibinfo
  {year} {2015})}\BibitemShut {NoStop}%
\bibitem [{\citenamefont {Ishida}\ \emph {et~al.}(2009)\citenamefont {Ishida},
  \citenamefont {Nakai},\ and\ \citenamefont {Hosono}}]{Ishida2009}%
  \BibitemOpen
  \bibfield  {author} {\bibinfo {author} {\bibfnamefont {K.}~\bibnamefont
  {Ishida}}, \bibinfo {author} {\bibfnamefont {Y.}~\bibnamefont {Nakai}}, \
  and\ \bibinfo {author} {\bibfnamefont {H.}~\bibnamefont {Hosono}},\ }\href
  {\doibase 10.1143/JPSJ.78.062001} {\bibfield  {journal} {\bibinfo  {journal}
  {J. Phys. Soc. Jpn.}\ }\textbf {\bibinfo {volume} {78}},\ \bibinfo {pages}
  {062001} (\bibinfo {year} {2009})}\BibitemShut {NoStop}%
\bibitem [{\citenamefont {Shein}\ and\ \citenamefont
  {Ivanovskii}(2011)}]{Shein2011}%
  \BibitemOpen
  \bibfield  {author} {\bibinfo {author} {\bibfnamefont {I.~R.}\ \bibnamefont
  {Shein}}\ and\ \bibinfo {author} {\bibfnamefont {A.~L.}\ \bibnamefont
  {Ivanovskii}},\ }\href {\doibase 10.1016/j.physc.2011.07.009} {\bibfield
  {journal} {\bibinfo  {journal} {Physica C}\ }\textbf {\bibinfo {volume}
  {471}},\ \bibinfo {pages} {594} (\bibinfo {year} {2011})}\BibitemShut
  {NoStop}%
\bibitem [{\citenamefont {Youn}\ \emph {et~al.}(2012)\citenamefont {Youn},
  \citenamefont {Fischer}, \citenamefont {Rhim}, \citenamefont {Sigrist},\ and\
  \citenamefont {Agterberg}}]{Youn2012}%
  \BibitemOpen
  \bibfield  {author} {\bibinfo {author} {\bibfnamefont {S.~J.}\ \bibnamefont
  {Youn}}, \bibinfo {author} {\bibfnamefont {M.~H.}\ \bibnamefont {Fischer}},
  \bibinfo {author} {\bibfnamefont {S.~H.}\ \bibnamefont {Rhim}}, \bibinfo
  {author} {\bibfnamefont {M.}~\bibnamefont {Sigrist}}, \ and\ \bibinfo
  {author} {\bibfnamefont {D.~F.}\ \bibnamefont {Agterberg}},\ }\href {\doibase
  10.1103/PhysRevB.85.220505} {\bibfield  {journal} {\bibinfo  {journal} {Phys.
  Rev. B}\ }\textbf {\bibinfo {volume} {85}},\ \bibinfo {pages} {220505}
  (\bibinfo {year} {2012})}\BibitemShut {NoStop}%
\bibitem [{\citenamefont {Youn}\ \emph {et~al.}()\citenamefont {Youn},
  \citenamefont {Rhim}, \citenamefont {Agterberg}, \citenamefont {Weinert},\
  and\ \citenamefont {Freeman}}]{Youn2012_arXiv}%
  \BibitemOpen
  \bibfield  {author} {\bibinfo {author} {\bibfnamefont {S.~J.}\ \bibnamefont
  {Youn}}, \bibinfo {author} {\bibfnamefont {S.~H.}\ \bibnamefont {Rhim}},
  \bibinfo {author} {\bibfnamefont {D.~F.}\ \bibnamefont {Agterberg}}, \bibinfo
  {author} {\bibfnamefont {M.}~\bibnamefont {Weinert}}, \ and\ \bibinfo
  {author} {\bibfnamefont {A.~J.}\ \bibnamefont {Freeman}},\ }\Eprint
  {http://arxiv.org/abs/1202.1604} {arXiv:1202.1604} \BibitemShut {NoStop}%
\bibitem [{\citenamefont {Biswas}\ \emph {et~al.}(2013)\citenamefont {Biswas},
  \citenamefont {Luetkens}, \citenamefont {Neupert}, \citenamefont {St\"urzer},
  \citenamefont {Baines}, \citenamefont {Pascua}, \citenamefont {Schnyder},
  \citenamefont {Fischer}, \citenamefont {Goryo}, \citenamefont {Lees},
  \citenamefont {Maeter}, \citenamefont {Br\"uckner}, \citenamefont {Klauss},
  \citenamefont {Nicklas}, \citenamefont {Baker}, \citenamefont {Hillier},
  \citenamefont {Sigrist}, \citenamefont {Amato},\ and\ \citenamefont
  {Johrendt}}]{Biswas2013}%
  \BibitemOpen
  \bibfield  {author} {\bibinfo {author} {\bibfnamefont {P.~K.}\ \bibnamefont
  {Biswas}}, \bibinfo {author} {\bibfnamefont {H.}~\bibnamefont {Luetkens}},
  \bibinfo {author} {\bibfnamefont {T.}~\bibnamefont {Neupert}}, \bibinfo
  {author} {\bibfnamefont {T.}~\bibnamefont {St\"urzer}}, \bibinfo {author}
  {\bibfnamefont {C.}~\bibnamefont {Baines}}, \bibinfo {author} {\bibfnamefont
  {G.}~\bibnamefont {Pascua}}, \bibinfo {author} {\bibfnamefont {A.~P.}\
  \bibnamefont {Schnyder}}, \bibinfo {author} {\bibfnamefont {M.~H.}\
  \bibnamefont {Fischer}}, \bibinfo {author} {\bibfnamefont {J.}~\bibnamefont
  {Goryo}}, \bibinfo {author} {\bibfnamefont {M.~R.}\ \bibnamefont {Lees}},
  \bibinfo {author} {\bibfnamefont {H.}~\bibnamefont {Maeter}}, \bibinfo
  {author} {\bibfnamefont {F.}~\bibnamefont {Br\"uckner}}, \bibinfo {author}
  {\bibfnamefont {H.-H.}\ \bibnamefont {Klauss}}, \bibinfo {author}
  {\bibfnamefont {M.}~\bibnamefont {Nicklas}}, \bibinfo {author} {\bibfnamefont
  {P.~J.}\ \bibnamefont {Baker}}, \bibinfo {author} {\bibfnamefont {A.~D.}\
  \bibnamefont {Hillier}}, \bibinfo {author} {\bibfnamefont {M.}~\bibnamefont
  {Sigrist}}, \bibinfo {author} {\bibfnamefont {A.}~\bibnamefont {Amato}}, \
  and\ \bibinfo {author} {\bibfnamefont {D.}~\bibnamefont {Johrendt}},\ }\href
  {\doibase 10.1103/PhysRevB.87.180503} {\bibfield  {journal} {\bibinfo
  {journal} {Phys. Rev. B}\ }\textbf {\bibinfo {volume} {87}},\ \bibinfo
  {pages} {180503} (\bibinfo {year} {2013})}\BibitemShut {NoStop}%
\bibitem [{\citenamefont {Matano}\ \emph {et~al.}(2014)\citenamefont {Matano},
  \citenamefont {Arima}, \citenamefont {Maeda}, \citenamefont {Nishikubo},
  \citenamefont {Kudo}, \citenamefont {Nohara},\ and\ \citenamefont
  {Zheng}}]{Matano2014}%
  \BibitemOpen
  \bibfield  {author} {\bibinfo {author} {\bibfnamefont {K.}~\bibnamefont
  {Matano}}, \bibinfo {author} {\bibfnamefont {K.}~\bibnamefont {Arima}},
  \bibinfo {author} {\bibfnamefont {S.}~\bibnamefont {Maeda}}, \bibinfo
  {author} {\bibfnamefont {Y.}~\bibnamefont {Nishikubo}}, \bibinfo {author}
  {\bibfnamefont {K.}~\bibnamefont {Kudo}}, \bibinfo {author} {\bibfnamefont
  {M.}~\bibnamefont {Nohara}}, \ and\ \bibinfo {author} {\bibfnamefont {G.-q.}\
  \bibnamefont {Zheng}},\ }\href {\doibase 10.1103/PhysRevB.89.140504}
  {\bibfield  {journal} {\bibinfo  {journal} {Phys. Rev. B}\ }\textbf {\bibinfo
  {volume} {89}},\ \bibinfo {pages} {140504} (\bibinfo {year}
  {2014})}\BibitemShut {NoStop}%
\bibitem [{\citenamefont {Nakamura}\ and\ \citenamefont
  {Yanase}(2017)}]{Nakamura2017}%
  \BibitemOpen
  \bibfield  {author} {\bibinfo {author} {\bibfnamefont {Y.}~\bibnamefont
  {Nakamura}}\ and\ \bibinfo {author} {\bibfnamefont {Y.}~\bibnamefont
  {Yanase}},\ }\href {\doibase 10.1103/PhysRevB.96.054501} {\bibfield
  {journal} {\bibinfo  {journal} {Phys. Rev. B}\ }\textbf {\bibinfo {volume}
  {96}},\ \bibinfo {pages} {054501} (\bibinfo {year} {2017})}\BibitemShut
  {NoStop}%
\bibitem [{\citenamefont {Ge}\ and\ \citenamefont {Liu}(2013)}]{Ge2013}%
  \BibitemOpen
  \bibfield  {author} {\bibinfo {author} {\bibfnamefont {Y.}~\bibnamefont
  {Ge}}\ and\ \bibinfo {author} {\bibfnamefont {A.~Y.}\ \bibnamefont {Liu}},\
  }\href {\doibase 10.1103/PhysRevB.87.241408} {\bibfield  {journal} {\bibinfo
  {journal} {Phys. Rev. B}\ }\textbf {\bibinfo {volume} {87}},\ \bibinfo
  {pages} {241408} (\bibinfo {year} {2013})}\BibitemShut {NoStop}%
\bibitem [{\citenamefont {R\"osner}\ \emph {et~al.}(2014)\citenamefont
  {R\"osner}, \citenamefont {Haas},\ and\ \citenamefont
  {Wehling}}]{Rosner2014}%
  \BibitemOpen
  \bibfield  {author} {\bibinfo {author} {\bibfnamefont {M.}~\bibnamefont
  {R\"osner}}, \bibinfo {author} {\bibfnamefont {S.}~\bibnamefont {Haas}}, \
  and\ \bibinfo {author} {\bibfnamefont {T.~O.}\ \bibnamefont {Wehling}},\
  }\href {\doibase 10.1103/PhysRevB.90.245105} {\bibfield  {journal} {\bibinfo
  {journal} {Phys. Rev. B}\ }\textbf {\bibinfo {volume} {90}},\ \bibinfo
  {pages} {245105} (\bibinfo {year} {2014})}\BibitemShut {NoStop}%
\bibitem [{\citenamefont {Das}\ and\ \citenamefont {Dolui}(2015)}]{Das2015}%
  \BibitemOpen
  \bibfield  {author} {\bibinfo {author} {\bibfnamefont {T.}~\bibnamefont
  {Das}}\ and\ \bibinfo {author} {\bibfnamefont {K.}~\bibnamefont {Dolui}},\
  }\href {\doibase 10.1103/PhysRevB.91.094510} {\bibfield  {journal} {\bibinfo
  {journal} {Phys. Rev. B}\ }\textbf {\bibinfo {volume} {91}},\ \bibinfo
  {pages} {094510} (\bibinfo {year} {2015})}\BibitemShut {NoStop}%
\bibitem [{\citenamefont {Yuan}\ \emph {et~al.}(2014)\citenamefont {Yuan},
  \citenamefont {Mak},\ and\ \citenamefont {Law}}]{Yuan2014}%
  \BibitemOpen
  \bibfield  {author} {\bibinfo {author} {\bibfnamefont {N.~F.~Q.}\
  \bibnamefont {Yuan}}, \bibinfo {author} {\bibfnamefont {K.~F.}\ \bibnamefont
  {Mak}}, \ and\ \bibinfo {author} {\bibfnamefont {K.~T.}\ \bibnamefont
  {Law}},\ }\href {\doibase 10.1103/PhysRevLett.113.097001} {\bibfield
  {journal} {\bibinfo  {journal} {Phys. Rev. Lett.}\ }\textbf {\bibinfo
  {volume} {113}},\ \bibinfo {pages} {097001} (\bibinfo {year}
  {2014})}\BibitemShut {NoStop}%
\bibitem [{\citenamefont {Hsu}\ \emph {et~al.}(2017)\citenamefont {Hsu},
  \citenamefont {Vaezi}, \citenamefont {Fischer},\ and\ \citenamefont
  {Kim}}]{Hsu2017}%
  \BibitemOpen
  \bibfield  {author} {\bibinfo {author} {\bibfnamefont {Y.-T.}\ \bibnamefont
  {Hsu}}, \bibinfo {author} {\bibfnamefont {A.}~\bibnamefont {Vaezi}}, \bibinfo
  {author} {\bibfnamefont {M.~H.}\ \bibnamefont {Fischer}}, \ and\ \bibinfo
  {author} {\bibfnamefont {E.-A.}\ \bibnamefont {Kim}},\ }\href@noop {}
  {\bibfield  {journal} {\bibinfo  {journal} {Nat. Commun.}\ }\textbf {\bibinfo
  {volume} {8}},\ \bibinfo {pages} {14985} (\bibinfo {year}
  {2017})}\BibitemShut {NoStop}%
\bibitem [{\citenamefont {Goryo}\ \emph {et~al.}(2012)\citenamefont {Goryo},
  \citenamefont {Fischer},\ and\ \citenamefont {Sigrist}}]{Goryo2012}%
  \BibitemOpen
  \bibfield  {author} {\bibinfo {author} {\bibfnamefont {J.}~\bibnamefont
  {Goryo}}, \bibinfo {author} {\bibfnamefont {M.~H.}\ \bibnamefont {Fischer}},
  \ and\ \bibinfo {author} {\bibfnamefont {M.}~\bibnamefont {Sigrist}},\ }\href
  {\doibase 10.1103/PhysRevB.86.100507} {\bibfield  {journal} {\bibinfo
  {journal} {Phys. Rev. B}\ }\textbf {\bibinfo {volume} {86}},\ \bibinfo
  {pages} {100507} (\bibinfo {year} {2012})}\BibitemShut {NoStop}%
\bibitem [{\citenamefont {Wang}\ \emph {et~al.}(2014)\citenamefont {Wang},
  \citenamefont {Yang},\ and\ \citenamefont {Wang}}]{WangW2014}%
  \BibitemOpen
  \bibfield  {author} {\bibinfo {author} {\bibfnamefont {W.-S.}\ \bibnamefont
  {Wang}}, \bibinfo {author} {\bibfnamefont {Y.}~\bibnamefont {Yang}}, \ and\
  \bibinfo {author} {\bibfnamefont {Q.-H.}\ \bibnamefont {Wang}},\ }\href
  {\doibase 10.1103/PhysRevB.90.094514} {\bibfield  {journal} {\bibinfo
  {journal} {Phys. Rev. B}\ }\textbf {\bibinfo {volume} {90}},\ \bibinfo
  {pages} {094514} (\bibinfo {year} {2014})}\BibitemShut {NoStop}%
\bibitem [{\citenamefont {Br\"uckner}\ \emph {et~al.}(2014)\citenamefont
  {Br\"uckner}, \citenamefont {Sarkar}, \citenamefont {G\"unther},
  \citenamefont {K\"uhne}, \citenamefont {Luetkens}, \citenamefont {Neupert},
  \citenamefont {Reyes}, \citenamefont {Kuhns}, \citenamefont {Biswas},
  \citenamefont {St\"urzer}, \citenamefont {Johrendt},\ and\ \citenamefont
  {Klauss}}]{Bruckner2014}%
  \BibitemOpen
  \bibfield  {author} {\bibinfo {author} {\bibfnamefont {F.}~\bibnamefont
  {Br\"uckner}}, \bibinfo {author} {\bibfnamefont {R.}~\bibnamefont {Sarkar}},
  \bibinfo {author} {\bibfnamefont {M.}~\bibnamefont {G\"unther}}, \bibinfo
  {author} {\bibfnamefont {H.}~\bibnamefont {K\"uhne}}, \bibinfo {author}
  {\bibfnamefont {H.}~\bibnamefont {Luetkens}}, \bibinfo {author}
  {\bibfnamefont {T.}~\bibnamefont {Neupert}}, \bibinfo {author} {\bibfnamefont
  {A.~P.}\ \bibnamefont {Reyes}}, \bibinfo {author} {\bibfnamefont {P.~L.}\
  \bibnamefont {Kuhns}}, \bibinfo {author} {\bibfnamefont {P.~K.}\ \bibnamefont
  {Biswas}}, \bibinfo {author} {\bibfnamefont {T.}~\bibnamefont {St\"urzer}},
  \bibinfo {author} {\bibfnamefont {D.}~\bibnamefont {Johrendt}}, \ and\
  \bibinfo {author} {\bibfnamefont {H.-H.}\ \bibnamefont {Klauss}},\ }\href
  {\doibase 10.1103/PhysRevB.90.220503} {\bibfield  {journal} {\bibinfo
  {journal} {Phys. Rev. B}\ }\textbf {\bibinfo {volume} {90}},\ \bibinfo
  {pages} {220503} (\bibinfo {year} {2014})}\BibitemShut {NoStop}%
\bibitem [{\citenamefont {Fischer}\ and\ \citenamefont
  {Goryo}(2015)}]{Fischer2015}%
  \BibitemOpen
  \bibfield  {author} {\bibinfo {author} {\bibfnamefont {M.~H.}\ \bibnamefont
  {Fischer}}\ and\ \bibinfo {author} {\bibfnamefont {J.}~\bibnamefont
  {Goryo}},\ }\href {\doibase 10.7566/JPSJ.84.054705} {\bibfield  {journal}
  {\bibinfo  {journal} {J. Phys. Soc. Jpn.}\ }\textbf {\bibinfo {volume}
  {84}},\ \bibinfo {pages} {054705} (\bibinfo {year} {2015})}\BibitemShut
  {NoStop}%
\bibitem [{\citenamefont {Ott}\ \emph {et~al.}(1983)\citenamefont {Ott},
  \citenamefont {Rudigier}, \citenamefont {Fisk},\ and\ \citenamefont
  {Smith}}]{Ott1983}%
  \BibitemOpen
  \bibfield  {author} {\bibinfo {author} {\bibfnamefont {H.~R.}\ \bibnamefont
  {Ott}}, \bibinfo {author} {\bibfnamefont {H.}~\bibnamefont {Rudigier}},
  \bibinfo {author} {\bibfnamefont {Z.}~\bibnamefont {Fisk}}, \ and\ \bibinfo
  {author} {\bibfnamefont {J.~L.}\ \bibnamefont {Smith}},\ }\href {\doibase
  10.1103/PhysRevLett.50.1595} {\bibfield  {journal} {\bibinfo  {journal}
  {Phys. Rev. Lett.}\ }\textbf {\bibinfo {volume} {50}},\ \bibinfo {pages}
  {1595} (\bibinfo {year} {1983})}\BibitemShut {NoStop}%
\bibitem [{\citenamefont {Ott}\ \emph {et~al.}(1984)\citenamefont {Ott},
  \citenamefont {Rudigier}, \citenamefont {Rice}, \citenamefont {Ueda},
  \citenamefont {Fisk},\ and\ \citenamefont {Smith}}]{Ott1984}%
  \BibitemOpen
  \bibfield  {author} {\bibinfo {author} {\bibfnamefont {H.~R.}\ \bibnamefont
  {Ott}}, \bibinfo {author} {\bibfnamefont {H.}~\bibnamefont {Rudigier}},
  \bibinfo {author} {\bibfnamefont {T.~M.}\ \bibnamefont {Rice}}, \bibinfo
  {author} {\bibfnamefont {K.}~\bibnamefont {Ueda}}, \bibinfo {author}
  {\bibfnamefont {Z.}~\bibnamefont {Fisk}}, \ and\ \bibinfo {author}
  {\bibfnamefont {J.~L.}\ \bibnamefont {Smith}},\ }\href {\doibase
  10.1103/PhysRevLett.52.1915} {\bibfield  {journal} {\bibinfo  {journal}
  {Phys. Rev. Lett.}\ }\textbf {\bibinfo {volume} {52}},\ \bibinfo {pages}
  {1915} (\bibinfo {year} {1984})}\BibitemShut {NoStop}%
\bibitem [{\citenamefont {MacLaughlin}\ \emph {et~al.}(1984)\citenamefont
  {MacLaughlin}, \citenamefont {Tien}, \citenamefont {Clark}, \citenamefont
  {Lan}, \citenamefont {Fisk}, \citenamefont {Smith},\ and\ \citenamefont
  {Ott}}]{MacLaughlin1984}%
  \BibitemOpen
  \bibfield  {author} {\bibinfo {author} {\bibfnamefont {D.~E.}\ \bibnamefont
  {MacLaughlin}}, \bibinfo {author} {\bibfnamefont {C.}~\bibnamefont {Tien}},
  \bibinfo {author} {\bibfnamefont {W.~G.}\ \bibnamefont {Clark}}, \bibinfo
  {author} {\bibfnamefont {M.~D.}\ \bibnamefont {Lan}}, \bibinfo {author}
  {\bibfnamefont {Z.}~\bibnamefont {Fisk}}, \bibinfo {author} {\bibfnamefont
  {J.~L.}\ \bibnamefont {Smith}}, \ and\ \bibinfo {author} {\bibfnamefont
  {H.~R.}\ \bibnamefont {Ott}},\ }\href {\doibase 10.1103/PhysRevLett.53.1833}
  {\bibfield  {journal} {\bibinfo  {journal} {Phys. Rev. Lett.}\ }\textbf
  {\bibinfo {volume} {53}},\ \bibinfo {pages} {1833} (\bibinfo {year}
  {1984})}\BibitemShut {NoStop}%
\bibitem [{\citenamefont {Shimizu}\ \emph {et~al.}(2015)\citenamefont
  {Shimizu}, \citenamefont {Kittaka}, \citenamefont {Sakakibara}, \citenamefont
  {Haga}, \citenamefont {Yamamoto}, \citenamefont {Amitsuka}, \citenamefont
  {Tsutsumi},\ and\ \citenamefont {Machida}}]{Shimizu2015}%
  \BibitemOpen
  \bibfield  {author} {\bibinfo {author} {\bibfnamefont {Y.}~\bibnamefont
  {Shimizu}}, \bibinfo {author} {\bibfnamefont {S.}~\bibnamefont {Kittaka}},
  \bibinfo {author} {\bibfnamefont {T.}~\bibnamefont {Sakakibara}}, \bibinfo
  {author} {\bibfnamefont {Y.}~\bibnamefont {Haga}}, \bibinfo {author}
  {\bibfnamefont {E.}~\bibnamefont {Yamamoto}}, \bibinfo {author}
  {\bibfnamefont {H.}~\bibnamefont {Amitsuka}}, \bibinfo {author}
  {\bibfnamefont {Y.}~\bibnamefont {Tsutsumi}}, \ and\ \bibinfo {author}
  {\bibfnamefont {K.}~\bibnamefont {Machida}},\ }\href {\doibase
  10.1103/PhysRevLett.114.147002} {\bibfield  {journal} {\bibinfo  {journal}
  {Phys. Rev. Lett.}\ }\textbf {\bibinfo {volume} {114}},\ \bibinfo {pages}
  {147002} (\bibinfo {year} {2015})}\BibitemShut {NoStop}%
\bibitem [{\citenamefont {Smith}\ \emph {et~al.}(1985)\citenamefont {Smith},
  \citenamefont {Fisk}, \citenamefont {Willis}, \citenamefont {Giorgi},
  \citenamefont {Roof}, \citenamefont {Ott}, \citenamefont {Rudigier},\ and\
  \citenamefont {Felder}}]{Smith1985}%
  \BibitemOpen
  \bibfield  {author} {\bibinfo {author} {\bibfnamefont {J.~L.}\ \bibnamefont
  {Smith}}, \bibinfo {author} {\bibfnamefont {Z.}~\bibnamefont {Fisk}},
  \bibinfo {author} {\bibfnamefont {J.~O.}\ \bibnamefont {Willis}}, \bibinfo
  {author} {\bibfnamefont {A.~L.}\ \bibnamefont {Giorgi}}, \bibinfo {author}
  {\bibfnamefont {R.~B.}\ \bibnamefont {Roof}}, \bibinfo {author}
  {\bibfnamefont {H.~R.}\ \bibnamefont {Ott}}, \bibinfo {author} {\bibfnamefont
  {H.}~\bibnamefont {Rudigier}}, \ and\ \bibinfo {author} {\bibfnamefont
  {E.}~\bibnamefont {Felder}},\ }\href {\doibase
  https://doi.org/10.1016/0378-4363(85)90420-6} {\bibfield  {journal} {\bibinfo
   {journal} {Physica B}\ }\textbf {\bibinfo {volume} {135}},\ \bibinfo {pages}
  {3} (\bibinfo {year} {1985})}\BibitemShut {NoStop}%
\bibitem [{\citenamefont {Ott}\ \emph {et~al.}(1985)\citenamefont {Ott},
  \citenamefont {Rudigier}, \citenamefont {Fisk},\ and\ \citenamefont
  {Smith}}]{Ott1985}%
  \BibitemOpen
  \bibfield  {author} {\bibinfo {author} {\bibfnamefont {H.~R.}\ \bibnamefont
  {Ott}}, \bibinfo {author} {\bibfnamefont {H.}~\bibnamefont {Rudigier}},
  \bibinfo {author} {\bibfnamefont {Z.}~\bibnamefont {Fisk}}, \ and\ \bibinfo
  {author} {\bibfnamefont {J.~L.}\ \bibnamefont {Smith}},\ }\href {\doibase
  10.1103/PhysRevB.31.1651} {\bibfield  {journal} {\bibinfo  {journal} {Phys.
  Rev. B}\ }\textbf {\bibinfo {volume} {31}},\ \bibinfo {pages} {1651}
  (\bibinfo {year} {1985})}\BibitemShut {NoStop}%
\bibitem [{\citenamefont {Heffner}\ \emph {et~al.}(1990)\citenamefont
  {Heffner}, \citenamefont {Smith}, \citenamefont {Willis}, \citenamefont
  {Birrer}, \citenamefont {Baines}, \citenamefont {Gygax}, \citenamefont
  {Hitti}, \citenamefont {Lippelt}, \citenamefont {Ott}, \citenamefont
  {Schenck}, \citenamefont {Knetsch}, \citenamefont {Mydosh},\ and\
  \citenamefont {MacLaughlin}}]{Heffner1990}%
  \BibitemOpen
  \bibfield  {author} {\bibinfo {author} {\bibfnamefont {R.~H.}\ \bibnamefont
  {Heffner}}, \bibinfo {author} {\bibfnamefont {J.~L.}\ \bibnamefont {Smith}},
  \bibinfo {author} {\bibfnamefont {J.~O.}\ \bibnamefont {Willis}}, \bibinfo
  {author} {\bibfnamefont {P.}~\bibnamefont {Birrer}}, \bibinfo {author}
  {\bibfnamefont {C.}~\bibnamefont {Baines}}, \bibinfo {author} {\bibfnamefont
  {F.~N.}\ \bibnamefont {Gygax}}, \bibinfo {author} {\bibfnamefont
  {B.}~\bibnamefont {Hitti}}, \bibinfo {author} {\bibfnamefont
  {E.}~\bibnamefont {Lippelt}}, \bibinfo {author} {\bibfnamefont {H.~R.}\
  \bibnamefont {Ott}}, \bibinfo {author} {\bibfnamefont {A.}~\bibnamefont
  {Schenck}}, \bibinfo {author} {\bibfnamefont {E.~A.}\ \bibnamefont
  {Knetsch}}, \bibinfo {author} {\bibfnamefont {J.~A.}\ \bibnamefont {Mydosh}},
  \ and\ \bibinfo {author} {\bibfnamefont {D.~E.}\ \bibnamefont
  {MacLaughlin}},\ }\href {\doibase 10.1103/PhysRevLett.65.2816} {\bibfield
  {journal} {\bibinfo  {journal} {Phys. Rev. Lett.}\ }\textbf {\bibinfo
  {volume} {65}},\ \bibinfo {pages} {2816} (\bibinfo {year}
  {1990})}\BibitemShut {NoStop}%
\bibitem [{\citenamefont {Kromer}\ \emph {et~al.}(1998)\citenamefont {Kromer},
  \citenamefont {Helfrich}, \citenamefont {Lang}, \citenamefont {Steglich},
  \citenamefont {Langhammer}, \citenamefont {Bach}, \citenamefont {Michels},
  \citenamefont {Kim},\ and\ \citenamefont {Stewart}}]{Kromer1998}%
  \BibitemOpen
  \bibfield  {author} {\bibinfo {author} {\bibfnamefont {F.}~\bibnamefont
  {Kromer}}, \bibinfo {author} {\bibfnamefont {R.}~\bibnamefont {Helfrich}},
  \bibinfo {author} {\bibfnamefont {M.}~\bibnamefont {Lang}}, \bibinfo {author}
  {\bibfnamefont {F.}~\bibnamefont {Steglich}}, \bibinfo {author}
  {\bibfnamefont {C.}~\bibnamefont {Langhammer}}, \bibinfo {author}
  {\bibfnamefont {A.}~\bibnamefont {Bach}}, \bibinfo {author} {\bibfnamefont
  {T.}~\bibnamefont {Michels}}, \bibinfo {author} {\bibfnamefont {J.~S.}\
  \bibnamefont {Kim}}, \ and\ \bibinfo {author} {\bibfnamefont {G.~R.}\
  \bibnamefont {Stewart}},\ }\href {\doibase 10.1103/PhysRevLett.81.4476}
  {\bibfield  {journal} {\bibinfo  {journal} {Phys. Rev. Lett.}\ }\textbf
  {\bibinfo {volume} {81}},\ \bibinfo {pages} {4476} (\bibinfo {year}
  {1998})}\BibitemShut {NoStop}%
\bibitem [{\citenamefont {Takegahara}\ and\ \citenamefont
  {Harima}(2000)}]{Takegahara2000}%
  \BibitemOpen
  \bibfield  {author} {\bibinfo {author} {\bibfnamefont {K.}~\bibnamefont
  {Takegahara}}\ and\ \bibinfo {author} {\bibfnamefont {H.}~\bibnamefont
  {Harima}},\ }\href {\doibase https://doi.org/10.1016/S0921-4526(99)01132-1}
  {\bibfield  {journal} {\bibinfo  {journal} {Physica B}\ }\textbf {\bibinfo
  {volume} {281-282}},\ \bibinfo {pages} {764 } (\bibinfo {year}
  {2000})}\BibitemShut {NoStop}%
\bibitem [{\citenamefont {Maehira}\ \emph {et~al.}(2002)\citenamefont
  {Maehira}, \citenamefont {Higashiya}, \citenamefont {Higuchi}, \citenamefont
  {Yasuhara},\ and\ \citenamefont {Hasegawa}}]{Maehira2002}%
  \BibitemOpen
  \bibfield  {author} {\bibinfo {author} {\bibfnamefont {T.}~\bibnamefont
  {Maehira}}, \bibinfo {author} {\bibfnamefont {A.}~\bibnamefont {Higashiya}},
  \bibinfo {author} {\bibfnamefont {M.}~\bibnamefont {Higuchi}}, \bibinfo
  {author} {\bibfnamefont {H.}~\bibnamefont {Yasuhara}}, \ and\ \bibinfo
  {author} {\bibfnamefont {A.}~\bibnamefont {Hasegawa}},\ }\href {\doibase
  https://doi.org/10.1016/S0921-4526(01)01073-0} {\bibfield  {journal}
  {\bibinfo  {journal} {Physica B}\ }\textbf {\bibinfo {volume} {312-313}},\
  \bibinfo {pages} {103} (\bibinfo {year} {2002})}\BibitemShut {NoStop}%
\bibitem [{\citenamefont {Shimizu}\ \emph {et~al.}(2017)\citenamefont
  {Shimizu}, \citenamefont {Kittaka}, \citenamefont {Nakamura}, \citenamefont
  {Sakakibara}, \citenamefont {Aoki}, \citenamefont {Homma}, \citenamefont
  {Nakamura},\ and\ \citenamefont {Machida}}]{Shimizu2017}%
  \BibitemOpen
  \bibfield  {author} {\bibinfo {author} {\bibfnamefont {Y.}~\bibnamefont
  {Shimizu}}, \bibinfo {author} {\bibfnamefont {S.}~\bibnamefont {Kittaka}},
  \bibinfo {author} {\bibfnamefont {S.}~\bibnamefont {Nakamura}}, \bibinfo
  {author} {\bibfnamefont {T.}~\bibnamefont {Sakakibara}}, \bibinfo {author}
  {\bibfnamefont {D.}~\bibnamefont {Aoki}}, \bibinfo {author} {\bibfnamefont
  {Y.}~\bibnamefont {Homma}}, \bibinfo {author} {\bibfnamefont
  {A.}~\bibnamefont {Nakamura}}, \ and\ \bibinfo {author} {\bibfnamefont
  {K.}~\bibnamefont {Machida}},\ }\href {\doibase 10.1103/PhysRevB.96.100505}
  {\bibfield  {journal} {\bibinfo  {journal} {Phys. Rev. B}\ }\textbf {\bibinfo
  {volume} {96}},\ \bibinfo {pages} {100505} (\bibinfo {year}
  {2017})}\BibitemShut {NoStop}%
\bibitem [{\citenamefont {Machida}(2018)}]{Machida2018}%
  \BibitemOpen
  \bibfield  {author} {\bibinfo {author} {\bibfnamefont {K.}~\bibnamefont
  {Machida}},\ }\href {\doibase 10.7566/JPSJ.87.033703} {\bibfield  {journal}
  {\bibinfo  {journal} {J. Phys. Soc. Jpn.}\ }\textbf {\bibinfo {volume}
  {87}},\ \bibinfo {pages} {033703} (\bibinfo {year} {2018})}\BibitemShut
  {NoStop}%
\bibitem [{\citenamefont {Maple}\ \emph {et~al.}(2006)\citenamefont {Maple},
  \citenamefont {Frederick}, \citenamefont {Ho}, \citenamefont {Yuhasz},\ and\
  \citenamefont {Yanagisawa}}]{Maple2006}%
  \BibitemOpen
  \bibfield  {author} {\bibinfo {author} {\bibfnamefont {M.~B.}\ \bibnamefont
  {Maple}}, \bibinfo {author} {\bibfnamefont {N.~A.}\ \bibnamefont
  {Frederick}}, \bibinfo {author} {\bibfnamefont {P.-C.}\ \bibnamefont {Ho}},
  \bibinfo {author} {\bibfnamefont {W.~M.}\ \bibnamefont {Yuhasz}}, \ and\
  \bibinfo {author} {\bibfnamefont {T.}~\bibnamefont {Yanagisawa}},\ }\href
  {\doibase 10.1007/s10948-006-0165-8} {\bibfield  {journal} {\bibinfo
  {journal} {J. Supercond. Novel Magn.}\ }\textbf {\bibinfo {volume} {19}},\
  \bibinfo {pages} {299} (\bibinfo {year} {2006})}\BibitemShut {NoStop}%
\bibitem [{\citenamefont {Aoki}\ \emph {et~al.}(2007)\citenamefont {Aoki},
  \citenamefont {Tayama}, \citenamefont {Sakakibara}, \citenamefont {Kuwahara},
  \citenamefont {Iwasa}, \citenamefont {Kohgi}, \citenamefont {Higemoto},
  \citenamefont {MacLaughlin}, \citenamefont {Sugawara},\ and\ \citenamefont
  {Sato}}]{Aoki2007}%
  \BibitemOpen
  \bibfield  {author} {\bibinfo {author} {\bibfnamefont {Y.}~\bibnamefont
  {Aoki}}, \bibinfo {author} {\bibfnamefont {T.}~\bibnamefont {Tayama}},
  \bibinfo {author} {\bibfnamefont {T.}~\bibnamefont {Sakakibara}}, \bibinfo
  {author} {\bibfnamefont {K.}~\bibnamefont {Kuwahara}}, \bibinfo {author}
  {\bibfnamefont {K.}~\bibnamefont {Iwasa}}, \bibinfo {author} {\bibfnamefont
  {M.}~\bibnamefont {Kohgi}}, \bibinfo {author} {\bibfnamefont
  {W.}~\bibnamefont {Higemoto}}, \bibinfo {author} {\bibfnamefont {D.~E.}\
  \bibnamefont {MacLaughlin}}, \bibinfo {author} {\bibfnamefont
  {H.}~\bibnamefont {Sugawara}}, \ and\ \bibinfo {author} {\bibfnamefont
  {H.}~\bibnamefont {Sato}},\ }\href {\doibase 10.1143/JPSJ.76.051006}
  {\bibfield  {journal} {\bibinfo  {journal} {J. Phys. Soc. Jpn.}\ }\textbf
  {\bibinfo {volume} {76}},\ \bibinfo {pages} {051006} (\bibinfo {year}
  {2007})}\BibitemShut {NoStop}%
\bibitem [{\citenamefont {Maple}\ \emph {et~al.}(2002)\citenamefont {Maple},
  \citenamefont {Ho}, \citenamefont {Zapf}, \citenamefont {Frederick},
  \citenamefont {Bauer}, \citenamefont {Yuhasz}, \citenamefont {Woodward},\
  and\ \citenamefont {Lynn}}]{Maple2002}%
  \BibitemOpen
  \bibfield  {author} {\bibinfo {author} {\bibfnamefont {M.~B.}\ \bibnamefont
  {Maple}}, \bibinfo {author} {\bibfnamefont {P.-C.}\ \bibnamefont {Ho}},
  \bibinfo {author} {\bibfnamefont {V.~S.}\ \bibnamefont {Zapf}}, \bibinfo
  {author} {\bibfnamefont {N.~A.}\ \bibnamefont {Frederick}}, \bibinfo {author}
  {\bibfnamefont {E.~D.}\ \bibnamefont {Bauer}}, \bibinfo {author}
  {\bibfnamefont {W.~M.}\ \bibnamefont {Yuhasz}}, \bibinfo {author}
  {\bibfnamefont {F.~M.}\ \bibnamefont {Woodward}}, \ and\ \bibinfo {author}
  {\bibfnamefont {J.~W.}\ \bibnamefont {Lynn}},\ }\href {\doibase
  10.1143/JPSJS.71S.23} {\bibfield  {journal} {\bibinfo  {journal} {J. Phys.
  Soc. Jpn.}\ }\textbf {\bibinfo {volume} {71}},\ \bibinfo {pages} {23}
  (\bibinfo {year} {2002})}\BibitemShut {NoStop}%
\bibitem [{\citenamefont {Vollmer}\ \emph {et~al.}(2003)\citenamefont
  {Vollmer}, \citenamefont {Fai\ss{}t}, \citenamefont {Pfleiderer},
  \citenamefont {v.~L\"ohneysen}, \citenamefont {Bauer}, \citenamefont {Ho},
  \citenamefont {Zapf},\ and\ \citenamefont {Maple}}]{Vollmer2003}%
  \BibitemOpen
  \bibfield  {author} {\bibinfo {author} {\bibfnamefont {R.}~\bibnamefont
  {Vollmer}}, \bibinfo {author} {\bibfnamefont {A.}~\bibnamefont {Fai\ss{}t}},
  \bibinfo {author} {\bibfnamefont {C.}~\bibnamefont {Pfleiderer}}, \bibinfo
  {author} {\bibfnamefont {H.}~\bibnamefont {v.~L\"ohneysen}}, \bibinfo
  {author} {\bibfnamefont {E.~D.}\ \bibnamefont {Bauer}}, \bibinfo {author}
  {\bibfnamefont {P.-C.}\ \bibnamefont {Ho}}, \bibinfo {author} {\bibfnamefont
  {V.}~\bibnamefont {Zapf}}, \ and\ \bibinfo {author} {\bibfnamefont {M.~B.}\
  \bibnamefont {Maple}},\ }\href {\doibase 10.1103/PhysRevLett.90.057001}
  {\bibfield  {journal} {\bibinfo  {journal} {Phys. Rev. Lett.}\ }\textbf
  {\bibinfo {volume} {90}},\ \bibinfo {pages} {057001} (\bibinfo {year}
  {2003})}\BibitemShut {NoStop}%
\bibitem [{\citenamefont {Izawa}\ \emph {et~al.}(2003)\citenamefont {Izawa},
  \citenamefont {Nakajima}, \citenamefont {Goryo}, \citenamefont {Matsuda},
  \citenamefont {Osaki}, \citenamefont {Sugawara}, \citenamefont {Sato},
  \citenamefont {Thalmeier},\ and\ \citenamefont {Maki}}]{Izawa2003}%
  \BibitemOpen
  \bibfield  {author} {\bibinfo {author} {\bibfnamefont {K.}~\bibnamefont
  {Izawa}}, \bibinfo {author} {\bibfnamefont {Y.}~\bibnamefont {Nakajima}},
  \bibinfo {author} {\bibfnamefont {J.}~\bibnamefont {Goryo}}, \bibinfo
  {author} {\bibfnamefont {Y.}~\bibnamefont {Matsuda}}, \bibinfo {author}
  {\bibfnamefont {S.}~\bibnamefont {Osaki}}, \bibinfo {author} {\bibfnamefont
  {H.}~\bibnamefont {Sugawara}}, \bibinfo {author} {\bibfnamefont
  {H.}~\bibnamefont {Sato}}, \bibinfo {author} {\bibfnamefont {P.}~\bibnamefont
  {Thalmeier}}, \ and\ \bibinfo {author} {\bibfnamefont {K.}~\bibnamefont
  {Maki}},\ }\href {\doibase 10.1103/PhysRevLett.90.117001} {\bibfield
  {journal} {\bibinfo  {journal} {Phys. Rev. Lett.}\ }\textbf {\bibinfo
  {volume} {90}},\ \bibinfo {pages} {117001} (\bibinfo {year}
  {2003})}\BibitemShut {NoStop}%
\bibitem [{\citenamefont {Katayama}\ \emph {et~al.}(2007)\citenamefont
  {Katayama}, \citenamefont {Kawasaki}, \citenamefont {Nishiyama},
  \citenamefont {Sugawara}, \citenamefont {Kikuchi}, \citenamefont {Sato},\
  and\ \citenamefont {Zheng}}]{Katayama2007}%
  \BibitemOpen
  \bibfield  {author} {\bibinfo {author} {\bibfnamefont {K.}~\bibnamefont
  {Katayama}}, \bibinfo {author} {\bibfnamefont {S.}~\bibnamefont {Kawasaki}},
  \bibinfo {author} {\bibfnamefont {M.}~\bibnamefont {Nishiyama}}, \bibinfo
  {author} {\bibfnamefont {H.}~\bibnamefont {Sugawara}}, \bibinfo {author}
  {\bibfnamefont {D.}~\bibnamefont {Kikuchi}}, \bibinfo {author} {\bibfnamefont
  {H.}~\bibnamefont {Sato}}, \ and\ \bibinfo {author} {\bibfnamefont {G.-q.}\
  \bibnamefont {Zheng}},\ }\href {\doibase 10.1143/JPSJ.76.023701} {\bibfield
  {journal} {\bibinfo  {journal} {J Phys. Soc. Jpn.}\ }\textbf {\bibinfo
  {volume} {76}},\ \bibinfo {pages} {023701} (\bibinfo {year}
  {2007})}\BibitemShut {NoStop}%
\bibitem [{\citenamefont {Chia}\ \emph {et~al.}(2003)\citenamefont {Chia},
  \citenamefont {Salamon}, \citenamefont {Sugawara},\ and\ \citenamefont
  {Sato}}]{Chia2003}%
  \BibitemOpen
  \bibfield  {author} {\bibinfo {author} {\bibfnamefont {E.~E.~M.}\
  \bibnamefont {Chia}}, \bibinfo {author} {\bibfnamefont {M.~B.}\ \bibnamefont
  {Salamon}}, \bibinfo {author} {\bibfnamefont {H.}~\bibnamefont {Sugawara}}, \
  and\ \bibinfo {author} {\bibfnamefont {H.}~\bibnamefont {Sato}},\ }\href
  {\doibase 10.1103/PhysRevLett.91.247003} {\bibfield  {journal} {\bibinfo
  {journal} {Phys. Rev. Lett.}\ }\textbf {\bibinfo {volume} {91}},\ \bibinfo
  {pages} {247003} (\bibinfo {year} {2003})}\BibitemShut {NoStop}%
\bibitem [{\citenamefont {Seyfarth}\ \emph {et~al.}(2006)\citenamefont
  {Seyfarth}, \citenamefont {Brison}, \citenamefont {M\'easson}, \citenamefont
  {Braithwaite}, \citenamefont {Lapertot},\ and\ \citenamefont
  {Flouquet}}]{Seyfarth2006}%
  \BibitemOpen
  \bibfield  {author} {\bibinfo {author} {\bibfnamefont {G.}~\bibnamefont
  {Seyfarth}}, \bibinfo {author} {\bibfnamefont {J.~P.}\ \bibnamefont
  {Brison}}, \bibinfo {author} {\bibfnamefont {M.-A.}\ \bibnamefont
  {M\'easson}}, \bibinfo {author} {\bibfnamefont {D.}~\bibnamefont
  {Braithwaite}}, \bibinfo {author} {\bibfnamefont {G.}~\bibnamefont
  {Lapertot}}, \ and\ \bibinfo {author} {\bibfnamefont {J.}~\bibnamefont
  {Flouquet}},\ }\href {\doibase 10.1103/PhysRevLett.97.236403} {\bibfield
  {journal} {\bibinfo  {journal} {Phys. Rev. Lett.}\ }\textbf {\bibinfo
  {volume} {97}},\ \bibinfo {pages} {236403} (\bibinfo {year}
  {2006})}\BibitemShut {NoStop}%
\bibitem [{\citenamefont {MacLaughlin}\ \emph {et~al.}(2002)\citenamefont
  {MacLaughlin}, \citenamefont {Sonier}, \citenamefont {Heffner}, \citenamefont
  {Bernal}, \citenamefont {Young}, \citenamefont {Rose}, \citenamefont
  {Morris}, \citenamefont {Bauer}, \citenamefont {Do},\ and\ \citenamefont
  {Maple}}]{MacLaughlin2002}%
  \BibitemOpen
  \bibfield  {author} {\bibinfo {author} {\bibfnamefont {D.~E.}\ \bibnamefont
  {MacLaughlin}}, \bibinfo {author} {\bibfnamefont {J.~E.}\ \bibnamefont
  {Sonier}}, \bibinfo {author} {\bibfnamefont {R.~H.}\ \bibnamefont {Heffner}},
  \bibinfo {author} {\bibfnamefont {O.~O.}\ \bibnamefont {Bernal}}, \bibinfo
  {author} {\bibfnamefont {B.-L.}\ \bibnamefont {Young}}, \bibinfo {author}
  {\bibfnamefont {M.~S.}\ \bibnamefont {Rose}}, \bibinfo {author}
  {\bibfnamefont {G.~D.}\ \bibnamefont {Morris}}, \bibinfo {author}
  {\bibfnamefont {E.~D.}\ \bibnamefont {Bauer}}, \bibinfo {author}
  {\bibfnamefont {T.~D.}\ \bibnamefont {Do}}, \ and\ \bibinfo {author}
  {\bibfnamefont {M.~B.}\ \bibnamefont {Maple}},\ }\href {\doibase
  10.1103/PhysRevLett.89.157001} {\bibfield  {journal} {\bibinfo  {journal}
  {Phys. Rev. Lett.}\ }\textbf {\bibinfo {volume} {89}},\ \bibinfo {pages}
  {157001} (\bibinfo {year} {2002})}\BibitemShut {NoStop}%
\bibitem [{\citenamefont {Kotegawa}\ \emph {et~al.}(2003)\citenamefont
  {Kotegawa}, \citenamefont {Yogi}, \citenamefont {Imamura}, \citenamefont
  {Kawasaki}, \citenamefont {Zheng}, \citenamefont {Kitaoka}, \citenamefont
  {Ohsaki}, \citenamefont {Sugawara}, \citenamefont {Aoki},\ and\ \citenamefont
  {Sato}}]{Kotegawa2003}%
  \BibitemOpen
  \bibfield  {author} {\bibinfo {author} {\bibfnamefont {H.}~\bibnamefont
  {Kotegawa}}, \bibinfo {author} {\bibfnamefont {M.}~\bibnamefont {Yogi}},
  \bibinfo {author} {\bibfnamefont {Y.}~\bibnamefont {Imamura}}, \bibinfo
  {author} {\bibfnamefont {Y.}~\bibnamefont {Kawasaki}}, \bibinfo {author}
  {\bibfnamefont {G.-q.}\ \bibnamefont {Zheng}}, \bibinfo {author}
  {\bibfnamefont {Y.}~\bibnamefont {Kitaoka}}, \bibinfo {author} {\bibfnamefont
  {S.}~\bibnamefont {Ohsaki}}, \bibinfo {author} {\bibfnamefont
  {H.}~\bibnamefont {Sugawara}}, \bibinfo {author} {\bibfnamefont
  {Y.}~\bibnamefont {Aoki}}, \ and\ \bibinfo {author} {\bibfnamefont
  {H.}~\bibnamefont {Sato}},\ }\href {\doibase 10.1103/PhysRevLett.90.027001}
  {\bibfield  {journal} {\bibinfo  {journal} {Phys. Rev. Lett.}\ }\textbf
  {\bibinfo {volume} {90}},\ \bibinfo {pages} {027001} (\bibinfo {year}
  {2003})}\BibitemShut {NoStop}%
\bibitem [{\citenamefont {Aoki}\ \emph {et~al.}(2003)\citenamefont {Aoki},
  \citenamefont {Tsuchiya}, \citenamefont {Kanayama}, \citenamefont {Saha},
  \citenamefont {Sugawara}, \citenamefont {Sato}, \citenamefont {Higemoto},
  \citenamefont {Koda}, \citenamefont {Ohishi}, \citenamefont {Nishiyama},\
  and\ \citenamefont {Kadono}}]{Aoki2003}%
  \BibitemOpen
  \bibfield  {author} {\bibinfo {author} {\bibfnamefont {Y.}~\bibnamefont
  {Aoki}}, \bibinfo {author} {\bibfnamefont {A.}~\bibnamefont {Tsuchiya}},
  \bibinfo {author} {\bibfnamefont {T.}~\bibnamefont {Kanayama}}, \bibinfo
  {author} {\bibfnamefont {S.~R.}\ \bibnamefont {Saha}}, \bibinfo {author}
  {\bibfnamefont {H.}~\bibnamefont {Sugawara}}, \bibinfo {author}
  {\bibfnamefont {H.}~\bibnamefont {Sato}}, \bibinfo {author} {\bibfnamefont
  {W.}~\bibnamefont {Higemoto}}, \bibinfo {author} {\bibfnamefont
  {A.}~\bibnamefont {Koda}}, \bibinfo {author} {\bibfnamefont {K.}~\bibnamefont
  {Ohishi}}, \bibinfo {author} {\bibfnamefont {K.}~\bibnamefont {Nishiyama}}, \
  and\ \bibinfo {author} {\bibfnamefont {R.}~\bibnamefont {Kadono}},\ }\href
  {\doibase 10.1103/PhysRevLett.91.067003} {\bibfield  {journal} {\bibinfo
  {journal} {Phys. Rev. Lett.}\ }\textbf {\bibinfo {volume} {91}},\ \bibinfo
  {pages} {067003} (\bibinfo {year} {2003})}\BibitemShut {NoStop}%
\bibitem [{\citenamefont {Levenson-Falk}\ \emph {et~al.}()\citenamefont
  {Levenson-Falk}, \citenamefont {Schemm}, \citenamefont {Maple},\ and\
  \citenamefont {Kapitulnik}}]{Levenson2016_arXiv}%
  \BibitemOpen
  \bibfield  {author} {\bibinfo {author} {\bibfnamefont {E.~M.}\ \bibnamefont
  {Levenson-Falk}}, \bibinfo {author} {\bibfnamefont {E.~R.}\ \bibnamefont
  {Schemm}}, \bibinfo {author} {\bibfnamefont {M.~B.}\ \bibnamefont {Maple}}, \
  and\ \bibinfo {author} {\bibfnamefont {A.}~\bibnamefont {Kapitulnik}},\
  }\Eprint {http://arxiv.org/abs/1609.07535} {arXiv:1609.07535} \BibitemShut
  {NoStop}%
\bibitem [{\citenamefont {Sugawara}\ \emph {et~al.}(2002)\citenamefont
  {Sugawara}, \citenamefont {Osaki}, \citenamefont {Saha}, \citenamefont
  {Aoki}, \citenamefont {Sato}, \citenamefont {Inada}, \citenamefont
  {Shishido}, \citenamefont {Settai}, \citenamefont {\={O}nuki}, \citenamefont
  {Harima},\ and\ \citenamefont {Oikawa}}]{Sugawara2002}%
  \BibitemOpen
  \bibfield  {author} {\bibinfo {author} {\bibfnamefont {H.}~\bibnamefont
  {Sugawara}}, \bibinfo {author} {\bibfnamefont {S.}~\bibnamefont {Osaki}},
  \bibinfo {author} {\bibfnamefont {S.~R.}\ \bibnamefont {Saha}}, \bibinfo
  {author} {\bibfnamefont {Y.}~\bibnamefont {Aoki}}, \bibinfo {author}
  {\bibfnamefont {H.}~\bibnamefont {Sato}}, \bibinfo {author} {\bibfnamefont
  {Y.}~\bibnamefont {Inada}}, \bibinfo {author} {\bibfnamefont
  {H.}~\bibnamefont {Shishido}}, \bibinfo {author} {\bibfnamefont
  {R.}~\bibnamefont {Settai}}, \bibinfo {author} {\bibfnamefont
  {Y.}~\bibnamefont {\={O}nuki}}, \bibinfo {author} {\bibfnamefont
  {H.}~\bibnamefont {Harima}}, \ and\ \bibinfo {author} {\bibfnamefont
  {K.}~\bibnamefont {Oikawa}},\ }\href {\doibase 10.1103/PhysRevB.66.220504}
  {\bibfield  {journal} {\bibinfo  {journal} {Phys. Rev. B}\ }\textbf {\bibinfo
  {volume} {66}},\ \bibinfo {pages} {220504} (\bibinfo {year}
  {2002})}\BibitemShut {NoStop}%
\bibitem [{\citenamefont {Sergienko}\ and\ \citenamefont
  {Curnoe}(2004)}]{Sergienko2004}%
  \BibitemOpen
  \bibfield  {author} {\bibinfo {author} {\bibfnamefont {I.~A.}\ \bibnamefont
  {Sergienko}}\ and\ \bibinfo {author} {\bibfnamefont {S.~H.}\ \bibnamefont
  {Curnoe}},\ }\href {\doibase 10.1103/PhysRevB.70.144522} {\bibfield
  {journal} {\bibinfo  {journal} {Phys. Rev. B}\ }\textbf {\bibinfo {volume}
  {70}},\ \bibinfo {pages} {144522} (\bibinfo {year} {2004})}\BibitemShut
  {NoStop}%
\bibitem [{\citenamefont {Maki}\ \emph {et~al.}(2003)\citenamefont {Maki},
  \citenamefont {Won}, \citenamefont {Thalmeier}, \citenamefont {Yuan},
  \citenamefont {Izawa},\ and\ \citenamefont {Matsuda}}]{Maki2003}%
  \BibitemOpen
  \bibfield  {author} {\bibinfo {author} {\bibfnamefont {K.}~\bibnamefont
  {Maki}}, \bibinfo {author} {\bibfnamefont {H.}~\bibnamefont {Won}}, \bibinfo
  {author} {\bibfnamefont {P.}~\bibnamefont {Thalmeier}}, \bibinfo {author}
  {\bibfnamefont {Q.}~\bibnamefont {Yuan}}, \bibinfo {author} {\bibfnamefont
  {K.}~\bibnamefont {Izawa}}, \ and\ \bibinfo {author} {\bibfnamefont
  {Y.}~\bibnamefont {Matsuda}},\ }\href
  {http://stacks.iop.org/0295-5075/64/i=4/a=496} {\bibfield  {journal}
  {\bibinfo  {journal} {Europhys. Lett.}\ }\textbf {\bibinfo {volume} {64}},\
  \bibinfo {pages} {496} (\bibinfo {year} {2003})}\BibitemShut {NoStop}%
\bibitem [{\citenamefont {Maki}\ \emph {et~al.}(2004)\citenamefont {Maki},
  \citenamefont {Haas}, \citenamefont {Parker}, \citenamefont {Won},
  \citenamefont {Izawa},\ and\ \citenamefont {Matsuda}}]{Maki2004}%
  \BibitemOpen
  \bibfield  {author} {\bibinfo {author} {\bibfnamefont {K.}~\bibnamefont
  {Maki}}, \bibinfo {author} {\bibfnamefont {S.}~\bibnamefont {Haas}}, \bibinfo
  {author} {\bibfnamefont {D.}~\bibnamefont {Parker}}, \bibinfo {author}
  {\bibfnamefont {H.}~\bibnamefont {Won}}, \bibinfo {author} {\bibfnamefont
  {K.}~\bibnamefont {Izawa}}, \ and\ \bibinfo {author} {\bibfnamefont
  {Y.}~\bibnamefont {Matsuda}},\ }\href
  {http://stacks.iop.org/0295-5075/68/i=5/a=720} {\bibfield  {journal}
  {\bibinfo  {journal} {Europhys. Lett.}\ }\textbf {\bibinfo {volume} {68}},\
  \bibinfo {pages} {720} (\bibinfo {year} {2004})}\BibitemShut {NoStop}%
\bibitem [{\citenamefont {Goryo}(2003)}]{Goryo2003}%
  \BibitemOpen
  \bibfield  {author} {\bibinfo {author} {\bibfnamefont {J.}~\bibnamefont
  {Goryo}},\ }\href {\doibase 10.1103/PhysRevB.67.184511} {\bibfield  {journal}
  {\bibinfo  {journal} {Phys. Rev. B}\ }\textbf {\bibinfo {volume} {67}},\
  \bibinfo {pages} {184511} (\bibinfo {year} {2003})}\BibitemShut {NoStop}%
\bibitem [{\citenamefont {Miyake}\ \emph {et~al.}(2003)\citenamefont {Miyake},
  \citenamefont {Kohno},\ and\ \citenamefont {Harima}}]{Miyake2003}%
  \BibitemOpen
  \bibfield  {author} {\bibinfo {author} {\bibfnamefont {K.}~\bibnamefont
  {Miyake}}, \bibinfo {author} {\bibfnamefont {H.}~\bibnamefont {Kohno}}, \
  and\ \bibinfo {author} {\bibfnamefont {H.}~\bibnamefont {Harima}},\ }\href
  {http://stacks.iop.org/0953-8984/15/i=19/a=102} {\bibfield  {journal}
  {\bibinfo  {journal} {J. Phys.: Condens. Matter}\ }\textbf {\bibinfo {volume}
  {15}},\ \bibinfo {pages} {L275} (\bibinfo {year} {2003})}\BibitemShut
  {NoStop}%
\bibitem [{\citenamefont {Ichioka}\ \emph {et~al.}(2003)\citenamefont
  {Ichioka}, \citenamefont {Nakai},\ and\ \citenamefont
  {Machida}}]{Ichioka2003}%
  \BibitemOpen
  \bibfield  {author} {\bibinfo {author} {\bibfnamefont {M.}~\bibnamefont
  {Ichioka}}, \bibinfo {author} {\bibfnamefont {N.}~\bibnamefont {Nakai}}, \
  and\ \bibinfo {author} {\bibfnamefont {K.}~\bibnamefont {Machida}},\ }\href
  {\doibase 10.1143/JPSJ.72.1322} {\bibfield  {journal} {\bibinfo  {journal}
  {J. Phys. Soc. Jpn.}\ }\textbf {\bibinfo {volume} {72}},\ \bibinfo {pages}
  {1322} (\bibinfo {year} {2003})}\BibitemShut {NoStop}%
\bibitem [{\citenamefont {Chadov}\ \emph {et~al.}(2010)\citenamefont {Chadov},
  \citenamefont {Qi}, \citenamefont {K\"{u}bler}, \citenamefont {Fecher},
  \citenamefont {Felser},\ and\ \citenamefont {Zhang}}]{Chadov2010}%
  \BibitemOpen
  \bibfield  {author} {\bibinfo {author} {\bibfnamefont {S.}~\bibnamefont
  {Chadov}}, \bibinfo {author} {\bibfnamefont {X.}~\bibnamefont {Qi}}, \bibinfo
  {author} {\bibfnamefont {J.}~\bibnamefont {K\"{u}bler}}, \bibinfo {author}
  {\bibfnamefont {G.~H.}\ \bibnamefont {Fecher}}, \bibinfo {author}
  {\bibfnamefont {C.}~\bibnamefont {Felser}}, \ and\ \bibinfo {author}
  {\bibfnamefont {S.~C.}\ \bibnamefont {Zhang}},\ }\href@noop {} {\bibfield
  {journal} {\bibinfo  {journal} {Nat. Mater.}\ }\textbf {\bibinfo {volume}
  {9}},\ \bibinfo {pages} {541} (\bibinfo {year} {2010})}\BibitemShut {NoStop}%
\bibitem [{\citenamefont {Lin}\ \emph {et~al.}(2010)\citenamefont {Lin},
  \citenamefont {Wray}, \citenamefont {Xia}, \citenamefont {Xu}, \citenamefont
  {Jia}, \citenamefont {Cava}, \citenamefont {Bansil},\ and\ \citenamefont
  {Hasan}}]{Lin2010}%
  \BibitemOpen
  \bibfield  {author} {\bibinfo {author} {\bibfnamefont {H.}~\bibnamefont
  {Lin}}, \bibinfo {author} {\bibfnamefont {L.~A.}\ \bibnamefont {Wray}},
  \bibinfo {author} {\bibfnamefont {Y.}~\bibnamefont {Xia}}, \bibinfo {author}
  {\bibfnamefont {S.}~\bibnamefont {Xu}}, \bibinfo {author} {\bibfnamefont
  {S.}~\bibnamefont {Jia}}, \bibinfo {author} {\bibfnamefont {R.~J.}\
  \bibnamefont {Cava}}, \bibinfo {author} {\bibfnamefont {A.}~\bibnamefont
  {Bansil}}, \ and\ \bibinfo {author} {\bibfnamefont {M.~Z.}\ \bibnamefont
  {Hasan}},\ }\href@noop {} {\bibfield  {journal} {\bibinfo  {journal} {Nat.
  Mater.}\ }\textbf {\bibinfo {volume} {9}},\ \bibinfo {pages} {546} (\bibinfo
  {year} {2010})}\BibitemShut {NoStop}%
\bibitem [{\citenamefont {Xiao}\ \emph {et~al.}(2010)\citenamefont {Xiao},
  \citenamefont {Yao}, \citenamefont {Feng}, \citenamefont {Wen}, \citenamefont
  {Zhu}, \citenamefont {Chen}, \citenamefont {Stocks},\ and\ \citenamefont
  {Zhang}}]{Xiao2010}%
  \BibitemOpen
  \bibfield  {author} {\bibinfo {author} {\bibfnamefont {D.}~\bibnamefont
  {Xiao}}, \bibinfo {author} {\bibfnamefont {Y.}~\bibnamefont {Yao}}, \bibinfo
  {author} {\bibfnamefont {W.}~\bibnamefont {Feng}}, \bibinfo {author}
  {\bibfnamefont {J.}~\bibnamefont {Wen}}, \bibinfo {author} {\bibfnamefont
  {W.}~\bibnamefont {Zhu}}, \bibinfo {author} {\bibfnamefont {X.-Q.}\
  \bibnamefont {Chen}}, \bibinfo {author} {\bibfnamefont {G.~M.}\ \bibnamefont
  {Stocks}}, \ and\ \bibinfo {author} {\bibfnamefont {Z.}~\bibnamefont
  {Zhang}},\ }\href {\doibase 10.1103/PhysRevLett.105.096404} {\bibfield
  {journal} {\bibinfo  {journal} {Phys. Rev. Lett.}\ }\textbf {\bibinfo
  {volume} {105}},\ \bibinfo {pages} {096404} (\bibinfo {year}
  {2010})}\BibitemShut {NoStop}%
\bibitem [{\citenamefont {Al-Sawai}\ \emph {et~al.}(2010)\citenamefont
  {Al-Sawai}, \citenamefont {Lin}, \citenamefont {Markiewicz}, \citenamefont
  {Wray}, \citenamefont {Xia}, \citenamefont {Xu}, \citenamefont {Hasan},\ and\
  \citenamefont {Bansil}}]{Al-Sawai2010}%
  \BibitemOpen
  \bibfield  {author} {\bibinfo {author} {\bibfnamefont {W.}~\bibnamefont
  {Al-Sawai}}, \bibinfo {author} {\bibfnamefont {H.}~\bibnamefont {Lin}},
  \bibinfo {author} {\bibfnamefont {R.~S.}\ \bibnamefont {Markiewicz}},
  \bibinfo {author} {\bibfnamefont {L.~A.}\ \bibnamefont {Wray}}, \bibinfo
  {author} {\bibfnamefont {Y.}~\bibnamefont {Xia}}, \bibinfo {author}
  {\bibfnamefont {S.-Y.}\ \bibnamefont {Xu}}, \bibinfo {author} {\bibfnamefont
  {M.~Z.}\ \bibnamefont {Hasan}}, \ and\ \bibinfo {author} {\bibfnamefont
  {A.}~\bibnamefont {Bansil}},\ }\href {\doibase 10.1103/PhysRevB.82.125208}
  {\bibfield  {journal} {\bibinfo  {journal} {Phys. Rev. B}\ }\textbf {\bibinfo
  {volume} {82}},\ \bibinfo {pages} {125208} (\bibinfo {year}
  {2010})}\BibitemShut {NoStop}%
\bibitem [{\citenamefont {Brydon}\ \emph {et~al.}(2016)\citenamefont {Brydon},
  \citenamefont {Wang}, \citenamefont {Weinert},\ and\ \citenamefont
  {Agterberg}}]{Brydon2016}%
  \BibitemOpen
  \bibfield  {author} {\bibinfo {author} {\bibfnamefont {P.~M.~R.}\
  \bibnamefont {Brydon}}, \bibinfo {author} {\bibfnamefont {L.}~\bibnamefont
  {Wang}}, \bibinfo {author} {\bibfnamefont {M.}~\bibnamefont {Weinert}}, \
  and\ \bibinfo {author} {\bibfnamefont {D.~F.}\ \bibnamefont {Agterberg}},\
  }\href {\doibase 10.1103/PhysRevLett.116.177001} {\bibfield  {journal}
  {\bibinfo  {journal} {Phys. Rev. Lett.}\ }\textbf {\bibinfo {volume} {116}},\
  \bibinfo {pages} {177001} (\bibinfo {year} {2016})}\BibitemShut {NoStop}%
\bibitem [{\citenamefont {Kim}\ \emph {et~al.}()\citenamefont {Kim},
  \citenamefont {Wang}, \citenamefont {Nakajima}, \citenamefont {Hu},
  \citenamefont {Ziemak}, \citenamefont {Syers}, \citenamefont {Wang},
  \citenamefont {Hodovanets}, \citenamefont {Denlinger}, \citenamefont
  {Brydon}, \citenamefont {Agterberg}, \citenamefont {Tanatar}, \citenamefont
  {Prozorov},\ and\ \citenamefont {Johnpierre}}]{Kim2016_arXiv}%
  \BibitemOpen
  \bibfield  {author} {\bibinfo {author} {\bibfnamefont {H.}~\bibnamefont
  {Kim}}, \bibinfo {author} {\bibfnamefont {K.}~\bibnamefont {Wang}}, \bibinfo
  {author} {\bibfnamefont {Y.}~\bibnamefont {Nakajima}}, \bibinfo {author}
  {\bibfnamefont {R.}~\bibnamefont {Hu}}, \bibinfo {author} {\bibfnamefont
  {S.}~\bibnamefont {Ziemak}}, \bibinfo {author} {\bibfnamefont
  {P.}~\bibnamefont {Syers}}, \bibinfo {author} {\bibfnamefont
  {L.}~\bibnamefont {Wang}}, \bibinfo {author} {\bibfnamefont {H.}~\bibnamefont
  {Hodovanets}}, \bibinfo {author} {\bibfnamefont {J.~D.}\ \bibnamefont
  {Denlinger}}, \bibinfo {author} {\bibfnamefont {P.~M.~R.}\ \bibnamefont
  {Brydon}}, \bibinfo {author} {\bibfnamefont {D.~F.}\ \bibnamefont
  {Agterberg}}, \bibinfo {author} {\bibfnamefont {M.~A.}\ \bibnamefont
  {Tanatar}}, \bibinfo {author} {\bibfnamefont {R.}~\bibnamefont {Prozorov}}, \
  and\ \bibinfo {author} {\bibfnamefont {P.}~\bibnamefont {Johnpierre}},\
  }\Eprint {http://arxiv.org/abs/1603.03375} {arXiv:1603.03375} \BibitemShut
  {NoStop}%
\bibitem [{\citenamefont {Venderbos}\ \emph {et~al.}(2018)\citenamefont
  {Venderbos}, \citenamefont {Savary}, \citenamefont {Ruhman}, \citenamefont
  {Lee},\ and\ \citenamefont {Fu}}]{Venderbos2018}%
  \BibitemOpen
  \bibfield  {author} {\bibinfo {author} {\bibfnamefont {J.~W.~F.}\
  \bibnamefont {Venderbos}}, \bibinfo {author} {\bibfnamefont {L.}~\bibnamefont
  {Savary}}, \bibinfo {author} {\bibfnamefont {J.}~\bibnamefont {Ruhman}},
  \bibinfo {author} {\bibfnamefont {P.~A.}\ \bibnamefont {Lee}}, \ and\
  \bibinfo {author} {\bibfnamefont {L.}~\bibnamefont {Fu}},\ }\href {\doibase
  10.1103/PhysRevX.8.011029} {\bibfield  {journal} {\bibinfo  {journal} {Phys.
  Rev. X}\ }\textbf {\bibinfo {volume} {8}},\ \bibinfo {pages} {011029}
  (\bibinfo {year} {2018})}\BibitemShut {NoStop}%
\bibitem [{\citenamefont {Nomoto}\ \emph {et~al.}(2016)\citenamefont {Nomoto},
  \citenamefont {Hattori},\ and\ \citenamefont {Ikeda}}]{Nomoto2016_PRB}%
  \BibitemOpen
  \bibfield  {author} {\bibinfo {author} {\bibfnamefont {T.}~\bibnamefont
  {Nomoto}}, \bibinfo {author} {\bibfnamefont {K.}~\bibnamefont {Hattori}}, \
  and\ \bibinfo {author} {\bibfnamefont {H.}~\bibnamefont {Ikeda}},\ }\href
  {\doibase 10.1103/PhysRevB.94.174513} {\bibfield  {journal} {\bibinfo
  {journal} {Phys. Rev. B}\ }\textbf {\bibinfo {volume} {94}},\ \bibinfo
  {pages} {174513} (\bibinfo {year} {2016})}\BibitemShut {NoStop}%
\bibitem [{\citenamefont {Agterberg}\ \emph {et~al.}(2017)\citenamefont
  {Agterberg}, \citenamefont {Brydon},\ and\ \citenamefont
  {Timm}}]{Agterberg2017}%
  \BibitemOpen
  \bibfield  {author} {\bibinfo {author} {\bibfnamefont {D.~F.}\ \bibnamefont
  {Agterberg}}, \bibinfo {author} {\bibfnamefont {P.~M.~R.}\ \bibnamefont
  {Brydon}}, \ and\ \bibinfo {author} {\bibfnamefont {C.}~\bibnamefont
  {Timm}},\ }\href {\doibase 10.1103/PhysRevLett.118.127001} {\bibfield
  {journal} {\bibinfo  {journal} {Phys. Rev. Lett.}\ }\textbf {\bibinfo
  {volume} {118}},\ \bibinfo {pages} {127001} (\bibinfo {year}
  {2017})}\BibitemShut {NoStop}%
\bibitem [{\citenamefont {Timm}\ \emph {et~al.}(2017)\citenamefont {Timm},
  \citenamefont {Schnyder}, \citenamefont {Agterberg},\ and\ \citenamefont
  {Brydon}}]{Timm2017}%
  \BibitemOpen
  \bibfield  {author} {\bibinfo {author} {\bibfnamefont {C.}~\bibnamefont
  {Timm}}, \bibinfo {author} {\bibfnamefont {A.~P.}\ \bibnamefont {Schnyder}},
  \bibinfo {author} {\bibfnamefont {D.~F.}\ \bibnamefont {Agterberg}}, \ and\
  \bibinfo {author} {\bibfnamefont {P.~M.~R.}\ \bibnamefont {Brydon}},\ }\href
  {\doibase 10.1103/PhysRevB.96.094526} {\bibfield  {journal} {\bibinfo
  {journal} {Phys. Rev. B}\ }\textbf {\bibinfo {volume} {96}},\ \bibinfo
  {pages} {094526} (\bibinfo {year} {2017})}\BibitemShut {NoStop}%
\bibitem [{\citenamefont {Bzdu\v{s}ek}\ and\ \citenamefont
  {Sigrist}(2017)}]{Bzdusek2017}%
  \BibitemOpen
  \bibfield  {author} {\bibinfo {author} {\bibfnamefont {T.}~\bibnamefont
  {Bzdu\v{s}ek}}\ and\ \bibinfo {author} {\bibfnamefont {M.}~\bibnamefont
  {Sigrist}},\ }\href {\doibase 10.1103/PhysRevB.96.155105} {\bibfield
  {journal} {\bibinfo  {journal} {Phys. Rev. B}\ }\textbf {\bibinfo {volume}
  {96}},\ \bibinfo {pages} {155105} (\bibinfo {year} {2017})}\BibitemShut
  {NoStop}%
\bibitem [{\citenamefont {Savary}\ \emph {et~al.}(2017)\citenamefont {Savary},
  \citenamefont {Ruhman}, \citenamefont {Venderbos}, \citenamefont {Fu},\ and\
  \citenamefont {Lee}}]{Savary2017}%
  \BibitemOpen
  \bibfield  {author} {\bibinfo {author} {\bibfnamefont {L.}~\bibnamefont
  {Savary}}, \bibinfo {author} {\bibfnamefont {J.}~\bibnamefont {Ruhman}},
  \bibinfo {author} {\bibfnamefont {J.~W.~F.}\ \bibnamefont {Venderbos}},
  \bibinfo {author} {\bibfnamefont {L.}~\bibnamefont {Fu}}, \ and\ \bibinfo
  {author} {\bibfnamefont {P.~A.}\ \bibnamefont {Lee}},\ }\href {\doibase
  10.1103/PhysRevB.96.214514} {\bibfield  {journal} {\bibinfo  {journal} {Phys.
  Rev. B}\ }\textbf {\bibinfo {volume} {96}},\ \bibinfo {pages} {214514}
  (\bibinfo {year} {2017})}\BibitemShut {NoStop}%
\bibitem [{\citenamefont {Boettcher}\ and\ \citenamefont
  {Herbut}(2018)}]{Boettcher2018}%
  \BibitemOpen
  \bibfield  {author} {\bibinfo {author} {\bibfnamefont {I.}~\bibnamefont
  {Boettcher}}\ and\ \bibinfo {author} {\bibfnamefont {I.~F.}\ \bibnamefont
  {Herbut}},\ }\href {\doibase 10.1103/PhysRevLett.120.057002} {\bibfield
  {journal} {\bibinfo  {journal} {Phys. Rev. Lett.}\ }\textbf {\bibinfo
  {volume} {120}},\ \bibinfo {pages} {057002} (\bibinfo {year}
  {2018})}\BibitemShut {NoStop}%
\end{thebibliography}

%

\end{document}